\documentclass[11pt,draftclsnofoot,journal,letterpaper,onecolumn]{IEEEtran}
\IEEEoverridecommandlockouts

\usepackage[noadjust]{cite}

\ifCLASSINFOpdf
  \usepackage[pdftex]{graphicx}
\else
 
\fi

\usepackage[cmex10]{amsmath}
\usepackage{amssymb}   
\usepackage{url}

\usepackage[utf8]{inputenc} 
\usepackage[T1]{fontenc}

\usepackage{mleftright}       
\mleftright

\usepackage{booktabs}

\newtheorem{theorem}{Theorem}
\newtheorem{proposition}[theorem]{Proposition}

\newtheorem{definition}[theorem]{Definition}
\newtheorem{corollary}[theorem]{Corollary}
\newtheorem{lemma}[theorem]{Lemma}
\newtheorem{claim}[theorem]{Claim}

\newtheorem{remark}[theorem]{Remark}

\newcommand\independent{\protect\mathpalette{\protect\independenT}{\perp}}
\def\independenT#1#2{\mathrel{\rlap{$#1#2$}\mkern2mu{#1#2}}}

\begin{document}

\title{
Rate-Distortion-Perception Trade-off with Strong Realism Constraints:
Role of
Side Information and Common Randomness\\ 
\thanks{Y. Hamdi and D. G\"{u}nd\"{u}z are with the Department of Electrical and Electronic Engineering, Imperial College London, UK. Emails: \{y.hamdi, d.gunduz\}@imperial.ac.uk. A. B. Wagner is with the School of Electrical and Computer Engineering, Cornell University, USA. Email: wagner@cornell.edu}
\thanks{The present work has received funding from the European Union’s Horizon 2020 Marie Sk\l{}odowska Curie Innovative Training Network Greenedge (GA. No. 953775), from UKRI
for the projects
AIR (ERC-CoG) (EP/X030806/1), INFORMED-AI (EP/Y028732/1), and the SNS JU project 6G-GOALS (Grant Agreement No. 101139232),
and from the US National Science Foundation (Grant CCF-2306228). This paper was presented in part at the 2023 IEEE International Symposium on Information Theory [DOI 10.1109/ISIT54713.2023.10206643]. For the purpose of open access, the authors have applied a Creative Commons Attribution (CCBY) license to any Author Accepted Manuscript version arising from this submission.}
}

\author{%
  \IEEEauthorblockN{Yassine Hamdi,~\IEEEmembership{Graduate Student Member,~IEEE},
  Aaron B. Wagner,~\IEEEmembership{Fellow,~IEEE},\\
  Deniz G\"{u}nd\"{u}z,~\IEEEmembership{Fellow,~IEEE}
  }
}

\maketitle
\vspace{-0.4in}

\begin{abstract}
In image compression, with recent advances in generative modeling,
existence of a trade-off between the rate and
perceptual quality has been brought to light, where the perceptual quality is measured by the closeness of the output
and source distributions.
We consider the compression of a memoryless source sequence $X^n=(X_1, \ldots, X_n)$ in the presence of memoryless side information $Z^n=(Z_1, \ldots, Z_n),$ originally studied by Wyner and Ziv, but elucidate the impact of a \textit{strong perfect realism} constraint, which requires the joint distribution of output symbols $Y^n=(Y_1,...,Y_n)$ to match the distribution of the source sequence. We consider two cases: when $Z^n$ is available only at the decoder, or at both the encoder and decoder,
and characterize the information theoretic limits under various scenarios.
Previous works show the superiority of randomized codes under strong perceptual quality constraints.
When $Z^n$ is available at both terminals, we characterize its dual role, as a source of common randomness, and as a second look on the source
for the receiver.
We also study different notions of strong perfect realism which we call
\textit{marginal realism}, \textit{joint realism} and \textit{near-perfect realism}.
We derive explicit solutions when
$X$ and $Z$ are jointly Gaussian under the squared error distortion measure.
In
traditional lossy compression, having $Z$ only at the decoder imposes no rate penalty in the Gaussian scenario. We show that, when strong perfect realism constraints are imposed this holds only when sufficient common randomness is available.
\end{abstract}

\IEEEpeerreviewmaketitle

\section{Introduction}
\label{sec:intro}

\IEEEPARstart{I}{n} conventional
rate-distortion theory, the goal is to enable the decoder to reconstruct a representation $Y^n\triangleq (Y_1, ..., Y_n)$ of the source signal $X^n=(X_1, ..., X_n)$ that is close to the latter for some distortion measure $d(X^n,Y^n).$ Shannon characterized the optimal rate-distortion trade-off under an additive distortion measure, i.e.,
$d(x^n,y^n) = (1/n)\sum_{t=1}^n d(x_t,y_t)$. In \cite{Wyner&Ziv1977}, Wyner and Ziv generalized this result to the case where a side information $Z^n,$ correlated with $X^n,$ is available either only at the decoder or at both the encoder and the decoder.
While broadly successful (e.g.~\cite{Pearlman:Said,Sayood:Compression}), 
this approach does have certain limitations. One is that the 
reconstruction might be qualitatively quite different from the source
realization that generated it. For an i.i.d.\ Gaussian 
source with mean squared error (MSE) distortion measure,
the reconstruction is generally of lower 
power than the source. For stationary Gaussian sources,
the reverse waterfilling procedure~\cite{Berger:RD} generally
gives rise to reconstructions that have a null power
spectrum at high frequencies. Thus, JPEG images look 
blurry at low bit-rates.

Of course, all distortion measures used in theoretical
studies of multimedia compression are proxies for the
measure of real interest, namely how the reconstruction
would be perceived by the (usually human) end-user.
In some cases, this end-user will prefer a reconstruction 
that is more distorted according to conventional 
measures. For instance, in some cases, MPEG Advanced
Audio Coding (ACC) populates high-frequency bands 
with artificial noise instead of leaving them 
null~\cite[Sec.~17.4.2]{Sayood:Compression}, in order to
match the power spectrum of the source;
this is termed \emph{perceptual noise
substitution (PNS)}.

This general idea has acquired renewed 
interest
since methods based on deep neural networks
have been shown to outperform traditional image and video compression codecs 
\cite{ 2017BalleEndToEndCompression, 2018MentzerDeepImageCompr, ChannelWiseEntropyModelsforLearnedImageCompression2020, KankanahalliEndToEndSpeechCompression2018, KleijnLimWavenetSpeechCoding2018,PetermannBeackHarpNet2021, ZeghidourLuebsSoundStreamEndToEndNeuralAudioCodec2022, RN361, RN363, RN364, RN365, RN367, RN368}
under different distortion measures.
In \cite{2019AgustssonMentzerGANforExtremeCompression}, the authors used generative adversarial networks (GANs) to push the limits of image compression in very low bit-rates by synthesizing image content, such as facades of buildings, using a reference image database. This allows the receiver to generate images that resemble the source image semantically, although they may not match the details, providing visually pleasing reconstructions even at very low bit-rates. It has been observed (\cite{1991MomentPreservingQuantization,2010LiEtAlTheirFirstPaperOnDistributionPreservingQuantization,2011LiEtAlMainPaperOnDistributionPreservingQuantization,2012LiEtAlSpectralDensityPreservingQuantizationForAudio,2013LiEtAlMultipleDescriptionDistributionPreservingQuantization,2018GenerativeCompression,TschannenAgustssonNeurips2018,2019AgustssonMentzerGANforExtremeCompression}) that at such bit-rates the increase in perceptual quality comes at the cost of increased distortion --- see also \cite{2018PerceptionDistortionTradeOff} regarding image restoration ---, and the \textit{rate-distortion-perception} (RDP) trade-off was formalized in \cite{Aug2018MatsumotoRDPDeterministic,Nov2018MatsumotoRDPDeterministic,2019BlauMichaeliRethinkingLossyCompressionTheRDPTradeoff}.
Motivated by successful results in generative modeling, where the generated images
would exhibit the same statistical properties of the images in the dataset,
the formalism of
distribution-preserving compression \cite{1991MomentPreservingQuantization,2010LiEtAlTheirFirstPaperOnDistributionPreservingQuantization,2011LiEtAlMainPaperOnDistributionPreservingQuantization,2012LiEtAlSpectralDensityPreservingQuantizationForAudio,2013LiEtAlMultipleDescriptionDistributionPreservingQuantization}
was adopted
in \cite{TschannenAgustssonNeurips2018}, and extended in \cite{Aug2018MatsumotoRDPDeterministic,Nov2018MatsumotoRDPDeterministic,2019BlauMichaeliRethinkingLossyCompressionTheRDPTradeoff}. The problem is then to characterize the optimal rate for which both distortion constraint $d(X^n, Y^n) \leq \Delta$ and realism constraint $\mathcal{D}(P_{X^n}, P_{Y^n}) \leq \lambda$ are met,
where
$\mathcal{D}$ 
is a measure of discrepancy between distributions,
e.g., the total variation distance (TVD) or some other divergence. Instead of this \textit{strong realism constraint}, a weaker variant \cite{Dec2022WeakAndStrongPerceptionConstraintsAndRandomness} has been recently studied, as well as realization-based constraints \cite{Sep2015RDPLimitedCommonRandomnessSaldi,Dec2022WeakAndStrongPerceptionConstraintsAndRandomness,2023YangQiuAaronBWagnerUnifyingFidelityAndRealism}. In the latter, the proposed parametric measure interpolates between the notions of distortion and realism.\\

In conventional rate-distortion theory, it is known that deterministic encoders and decoders are sufficient to achieve the optimal asymptotic rate-distortion performance for a stationary source --- as well as the optimal performance for one-shot fixed-length codes.
This simplifies both the analysis and implementation of rate-distortion optimal codes. 
However, for the RDP trade-off with a strong realism constraint,
stochastic codes are necessary in general for lossy compression \cite{2011LiEtAlMainPaperOnDistributionPreservingQuantization}.
Moreover, it was shown in \cite{2021TheisAgustssonOnTheAdvantagesOfStochasticEncoders} and \cite{Sep2015RDPLimitedCommonRandomnessSaldi} that when privately randomized decoders are considered, common randomness improves performance, in both the one-shot setting and the infinite blocklength regime.
In the remainder of this paper, we consider the latter regime unless stated otherwise.
Stochasticity can come from private randomness at either terminal, or from a source of randomness common to both the encoder and the decoder.
The role of randomness for certain realism constraints other than strong realism has been determined in \cite{Sep2015RDPLimitedCommonRandomnessSaldi,TheisWagner2021VariableRateRDP,UniversalRDPNeurips2021,2022JunChenKhistiBinarySourcesRDPAndFixedEncoderAndSuccessiveRefinement,Dec2022WeakAndStrongPerceptionConstraintsAndRandomness,SalehkalaibarISIT2024RDPConditioningOnTheMessage}.
The case where unlimited common randomness is available is well-studied, as described in the remainder of this paragraph.
For strong realism, a characterization of the optimal RDP trade-off was proved in \cite{Jan2015SaldiEtAlDistributionPreservationMeasureTheoreticConsiderationsForContinuousAndDiscreteCommonRandomness} in the case of the \textit{perfect realism} constraint $P_{Y^n}\equiv p_X^{\otimes n}$ ($\lambda=0$), for general alphabets. A characterization in the general case ($\lambda \geq 0$) is proved in \cite{Dec2022WeakAndStrongPerceptionConstraintsAndRandomness} for a
class of similarity measures.
With a weaker notion of realism, universal (fixed) encoders and successive refinement are studied in \cite{UniversalRDPNeurips2021}. Closed form expressions for the latter results are derived in \cite{2022JunChenKhistiBinarySourcesRDPAndFixedEncoderAndSuccessiveRefinement} for binary sources and Hamming distortion measure.
One-shot results are also proved in \cite{2013LiEtAlMultipleDescriptionDistributionPreservingQuantization,TheisWagner2021VariableRateRDP,UniversalRDPNeurips2021,2022JunChenKhistiBinarySourcesRDPAndFixedEncoderAndSuccessiveRefinement}, \cite{SalehkalaibarKhistiNeurips2023RDPVideoCompressionChoiceOfPerceptionLoss}.
One-shot multiple description coding is considered in \cite{2013LiEtAlMultipleDescriptionDistributionPreservingQuantization}.
In \cite{SalehkalaibarKhistiNeurips2023RDPVideoCompressionChoiceOfPerceptionLoss}, sequential compression is considered, such as video compression, with MSE distortion measure and a perfect realism constraint either on the marginal distribution of each frame, or the joint distribution of all frames.
The formulation in \cite{Jan2015SaldiEtAlDistributionPreservationMeasureTheoreticConsiderationsForContinuousAndDiscreteCommonRandomness} is fixed rate, and those in \cite{2013LiEtAlMultipleDescriptionDistributionPreservingQuantization,TheisWagner2021VariableRateRDP, UniversalRDPNeurips2021,2022JunChenKhistiBinarySourcesRDPAndFixedEncoderAndSuccessiveRefinement,Dec2022WeakAndStrongPerceptionConstraintsAndRandomness}, \cite{SalehkalaibarKhistiNeurips2023RDPVideoCompressionChoiceOfPerceptionLoss} are variable-rate. 
Note that private sources of randomness are not useful when common randomness is
unconstrained, since private randomness can be extracted from common randomness.
In that case, the operational rate-distortion function for the strong perfect realism constraint equals \cite{Jan2015SaldiEtAlDistributionPreservationMeasureTheoreticConsiderationsForContinuousAndDiscreteCommonRandomness}, \cite{2022AaronWagnerRDPTradeoffTheRoleOfCommonRandomness}
\begin{IEEEeqnarray}{c}
R(\Delta) = \inf_{p_{Y|X}} I(X;Y),
\label{eq:BM}
\end{IEEEeqnarray}
where the infimum is over all conditional distributions $p_{Y|X}$ such that $\mathbb{E}[d(X,Y)] \le \Delta$ and $p_Y \equiv p_X.$
As shown in \cite{Sep2015RDPLimitedCommonRandomnessSaldi}, \cite{2022AaronWagnerRDPTradeoffTheRoleOfCommonRandomness}, it turns out that the above RDP function only applies when common randomness is available at a sufficiently large rate, even 
with unconstrained 
encoder and decoder private randomness.
Thus, some care is required when interpreting (\ref{eq:BM}) operationally.\\

The focus of this paper is on the problem with side information,
as depicted in Fig.~\ref{fig:general_setup}. Here the side
information $Z^n$ may be present at both terminals or at the
decoder only. The problem is of practical interest because it
models the setup in which compression is performed successively,
as in video~\cite{SalehkalaibarKhistiNeurips2023RDPVideoCompressionChoiceOfPerceptionLoss}; here the side information models a
block of data, such as a video frame, that was previously
compressed. The problem is of theoretical interest due
to the dual role that the side information $Z^n$ can play.
On the one hand, it provides the receiever with an 
independent look at the source. On the other hand, it can 
serve as a source of common randomness (as is apparent 
in the extreme case in which $Z^n$ is independent
of the source). 

We prove various coding theorems that elucidate these dual
roles and the tension between them. It turns out that the
ability of $Z^n$ to simultaneously play both roles
depends on the nature of the realism
constraint, specifically whether we require $Y^n \stackrel{d}{=} X^n$ (which we call a \emph{marginal realism constraint})
or $(Y^n,Z^n) \stackrel{d}{=} (X^n,Z^n)$ (which we call
a \emph{joint realism constraint}) --- the latter being similar to a constraint considered in a one-shot video compression setup with fixed encoder in the case of Euclidean distortion measure \cite{SalehkalaibarKhistiNeurips2023RDPVideoCompressionChoiceOfPerceptionLoss}. Under a marginal
realism constraint, when $Z^n$ is present at both terminals,
$Z^n$ can play both roles in a particularly strong sense:
the side information $Z^n$ reduces the rate required
for compression, as in conventional rate-distortion, while
simultaneously providing a common-randomness rate of
$H(Z|Y)$. Conversely, under a joint realism constraint,
the side information does not provide any common-randomness
at all. Under a marginal constraint with side information
available only at the decoder, we provide an inner bound on the achievable region. The challenge in obtaining a conclusive
result in this case seems to center around characterizing
the extent to which $Z^n$ can also serve as common randomness.
As evidence, we provide conclusive results in
two special cases for which the ability of $Z^n$ to provide
useful common randomness is clear. 

We examine the quadratic Gaussian case in detail. We
give an explicit characterization of the optimal rate-distortion
trade-off for a 1-D Gaussian source under the MSE distortion
measure and a marginal realism constraint, for any amount
of common randomness. Previously, only the achievability
half of this result was known~\cite{Sep2015RDPLimitedCommonRandomnessSaldi}. 
When side information $Z^n$ is available and $p_{X,Z}$ is  bivariate Gaussian, we show that for large enough common randomness rate, the rate-distortion trade-off is the same whether $Z^n$ is available only at the encoder or at both terminals.\\

Overall, our contributions are
\begin{itemize}
    \item We characterize the three-way trade-off between compression rate, common randomness rate and distortion in the presence of side information, for strong perfect and near-perfect realism constraints
    (Definition \ref{def:achievability} to follow).
    \item We do so for both the marginal realism and joint realism constraints.
    \item We characterize how the side information plays a dual role as a source of extra common randomness and a second look on the source, when it is available at both terminals, under a marginal realism constraint. We show that this does not hold when the latter is replaced with a joint realism constraint.
    \item We prove the equivalence between near-perfect marginal (resp. joint) realism and perfect marginal (resp. joint) realism.
    \item We propose a proof technique to handle general alphabets.
    \item We provide an explicit solution in the case of a Normal source in the absence of side information.
    \item When the source and side information are jointly Gaussian, we show that the trade-off is the same whether the side information is available at both terminals, or only at the decoder, if additional common randomness is available at a sufficiently large rate.
\end{itemize}

The paper is organized as follows. We give the problem formulation and state the equivalence between perfect and near-perfect realism in Section \ref{sec:problem_formulation_and_equivalence}. We present all other results in Section \ref{sec:main_result}.
We prove our main theorems in Sections \ref{sec:achievability_two_sided}, \ref{sec:achievability}, \ref{sec:achievability_joint_two_sided} and \ref{sec:achievability_joint_one_sided}. Section \ref{sec:special_cases} is dedicated to special cases and corollaries, and Section \ref{sec:gaussian} to the case of Gaussian distributions. Other proofs are deferred to the appendix.

\section{Problem formulation and a first property}\label{sec:problem_formulation_and_equivalence}

\subsection{Notation}
Calligraphic letters such as $\mathcal{X}$ denote sets, except in $p^{\mathcal{U}}_{\mathcal{J}},$ which denotes the uniform distribution over alphabet $\mathcal{J}.$ 
Random variables are denoted using upper case letters such as $X$ and their realizations using lower case letters such as $x.$
For a distribution $P,$ the expression $P_X$ denotes the marginal of variable $X$ while $P_X(x)$ denotes a real number, probability of the event $X=x.$ Similarly, $P_{X|Y{=}y}$ denotes a distribution over $\mathcal{X},$ and $P_{X|Y{=}y}(x)$ the corresponding real number.
We denote by $[a]$ the set $\{1, ..., \lfloor a \rfloor\},$ and by $x^n$ the finite sequence $(x_1, ..., x_n),$
and use $x_{[n] \backslash k}$ to denote $(x_1, ..., x_{k-1}, x_{k+1}, ..., x_n),$ and $x_{a:b}$ to denote $(x_a, ..., x_b).$
For two sequences $(a_n)_{n \geq 1}$ and $(b_n)_{n \geq 1}$ of positive real numbers, the notation $a_n$ $\substack{\raisebox{-4pt}{$\sim$} \\ \scalebox{0.5}{$n$$\to$$\infty$}}$ $b_n$ means that $\lim_{n \rightarrow \infty} \tfrac{a_n}{b_n} = 1.$
We denote by $\|p-q\|_{TV}$ the total variation distance between distributions $p$ and $q.$ The closure of a set is denoted by an over-lined calligraphic letter, e.g.,
$\overline{\mathcal{A}}.$
For random variables $X$ and $Y,$ we use $H(X),$ $ H(X|Y),$ and $I(X;Y)$ to denote the entropy, conditional entropy, and mutual information, respectively, defined for general alphabets as in \cite{2011GrayGeneralAlphabetsRelativeEntropy} (see Appendix \ref{app:subsubsec:info_theory_for_general_spaces}). Logarithms are in base 2.
Given a distribution $P_{X^n}$ on a $n$-fold Cartesian product $\mathcal{X}^n,$ the \textit{average empirical distribution} of random string $X^n,$ denoted by $\hat{P}_{X^n},$ is the distribution on $\mathcal{X}$ defined by: for any measurable $A \subseteq \mathcal{X},$
\begin{equation*}
\hat{P}_{X^n}(A) = \dfrac{1}{n} \sum_{t=1}^n P_{X_t}(A).
\end{equation*}

\subsection{Definitions}

In this section we formulate the information-theoretical problem for general alphabets. To have the existence of conditional distribution given a joint distribution, we assume all alphabets are Polish spaces. This includes discrete spaces and real vector spaces. We always endow a Polish alphabet with the corresponding Borel $\sigma$-algebra. We omit measure-theoretic justifications where reasonable, see Appendix \ref{app:subsec:measure_theory_general} for key basic definitions and results we rely on. In particular, the conditional probability kernels we consider are regular.

\begin{definition}
Given an alphabet $\mathcal{X},$ a distortion measure is a measurable function $d:\mathcal{X}\times\mathcal{X} \to [0,\infty)$ extending to sequences as $$d(x^n, y^n) =
\dfrac{1}{n}
\sum_{t=1}^n
d(x_t,y_t).$$ 
\end{definition}

\begin{figure}[t!]
    \centering\includegraphics[width=0.6\columnwidth]{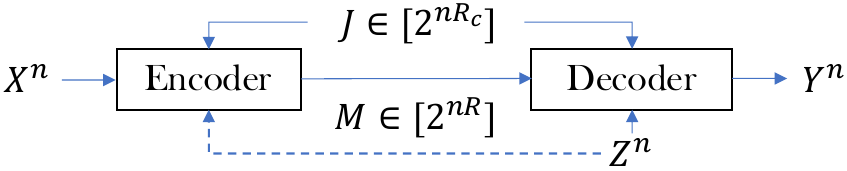}
    \caption{The system model. The side information $Z^n$ is always available at the decoder, but not necessarily at the encoder.}
\label{fig:general_setup}
\end{figure}

As shown in Figure \ref{fig:general_setup}, we consider two cases: side information is either only available at the decoder or available at both the encoder and the decoder. We also study the case where no side information is available at either terminal. Hereafter, we formalize the latter as the special case of constant side information. Throughout this paper, we do not impose any constraint on each terminal's private randomness, and study the role of common randomness.
The role of private randomness in the absence of side information was characterized in \cite{HamdiEtAl2024RDPPrivateRandomness}, where it is shown that encoder private randomness does not impact the optimal asymptotic trade-off, whatever the resources in terms of common randomness and decoder private randomness --- see also \cite{Dec2022WeakAndStrongPerceptionConstraintsAndRandomness} for results on weaker realism constraints.
\begin{definition}\label{def:code}
Given two measurable spaces with respective \textit{source} alphabet $\mathcal{X}$ and \textit{side-information} alphabet $\mathcal{Z},$
a $(n, R, R_c)$ D-code (resp. E-D-code) is a privately randomized encoder and decoder couple $(F^{(n)},G^{(n)})$ consisting of a conditional probability kernel $F^{(n)}_{M|X^n, J}$ (resp. $F^{(n)}_{M|X^n, Z^n, J}$) from $ \mathcal{X}^n \times [2^{nR_c}] $ (resp. $ \mathcal{X}^n \times \mathcal{Z}^n \times [2^{nR_c}] $) to $ [2^{nR}],$ and a conditional probability kernel $G^{(n)}_{Y^n|Z^n, J, M}$ from $ \mathcal{Z}^n \times [2^{nR_c}] \times [2^{nR}] $ to $ \mathcal{X}^n.$
Given a distribution $p_{X,Z}$ on $\mathcal{X}\times \mathcal{Z},$ the following distribution
\begin{IEEEeqnarray}{c}
P_{X^n, Z^n, J, M, Y^n} := p_{X,Z}^{\otimes n} \cdot p^{\mathcal{U}}_{[2^{nR_c}]} \cdot F^{(n)} \cdot G^{(n)}\nonumber
\end{IEEEeqnarray}is called \textit{the distribution induced by the code}. A $(n,R)$ D-code (resp. E-D-code) \textit{with unconstrained common randomness} is a triplet $(p_J,F^{(n)},G^{(n)}):$ the common randomness variable $J$ is allowed to be defined on any alphabet (instead of $[2^{nR_c}]$) --- possibly uncountable --- and to have any distribution $p_J$ (instead of a uniform one).
\end{definition}

The following lemma is a straightforward reformulation.
\begin{lemma}\label{lemma:conditions_for_P_to_define_a_code}
Consider a distribution $P_{X^n, Z^n, J, M, Y^n}$ 
 on $\mathcal{X}^n \times \mathcal{Z}^n \times \mathbb{N} \times \mathbb{N} \times \mathcal{X}^n$ having regular transition kernels $P_{M|X^n,J}$ (resp. $P_{M|X^n,Z^n,J}$) and $P_{Y^n|Z^n,J,M}.$ Suppose that there exist non-negative reals $R$ and $R_c$ such that $M \in [2^{nR}]$ and $J \in [2^{nR_c}],$ $P$-almost surely. Then, $P$ coincides with the distribution induced by $(n,R,R_c)$ D-code $(P_{M|X^n,J},P_{Y^n|Z^n,J,M})$ (resp. E-D code $(P_{M|X^n,Z^n,J},P_{Y^n|Z^n,J,M})$) iff $P$ satisfies all of the following properties:
\begin{itemize}
    \item $P_{X^n,Z^n,J} \equiv p_{X,Z}^{\otimes n} \cdot p^{\mathcal{U}}_{[2^{nR_c}]}$
    \item $X^n-(Z^n,J,M)-Y^n$
    \item (resp. $Z^n-(X^n,J)-M$ and $X^n-(Z^n,J,M)-Y^n$).
\end{itemize}
\end{lemma}
\vspace{10pt}
\begin{definition}\label{def:achievability}
Given source and side-information alphabets $\mathcal{X}$ and $\mathcal{Z}$ and a joint distribution $p_{X,Z}$ on $\mathcal{X} \times \mathcal{Z},$
a triplet $(R, R_c, \Delta)$ is said to be D-achievable (resp. E-D-achievable) with \textit{near-perfect marginal realism} if there exists a sequence $(F^{(n)},G^{(n)})_n$ of $(n, R, R_c)$ D-codes (resp. E-D-codes) such that 
\begin{IEEEeqnarray}{rCl}
&\limsup_{n \to \infty} \mathbb{E}_{P}[d(X^n,Y^n)] \leq \Delta, &\label{eq:def_achievability_distortion_constraint}\\*
&\|P_{Y^n} - p_{X}^{\otimes n}\|_{TV} \underset{n \to \infty}{\longrightarrow} 0.&\label{eq:def_achievability_marginal_perception_constraint}
\end{IEEEeqnarray}
When there exists an integer $N$ such that for all $n\geq N,$ $P_{Y^n} \equiv p_{X}^{\otimes n},$ then $(R, R_c, \Delta)$ is said to be D-achievable (resp. E-D-achievable) with \textit{perfect marginal realism.}
Each of the above notions has an analogue with \textit{joint realism} (instead of marginal realism), which is defined by replacing
$P_{Y^n}$ by $P_{Y^n,Z^n}$ and $p_{X}^{\otimes n}$ by $p_{X,Z}^{\otimes n}.$
Moreover, each notion of achievability extends to the case of large common randomness rate. A couple $(R,\Delta)$ is said to be \textit{achievable with finite-rate common randomness} if there exists a non-negative real $R_c$ such that $(R,R_c,\Delta)$ is achievable. More generally, a couple $(R,\Delta)$ is said to be \textit{achievable with unconstrained common randomness} if the above properties hold for some sequence of $(n,R)$ codes with unconstrained common randomness.
\end{definition}
\vspace{10pt}
\begin{table}[!t]
\caption{Main results}
\label{table:main_results}
\centering
\begin{tabular}{|c|c|c|}
\hline
 & \bfseries E-D-codes & \bfseries D-codes\\
\hline
&
Theorem \ref{theorem:E_D_rates_region_marginal_realism}
& Theorem \ref{theorem:D_rates_region_marginal_realism}\\
&\scalebox{1.0}{$
R$}\scalebox{1.0}{$\, \geq \,$}\scalebox{1.0}{$I_p(X;V|Z)
$}
&\scalebox{1.0}{$R$}\scalebox{1.0}{$\, \geq \,$}\scalebox{1.0}{$I_p(X;V)-I_p(Z;V)$}\\
\bfseries Marginal
&\scalebox{1.0}{$
R+R_c$}\scalebox{1.0}{$\, \geq \,$}\scalebox{1.0}{$I_p(Y;V|Z) - H_p(Z|Y)$}
&\scalebox{1.0}{$R+R_c$}\scalebox{1.0}{$\, \geq \,$}\scalebox{1.0}{$I_p(Y;V)-I_p(Z;V)$}\\
\bfseries realism
&\scalebox{1.0}{$
\Delta$}\scalebox{1.0}{$\, \geq \,$}\scalebox{1.0}{$\mathbb{E}_p[d(X, Y)]
$}
&\scalebox{1.0}{$\Delta$}\scalebox{1.0}{$\, \geq \,$}\scalebox{1.0}{$\mathbb{E}_p[d(X, Y)]$}\\
&\scalebox{1.0}{$
p_{Y}$}\scalebox{1.0}{$\, \equiv \, $}\scalebox{1.0}{$p_{X}
$}
&\scalebox{1.0}{$p_{Y}$}\scalebox{1.0}{$\, \equiv \, $}\scalebox{1.0}{$p_{X}$}\\
&\scalebox{1.0}{$
X - $}\scalebox{1.0}{$(Z, V)$}\scalebox{1.0}{$ - Y
$}
&\scalebox{1.0}{$X - $}\scalebox{1.0}{$(Z, V)$}\scalebox{1.0}{$ - Y$}
\text{and}
\scalebox{1.0}{$Z - $}\scalebox{1.0}{$X$}\scalebox{1.0}{$ - V$}\\
\hline
&
Theorem \ref{theorem:E_D_rates_region_joint_realism}
& Theorem \ref{theorem:D_rates_region_joint_realism}\\
&\scalebox{1.0}{$R$}\scalebox{1.0}{$\, \geq \,$}\scalebox{1.0}{$I_p(X;V|Z)$}
&\scalebox{1.0}{$R$}\scalebox{1.0}{$\, \geq \,$}\scalebox{1.0}{$I_p(X;V|Z)$}\\
\bfseries Joint
&\scalebox{1.0}{$R+R_c$}\scalebox{1.0}{$\, \geq \,$}\scalebox{1.0}{$I_p(Y;V|Z)$}
&\scalebox{1.0}{$R+R_c$}\scalebox{1.0}{$\, \geq \,$}\scalebox{1.0}{$I_p(Y;V|Z)$}\\
\bfseries realism
&\scalebox{1.0}{$\Delta$}\scalebox{1.0}{$\, \geq \,$}\scalebox{1.0}{$\mathbb{E}_p[d(X, Y)]$}
&\scalebox{1.0}{$\Delta$}\scalebox{1.0}{$\, \geq \,$}\scalebox{1.0}{$\mathbb{E}_p[d(X, Y)]$}\\
&\scalebox{1.0}{$p_{Y,Z}$}\scalebox{1.0}{$\, \equiv \, $}\scalebox{1.0}{$p_{X,Z}$}
&\scalebox{1.0}{$p_{Y,Z}$}\scalebox{1.0}{$\, \equiv \, $}\scalebox{1.0}{$p_{X,Z}$}\\
&\scalebox{1.0}{$X - $}\scalebox{1.0}{$(Z, V)$}\scalebox{1.0}{$ - Y$}
&\scalebox{1.0}{$X - $}\scalebox{1.0}{$(Z, V)$}\scalebox{1.0}{$ - Y$}
\text{and}
\scalebox{1.0}{$Z - $}\scalebox{1.0}{$X$}\scalebox{1.0}{$ - V$}\\
\hline
\end{tabular}
\end{table}

\begin{table}[!t]
    \caption{Corollaries}
    \label{table:special_cases}
    \centering
    \begin{tabular}{|c|c|c|}
    \hline
     & \bfseries E-D-codes & \bfseries D-codes\\
    \hline
    \bfseries Marginal realism & Corollaries \ref{cor:independent_side_info_two_sided}, \ref{cor:region_with_no_side_info}, and \ref{cor:E_D_infty_region_finite_alphabets} & Theorem \ref{thm:common_part} and \ref{thm:D_infty_rates_region_marginal_realism}\\
    \hline
    \end{tabular}
\end{table}
Our main results for the different notions of achievability are listed in Table \ref{table:main_results}. Main corollaries are listed in Table \ref{table:special_cases}.
The above formalization of the notion of near-perfect realism based on the total variation distance is related to the optimal performance of a critic (hypothesis test), as follows. Consider a fixed blocklength $n$ and a distribution $P$ induced by a code. A critic is given either a sample from $P_{Y^n},$ i.e., an \textit{artificial} sample, or from $p_X^{\otimes n},$ i.e., an \textit{original} sample, such that the distribution of the sample is $\tfrac{1}{2}P_{Y^n} + \tfrac{1}{2}p_X^{\otimes n}.$ Then, the probability for a critic (hypothesis test) to correctly classify a sample from that mixture, as original or artificial, is at most (e.g., \cite{TVinOptimalHypothesisTest})
\begin{IEEEeqnarray}{c}
    \scalebox{0.85}{$\dfrac{1}{2}$} + \scalebox{0.85}{$\dfrac{1}{2}$} \|P_{Y^n} - p_{X}^{\otimes n}\|_{TV}.
\end{IEEEeqnarray}Hence, when the total variation distance is very small, a critic's performance is essentially no better than that of a coin toss.
Our work has the merit of determining the
information-theoretic rate-distortion trade-off
under either a near-perfect or perfect realism constraint,
and of highlighting that from that lens, perfect realism is often not harder to achieve than near-perfect realism
(see Theorem \ref{theorem:equivalence_perfect_realism}).
To prove this and our other results for general alphabets, we propose a method in which the following property plays a central role.
\begin{definition}\label{def:uniform_integrability}
    Given a source alphabet $\mathcal{X},$ a probability distribution $p$ on $\mathcal{X}$ and a distortion measure $d,$ we say that $(d,p)$ is uniformly integrable if for every $\varepsilon >0$ there is a $\tau>0$ such that \begin{equation*}
        \sup_{\mathbb{P}_{X,Y,\xi}} \mathbb{E}[d(X,Y) \cdot \xi] \leq \varepsilon,
    \end{equation*} where the supremum is taken over all distributions $\mathbb{P}_{X,Y,\xi}$ on $\mathcal{X}^2 \times \{0,1\}$ satisfying $\mathbb{P}_X \equiv \mathbb{P}_Y \equiv p$ and $\mathbb{P}(\{\xi=1\}) \leq \tau.$
\end{definition}
\vspace{10pt}
The notion of uniform integrability is reminiscent of a notion in probability theory
of the same name that applies to a family of random variables defined over a common
probability space. The latter has arisen in rate-distortion theory before \cite[(5.3.2))]{2003BookHanInfoSpectrum}.
Following a preliminary version of this work \cite{2022AaronWagnerRDPTradeoffTheRoleOfCommonRandomness}, an extension of Definition
\ref{def:uniform_integrability} --- and of Claim \ref{claim:uniform_integrability_if_MSE} below --- was used in \cite{Dec2022WeakAndStrongPerceptionConstraintsAndRandomness}.
We shall assume uniform integrability throughout the paper. This incurs some loss of generality in that
there exist $(d,p)$ pairs that are not uniformly integrable yet have a well-defined and finite rate-distortion function 
(see Appendix~\ref{app:subsec_example_not_satisfying_uniform_integrability}). Naturally occurring examples are usually
uniformly integrable, however. To wit, $(d,p)$ is uniformly integrable if
$\mathcal{X}$ is finite and $d$ does not take infinite values. Moreover, we have:

\begin{claim}\label{claim:uniform_integrability_if_MSE}
If $\mathcal{X} \subseteq \mathbb{R}$ and $d$ is the MSE distortion measure, then $(d,p)$ is uniformly integrable if $p$ has a finite fourth moment, e.g., if $p$ is a Normal distribution.
\end{claim}

The proof is deferred to Appendix \ref{app:subsec:Gaussian}.

\subsection{Equivalence of the perfect / near-perfect realism problems}
Our first result shows that any sequence of codes that
achieves near-perfect realism can be upgraded to one
that achieves perfect realism with no asymptotic
change to the rates or the distortion.
\begin{theorem}\label{theorem:equivalence_perfect_realism}
Given Polish source and side-information alphabets $\mathcal{X}$ and $\mathcal{Z},$ a joint distribution $p_{X,Z}$ on $\mathcal{X} \times \mathcal{Z}$
and a distortion measure $d$ on $\mathcal{X}$ such that $(d,p_X)$ is uniformly integrable, then a triplet $(R,R_c,\Delta)$ is D-achievable (resp. E-D-achievable) with near-perfect marginal realism if and only if it is D-achievable (resp. E-D-achievable) with perfect marginal realism. The same applies for achievability with unconstrained common randomness,
and for joint realism.
\end{theorem}

See Appendix \ref{app:equivalence_perfect_realism} for
the proof,
which
does not rely on a 
single-letter characterization of a set of achievable 
rate-distortion tuples. Theorem \ref{theorem:equivalence_perfect_realism}
is
used directly in the converse proofs of our main results, stated in Section \ref{sec:main_result}.
For achievability proofs, we use a slightly stronger result (Propositions \ref{proposition:to_perfect_realism} and \ref{proposition:to_perfect_joint_realism}, Appendix \ref{app:equivalence_perfect_realism}), in conjunction with the uniform integrability assumption.

\section{Main results}\label{sec:main_result}

\subsection{Marginal realism when the side information is available at both encoder and decoder}

In all of Section \ref{sec:main_result}, the superscript $(m)$ (resp. $(j)$) refers to marginal (resp. joint) realism.
The proof of the following theorem is provided in Section \ref{sec:achievability_two_sided}.

\begin{theorem}\label{theorem:E_D_rates_region_marginal_realism}
Consider a Polish source alphabet $\mathcal{X},$ a finite side-information alphabet $\mathcal{Z},$ a joint distribution $p_{X,Z}$ on $\mathcal{X} \times \mathcal{Z}$
and a distortion measure $d$ such that $(d,p_X)$ is uniformly integrable. Define the region $\mathcal{S}^{(m)}_{E\text{-}D}$ as 
\begin{align}\label{eq:def_S_E_D}
      & 
    \left\{ \begin{array}{rcl}
        \scalebox{0.9}{$(R, R_c, \Delta) \in \mathbb{R}_{\geq 0}^3$} &:& \scalebox{0.9}{$\exists \ p_{X,Z,V,Y} \in \mathcal{D}^{(m)}_{E\text{-}D} \text{ s.t. } $}\\
        \scalebox{0.9}{$R$} &\geq& \scalebox{0.9}{$I_p(X;V|Z)$} \\
        \scalebox{0.9}{$R+R_c$} &\geq& \scalebox{0.9}{$I_p(Y;V|Z) - H_p(Z|Y)$} \\
        \scalebox{0.9}{$\Delta$} &\geq& \scalebox{0.9}{$\mathbb{E}_p[d(X, Y)]$}
    \end{array}\right\},
\end{align} with $\mathcal{D}^{(m)}_{E\text{-}D}$ defined as
\begin{align}\label{eq:def_D_E_D}
       & 
    \left\{ \begin{array}{rcl}
        &p_{X,Z,V,Y} : \scalebox{0.9}{$(X,Z) \sim$ } p_{X,Z}, \ p_{Y} \equiv p_{X}&\\
        &X - (Z, V) - Y&
    \end{array}\right\},
\end{align}where the alphabet of $V$ is constrained to be finite. Then, the closure of the set $\mathcal{A}^{(m)}_{E\text{-}D}$ of
E-D-achievable triplets $(R,R_c, \Delta)$ with perfect or near-perfect marginal realism is equal to the closure of $\mathcal{S}^{(m)}_{E\text{-}D}.$
\end{theorem}

\begin{corollary}\label{cor:independent_side_info_two_sided}
Consider the setting of Theorem \ref{theorem:E_D_rates_region_marginal_realism}. Assume that $p_{X,Z} \equiv p_X \cdot p_Z.$ Then, we have
\begin{IEEEeqnarray}{c}
\mathcal{S}^{(m)}_{E\text{-}D} + (0,H(Z),0) = \mathcal{S}^{(m)} \cap \mathbb{R}_{\geq 0} \times [H(Z),\infty] \times \mathbb{R}_{\geq 0},
\end{IEEEeqnarray}
where $\mathcal{S}^{(m)}$ is the optimal region in the absence of side information, as stated in Corollary \ref{cor:region_with_no_side_info} to follow, and is defined as
\begin{align}\label{eq:def_S}
      & 
    \left\{ \begin{array}{rcl}
        \scalebox{0.9}{$(R, R_c, \Delta) \in \mathbb{R}_{\geq 0}^3$} &:& \scalebox{0.9}{$\exists \ p_{X,V,Y} \in \mathcal{D}^{(m)} \text{ s.t. } $}\\
        \scalebox{0.9}{$R$} &\geq& \scalebox{0.9}{$I_p(X;V)$} \\
        \scalebox{0.9}{$R+R_c$} &\geq& \scalebox{0.9}{$I_p(Y;V)$} \\
        \scalebox{0.9}{$\Delta$} &\geq& \scalebox{0.9}{$\mathbb{E}_p[d(X, Y)]$}
    \end{array}\right\},
\end{align} with $\mathcal{D}^{(m)}$ defined as
\begin{align}\label{eq:def_D}
       & 
    \left\{ \begin{array}{rcl}
        &p_{X,V,Y} : \scalebox{0.9}{$X \sim$ } p_{X}, \ p_{Y} \equiv p_{X}&\\
        &X - V - Y&
    \end{array}\right\},
\end{align}where the alphabet of $V$ is constrained to be finite.
\end{corollary}

The proof is provided in Section \ref{sec:special_cases}.
When the side information is independent from the source, the former is equivalent to a uniform common randomness of rate $H(Z).$ In general (Theorem \ref{theorem:E_D_rates_region_marginal_realism}), the side information plays a dual role: it provides common randomness independent from the source, as well as a second look on the source.
The
conditioning with respect to the side information
is reminiscent of the result of Wyner and Ziv \cite{Wyner&Ziv1977} for lossy source coding.
We can also rewrite the
second
inequality
in \eqref{eq:def_S_E_D}
as
\begin{IEEEeqnarray}{c}
    R+R_c+H_p(Z) \geq I_p(Y;Z,V).
\end{IEEEeqnarray}The left hand side is the effective rate of the codebook used in our scheme (Section \ref{sec:achievability_two_sided}) and the right hand side corresponds to the requirement for the soft covering lemma with side information \cite[Corollary~VII.5]{2013PaulCuffDistributedChannelSynthesis}. The formal codebook rate is $R+R_c+\log|\mathcal{Z}|,$ but atypical strings $z^n$ have a negligible impact on the distortion and total variation distance.
The fact that we need only consider perfect (marginal) realism (owing to Theorem \ref{theorem:equivalence_perfect_realism})
allows us to sidestep the continuity argument 
in the proof of~\cite[Thm.~II.1]{2013PaulCuffDistributedChannelSynthesis}, which in 
turn allows us to establish the converse for general spaces.
In the absence of side information, this yields
an extension (Corollary \ref{cor:region_with_no_side_info} below) of \cite[Theorem~1~\&~Remark~2b]{Sep2015RDPLimitedCommonRandomnessSaldi}.
Indeed, said continuity argument, also invoked for the case of near-perfect realism \cite[Remark~2b]{Sep2015RDPLimitedCommonRandomnessSaldi}, strongly relies on the assumption of finite source alphabet. Moreover, regarding the case of perfect realism, \cite[Theorem~1]{Sep2015RDPLimitedCommonRandomnessSaldi}
was proved for
$\mathcal{X}=\mathbb{R}$ with $p_X$ square-integrable and $d$ the MSE,
and for
finite $\mathcal{X}$ with $d=\rho^\gamma,$ where $\rho$ is a metric on $\mathcal{X},$ and $\gamma>0.$
The following result, which is a direct consequence of Theorem \ref{theorem:E_D_rates_region_marginal_realism}, first appeared in a preliminary version of this work \cite{2022AaronWagnerRDPTradeoffTheRoleOfCommonRandomness}, and an extension appeared in \cite{2024arxivJunChenGaussianRDP}.
\begin{corollary}\label{cor:region_with_no_side_info}
Consider a Polish source alphabet $\mathcal{X},$ a distribution $p_{X}$ on $\mathcal{X}$ 
and a distortion measure $d$ such that $(d,p_X)$ is uniformly integrable.
Then, the closure of the set $\mathcal{A}^{(m)}$ of achievable triplets $(R,R_c, \Delta)$ with perfect or near-perfect marginal realism is equal to the closure of $\mathcal{S}^{(m)}$ (Eq. \ref{eq:def_S}).
\end{corollary}
\begin{remark}\label{remark:finite_mutual_info}
The rate gains allowed by the availability of a finite-valued side information are at most the following. In the setting of Theorem \ref{theorem:E_D_rates_region_marginal_realism}, by definition of $\mathcal{S}^{(m)}_{E\text{-}D}$ and $\mathcal{S}^{(m)},$ for any $(R,R_c,\Delta) \in \mathcal{S}^{(m)}_{E\text{-}D},$
\begin{IEEEeqnarray}{c}
    (R+I(X;Z), R_c+H(Z|X), \Delta) \in \mathcal{S}^{(m)}.\nonumber
\end{IEEEeqnarray}
\end{remark}

Note that
according to our definition (Definition \ref{def:achievability}) of
achievability of a triplet $(R,R_c,\Delta),$ 
then when $R_c=0$ absolutely no common randomness is available.
The other extreme, $R_c \to \infty,$ is also of interest.
We make a distinction between the case of large common randomness rate and that of unconstrained common randomness.
\begin{corollary}\label{cor:E_D_infty_region_finite_alphabets}
Consider a Polish source alphabet $\mathcal{X},$ a finite side-information alphabet $\mathcal{Z},$ a joint distribution $p_{X,Z}$ on $\mathcal{X} \times \mathcal{Z}$ 
and a distortion measure $d$ such that $(d,p_X)$ is uniformly integrable.
Then, the closure of the set $\mathcal{A}^{(m)}_{E\text{-}D,\infty}$ of couples $(R,\Delta)$ which are E-D-achievable with perfect or near-perfect marginal realism with finite-rate common randomness is the closure of the following region $\mathcal{S}^{(m)}_{E\text{-}D,\infty}:$
\begin{align}\label{eq:def_S_E_D_infty}
      & 
    \left\{ \begin{array}{rcl}
        \scalebox{0.9}{$(R, \Delta) \in \mathbb{R}_{\geq 0}^2$} &:& \scalebox{0.9}{$\exists \ p_{X,Z,V,Y} \in \mathcal{D}^{(m)}_{E\text{-}D} \text{ s.t. } $}\\
        \scalebox{0.9}{$R$} &\geq& \scalebox{0.9}{$I_p(X;V|Z)$}\\
        \scalebox{0.9}{$\Delta$} &\geq& \scalebox{0.9}{$\mathbb{E}_p[d(X, Y)]$}
    \end{array}\right\}.
\end{align}
If we also have that $\mathcal{X}$ is finite, then E-D-achievability with unconstrained common randomness is equivalent to E-D-achievability with finite-rate common randomness, and $\mathcal{S}^{(m)}_{E\text{-}D,\infty}$ rewrites as:
\begin{align}\label{eq:def_S_E_D_infty_finite_alphabet}
      & 
    \left\{ \begin{array}{rcl}
        \scalebox{0.9}{$(R, \Delta) \in \mathbb{R}_{\geq 0}^2$} &:& \scalebox{0.9}{$\exists \ p_{X,Z,Y} \in \mathcal{D}^{(m)}_{\infty} \text{ s.t. } $}\\
        \scalebox{0.9}{$R$} &\geq& \scalebox{0.9}{$I_p(X;Y|Z)$}\\
        \scalebox{0.9}{$\Delta$} &\geq& \scalebox{0.9}{$\mathbb{E}_p[d(X, Y)]$}
    \end{array}\right\},
\end{align}with $\mathcal{D}^{(m)}_{\infty}$ defined as
\begin{align}\label{eq:def_D_E_D_infty}
       & 
    \left\{ \begin{array}{rcl}
        &p_{X,Z,Y} : \scalebox{0.9}{$(X,Z) \sim$ } p_{X,Z}, \ p_{Y} \equiv p_{X}&
    \end{array}\right\}.
\end{align}
\end{corollary}

The proof is provided in Section \ref{subsec:proof_corollary_large_CR_two_sided}.
In the absence of side information,
region
\eqref{eq:def_S_E_D_infty_finite_alphabet} becomes
the region corresponding to \eqref{eq:BM} and we recover the result described at the beginning of \cite[Section~III.A]{Sep2015RDPLimitedCommonRandomnessSaldi} regarding the case of finite alphabets. See the latter for further remarks.
\begin{remark}\label{rem:continuous_vs_discrete_common_randomness}
If the target distribution has infinite entropy, it
is unclear whether it is
possible to achieve every point in (the interior of) region \eqref{eq:def_S_E_D_infty_finite_alphabet} --- nor \eqref{eq:BM} --- with finite-rate common randomness.
Whatever common randomness is available, region \eqref{eq:def_S_E_D_infty_finite_alphabet} is an outer bound of the set of achievable couples --- even for variable-length codes --- (see the proof of Corollary \ref{cor:E_D_infty_region_finite_alphabets}).
In the absence of side information, then as shown in \cite{Jan2015SaldiEtAlDistributionPreservationMeasureTheoreticConsiderationsForContinuousAndDiscreteCommonRandomness},
region \eqref{eq:BM} is achievable with perfect realism using some continuous common randomness variable --- and variable-length codes ---, for general source alphabets. As can be seen from our 
Proposition \ref{prop:standard_normal_source}, for 
a source with Normal distribution (see also \cite{Sep2015RDPLimitedCommonRandomnessSaldi}), 
every point in the interior of
\eqref{eq:BM}
is achievable with finite-rate common randomness.
Hence, in that case a discrete common randomness variable with infinite rate is sufficient.
Moreover, as stated in
Proposition \ref{prop:gaussian_infinite_common_randomness}, when $(X,Z)$ is a bi-dimensional Gaussian with non-zero correlation, region \eqref{eq:def_S_E_D_infty_finite_alphabet} is optimal and there is a finite common randomness rate with which one can achieve any point in its interior.
\end{remark}

\subsection{Marginal realism when the side information is only available at the decoder}

\begin{theorem}\label{theorem:D_rates_region_marginal_realism}
Consider Polish source and side-information alphabets $\mathcal{X}$ and $\mathcal{Z},$ a joint distribution $p_{X,Z}$ on $\mathcal{X} \times \mathcal{Z}$ 
and a distortion measure $d$ such that $(d,p_X)$ is uniformly integrable. Define the region $\mathcal{S}^{(m)}_D$ as 
\begin{align}\label{eq:def_S_D}
      & 
    \left\{ \begin{array}{rcl}
        \scalebox{0.9}{$(R, R_c, \Delta) \in \mathbb{R}_{\geq 0}^3$} &:& \scalebox{0.9}{$\exists \ p_{X,Z,V,Y} \in \mathcal{D}^{(m)}_D \text{ s.t. } $}\\
        \scalebox{0.9}{$R$} &\geq& \scalebox{0.9}{$I_p(X;V) - I_p(Z;V)$} \\
        \scalebox{0.9}{$R+R_c$} &\geq& \scalebox{0.9}{$I_p(Y;V) - I_p(Z;V)$} \\
        \scalebox{0.9}{$\Delta$} &\geq& \scalebox{0.9}{$\mathbb{E}_p[d(X, Y)]$}
    \end{array}\right\},
\end{align} with $\mathcal{D}^{(m)}_D$ defined as
\begin{align}\label{eq:def_D_D}
       & 
    \left\{ \begin{array}{rcl}
        &p_{X,Z,V,Y} : \scalebox{0.9}{$(X,Z) \sim$ } p_{X,Z}, \ p_{Y} \equiv p_{X}&\\
        &Z - X - V, \quad \ X - (Z, V) - Y& \\
        &I_p(Z;V) < \infty&
    \end{array}\right\},
\end{align}where the alphabet of $V$ is constrained to be Polish. Then, the closure of the set $\mathcal{A}^{(m)}_D$ of D-achievable triplets $(R,R_c, \Delta)$ with perfect or near-perfect marginal realism contains the closure of $\mathcal{S}^{(m)}_D.$
\end{theorem}
\vspace{10pt}

The proof is provided in Section \ref{sec:achievability}.
The first inequality in \eqref{eq:def_S_D} is reminiscent of the result of Wyner and Ziv \cite{Wyner&Ziv1977} for traditional lossy source coding with side information.
The difficulty in obtaining a conclusive characterization appears to be attributable to the
difficulty in determining the extent to which the side information
can be used to
produce
common randomness.
As evidence of this, we show that
the
region in \eqref{eq:def_S_D} is optimal
in two cases, one for which this question is clear, and one for which it is irrelevant.
The first is when the source and side information are conditionally independent with respect to a common component.
Previous work
has shown that making use of common components can strictly improve distributed compression schemes without realism constraints \cite{2011TITbyAaronBWagnerCommonComponentsInTwoEncoderDistributedLossyCompression}.
\begin{theorem}\label{thm:common_part}
Consider Polish source and side-information alphabets $\mathcal{X}$ and $\mathcal{Z},$ a joint distribution $p_{X,Z}$ on $\mathcal{X} \times \mathcal{Z},$ 
and a distortion measure $d$ such that $(d,p_X)$ is uniformly integrable. Assume that there exists (deterministic) functions $\phi,\psi$ such that, under $p_{X,Z},$ the Markov chain $X-\phi(X)-Z$ holds and $\phi(X)=\psi(Z)$ almost surely and $\phi(X)$ has finite support. Then, the achievable region of Theorem \ref{theorem:D_rates_region_marginal_realism} is optimal, i.e.,
$\overline{\mathcal{A}}^{(m)}_{D} = \overline{\mathcal{S}}^{(m)}_{D}.$ 
\end{theorem}
\vspace{10pt}

The proof is provided in Section \ref{subsec:proof_thm_common_part}.
The achievable region of Theorem \ref{theorem:D_rates_region_marginal_realism} is also optimal when common randomness is available at any rate,
and either $\mathcal{X}$ or $\mathcal{Z}$ is finite,
as follows. The proof is provided in Section \ref{subsec:proof_D_infty_marginal}.
\begin{theorem}\label{thm:D_infty_rates_region_marginal_realism}
Consider Polish source and side-information alphabets $\mathcal{X}$ and $\mathcal{Z},$ a joint distribution $p_{X,Z}$ on $\mathcal{X} \times \mathcal{Z}$ 
and a distortion measure $d$ such that $(d,p_X)$ is uniformly integrable. If either of $\mathcal{X}$ or $\mathcal{Z}$ is finite, then
the closure of the set $\mathcal{A}^{(m)}_{D,\infty}$ of D-achievable couples with finite-rate common randomness with perfect or near-perfect marginal realism is the closure of the following set $\mathcal{S}^{(m)}_{D,\infty}:$ 
\begin{align}\label{eq:def_S_D_infty}
      & 
    \left\{ \begin{array}{rcl}
        (R, \Delta) \in \mathbb{R}_{\geq 0}^2 &:& \exists \ p_{X,Z,V,Y} \in \mathcal{D}^{(m)}_D \text{ s.t. } \\
        R &\geq& I_p(X;V)-I_p(Z;V)\\
        \infty &>& I_p(Y;V) \\
        \Delta &\geq& \mathbb{E}_p[d(X, Y)]
    \end{array}\right\}. 
\end{align}
\end{theorem}

\subsection{When imposing joint realism of the reconstruction and the side information}

The perfect joint realism constraint is akin to a perfect marginal realism constraint with conditioning on the realization of the side information.
When the side information is available at both terminals, the optimal single-letter region corresponds to the one in the absence of side information with added conditioning on $Z,$ in each mutual information term, in the distributional constraint, and in the Markov chain constraint.
The codebook used in the corresponding scheme comprises of one sub-codebook per element of $\mathcal{Z}^n.$

\begin{theorem}\label{theorem:E_D_rates_region_joint_realism}
Consider a Polish source alphabet $\mathcal{X},$ a finite side-information alphabet $\mathcal{Z},$ a joint distribution $p_{X,Z}$ on $\mathcal{X} \times \mathcal{Z}$ 
and a distortion measure $d$ such that $(d,p_X)$ is uniformly integrable. Define the region $\mathcal{S}^{(j)}_{E\text{-}D}$ as 
\begin{align}\label{eq:def_S_E_D_joint}
      & 
    \left\{ \begin{array}{rcl}
        \scalebox{0.9}{$(R, R_c, \Delta) \in \mathbb{R}_{\geq 0}^3$} &:& \scalebox{0.9}{$\exists \ p_{X,Z,V,Y} \in \mathcal{D}^{(j)}_{E\text{-}D} \text{ s.t. } $}\\
        \scalebox{0.9}{$R$} &\geq& \scalebox{0.9}{$I_p(X;V|Z)$} \\
        \scalebox{0.9}{$R+R_c$} &\geq& \scalebox{0.9}{$I_p(Y;V|Z)$} \\
        \scalebox{0.9}{$\Delta$} &\geq& \scalebox{0.9}{$\mathbb{E}_p[d(X, Y)]$}
    \end{array}\right\},
\end{align} with $\mathcal{D}^{(j)}_{E\text{-}D}$ defined as
\begin{align}\label{eq:def_D_E_D_joint}
       & 
    \left\{ \begin{array}{rcl}
        &p_{X,Z,V,Y} : \scalebox{0.9}{$(X,Z) \sim$ } p_{X,Z}, \ p_{Y,Z} \equiv p_{X,Z}&\\
        &X - (Z, V) - Y&
    \end{array}\right\},
\end{align}where the alphabet of $V$ is constrained to be finite. Then, the closure of the set $\mathcal{A}^{(j)}_{E\text{-}D}$ of E-D-achievable triplets $(R,R_c, \Delta)$ with perfect or near-perfect joint realism is equal to the closure of $\mathcal{S}^{(j)}_{E\text{-}D}.$ 
\end{theorem}
The proof is provided in Section \ref{sec:achievability_joint_two_sided}.
We note the absence of entropy term contrary to the case of marginal realism (Theorem \ref{theorem:E_D_rates_region_marginal_realism}). Hence, in the case of joint realism the side information does not seem to act as a source of common randomness.
When the side information is only available at the decoder, we obtain a full characterization valid in all cases, contrary to the case of marginal realism (Theorems \ref{thm:common_part},\ref{thm:D_infty_rates_region_marginal_realism}).
\begin{theorem}\label{theorem:D_rates_region_joint_realism}
Consider Polish source and side information alphabets $\mathcal{X}$ and $\mathcal{Z},$ a joint distribution $p_{X,Z}$ on $\mathcal{X} \times \mathcal{Z}$ 
and a distortion measure $d$ such that $(d,p_X)$ is uniformly integrable. Define the region $\mathcal{S}^{(j)}_{D}$ as 
\begin{align}\label{eq:def_S_D_joint}
      & 
    \left\{ \begin{array}{rcl}
        \scalebox{0.9}{$(R, R_c, \Delta) \in \mathbb{R}_{\geq 0}^3$} &:& \scalebox{0.9}{$\exists \ p_{X,Z,V,Y} \in \mathcal{D}^{(j)}_{D} \text{ s.t. } $}\\
        \scalebox{0.9}{$R$} &\geq& \scalebox{0.9}{$I_p(X;V|Z)$} \\
        \scalebox{0.9}{$R+R_c$} &\geq& \scalebox{0.9}{$I_p(Y;V|Z)$} \\
        \scalebox{0.9}{$\Delta$} &\geq& \scalebox{0.9}{$\mathbb{E}_p[d(X, Y)]$}
    \end{array}\right\},
\end{align} with $\mathcal{D}^{(j)}_{D}$ defined as
\begin{align}\label{eq:def_D_D_joint}
       & 
    \left\{ \begin{array}{rcl}
        &p_{X,Z,V,Y} : \scalebox{0.9}{$(X,Z) \sim$ } p_{X,Z}, \ p_{Y,Z} \equiv p_{X,Z}&\\
        &Z - X - V, \quad \ X - (Z, V) - Y& \\
        &I_p(Z;V) < \infty&
    \end{array}\right\},
\end{align}where the alphabet of $V$ is constrained to be Polish.
Then, the closure of the set $\mathcal{A}^{(j)}_{D}$ of E-D-achievable triplets $(R,R_c, \Delta)$ with perfect or near-perfect joint realism is equal to the closure of $\mathcal{S}^{(j)}_{D}.$ 
\end{theorem}
The proof is provided in Section \ref{sec:achievability_joint_one_sided}.

\subsection{The Gaussian case}\label{subsec:Gaussian_statments_of_results}
The quadratic Gaussian scenario illustrates the difference between the case where no common randomness is available ($R_c=0$) and the case where it is available at a large or even infinite rate. We start with a detailed characterization in the absence of side information.
The following result first appeared in a preliminary version of this work \cite{2022AaronWagnerRDPTradeoffTheRoleOfCommonRandomness}, and an extension appeared in \cite{2024arxivJunChenGaussianRDP}.

\begin{proposition}\label{prop:standard_normal_source}
In the absence of side information, if the source $X$ is standard Normal and $d:(x,y)\mapsto (x-y)^2,$
then for $\Delta \in (0,2)$ and $R_c \geq 0,$ then the infimum of rates $R$ such that $(R,R_c,\Delta)$ is achievable with near-perfect or perfect marginal realism is
\begin{equation}\label{eq:Normal_R_min}
R(\Delta,R_c) = \frac{1}{2} \log \frac{1}{1 - \rho^2},
\end{equation}
where $\rho$ is such that
\begin{IEEEeqnarray}{c}
    2\rho^2+ 2^{2R_c}-1 = \sqrt{(2^{2R_c}-1)^2+2^{2R_c}(2-\Delta)^2}.\nonumber
\end{IEEEeqnarray}
\end{proposition}

\begin{figure}
    \begin{center}
    \scalebox{.8}{\includegraphics{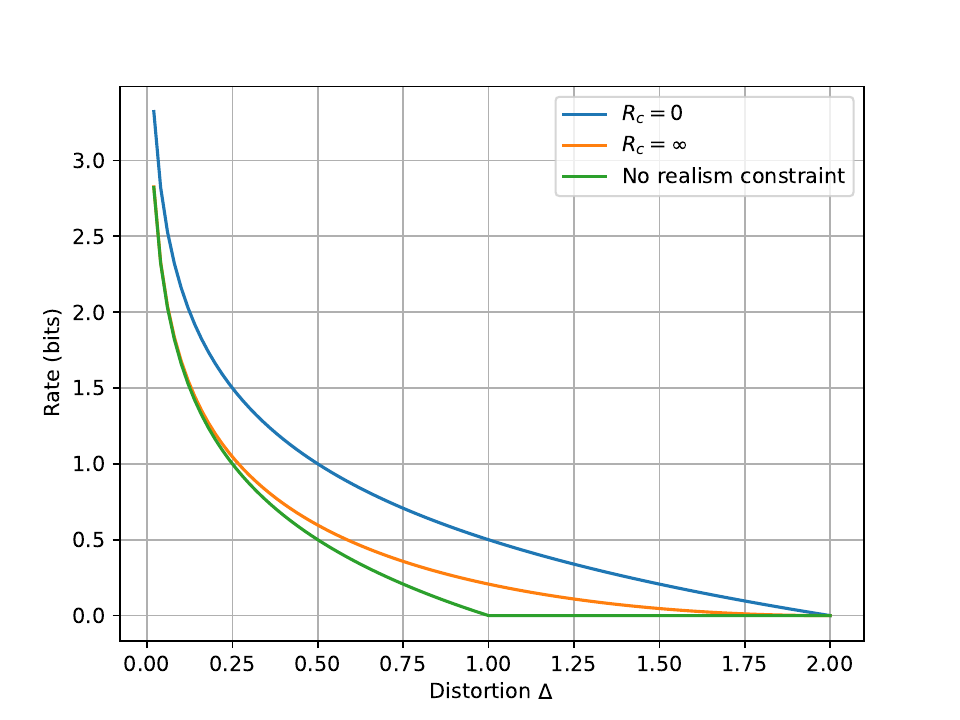}}
    \end{center}
    \caption{Rate-distortion trade-offs for a standard Normal source
    with MSE distortion for $R_c = 0$,
    $R_c \rightarrow \infty$, and the classical
    rate-distortion function without a realism
    constraint.}
    \label{fig:gauss}
\end{figure}

The proof is provided in Section \ref{subsec:normal_source_no_side_info}.
The achievability direction was proved 
in \cite{Sep2015RDPLimitedCommonRandomnessSaldi}, but not the converse. Taking $R_c \rightarrow \infty$ gives 
$R(\Delta,\infty) = \frac{1}{2} \log \frac{1}{\Delta(1 - \Delta/4)},$
which was previously derived by 
Li \emph{et al.}~\cite[Prop.~2]{2011LiEtAlMainPaperOnDistributionPreservingQuantization}.
However, a given $\Delta \in (0,2)$ cannot be achieved with finite common randomness rate $R_c$ and a compression rate $R$ arbitrarily close to $R(\Delta,\infty).$ 
On the other hand, taking $R_c = 0$ gives 
$R(\Delta,0) = \frac{1}{2} \log \frac{1}{\Delta/2}$,
which is consistent with the finding 
of Theis and Agustsson \cite{2021TheisAgustssonOnTheAdvantagesOfStochasticEncoders} and Blau
and Michael~\cite{2019BlauMichaeliRethinkingLossyCompressionTheRDPTradeoff} for general sources
that, under a MSE distortion constraint and in the absence of
common randomness, imposing perfect realism incurs
at most
a 3 dB penalty compared to the case without a realism 
constraint. Fig.~\ref{fig:gauss} illustrates the rate-distortion
trade-off for $R_c = 0$, $R_c \rightarrow \infty$, 
and the classical case with no realism constraint~\cite[Sec.~10.3.2]{Cover&Thomas2006}.
Note that, as also proved in \cite{2011LiEtAlMainPaperOnDistributionPreservingQuantization}, at small distortions, requiring perfect
realism incurs essentially no rate penalty, assuming
that
common randomness is available
at an infinite rate.\\

Interestingly, similarly to standard source coding with side information \cite[Example~11.2]{2011BookElGamalKimNetworkInformationTheory}, when the source and side information form a bi-variate Gaussian, then D-achievability is equivalent to E-D-achievability
if
common randomness
is available at a large enough
rate.
\begin{proposition}\label{prop:gaussian_infinite_common_randomness}
Consider a real-valued source and side information with Gaussian joint distribution
\begin{equation*}
p_{X,Z} = \mathcal{N}\Bigg(0, \scalebox{0.8}{$\begin{pmatrix}
1 & \eta \\
\eta & 1
\end{pmatrix}$} \Bigg),
\end{equation*}where $\eta \in (-1,1),$ and the distortion function
$d:(x,y)\mapsto (x-y)^2.$
Consider $\Delta$ in $(0, 2-2|\eta|],$ and denote $1-\Delta/2$ by $\rho.$ Then, the infimum of rates such that $(R,\Delta)$ is D- or E-D-achievable with finite-rate common randomness and (near-)perfect marginal realism is:
\begin{equation}
    \scalebox{0.93}{$R_{D}(\Delta) = R_{E\text{-}D}(\Delta) = \dfrac{1}{2}\log\Big(\dfrac{1-\eta^2}{1-\rho^2}\Big)
    .
    $}
    \label{eq:statement_R_E_D_equals_R_D_bi_dimensional_Gaussian}
\end{equation}
\end{proposition}
\vspace{5pt}

The proof is provided in Section \ref{subsec:bi_dimensional_gaussian}.
The numerator in the above logarithm is the same as in standard source coding with side information (see, e.g., \cite{Cover&Thomas2006}) and the denominator is the same as in \cite{2011LiEtAlMainPaperOnDistributionPreservingQuantization}, which we recover when the side-information $Z$ is independent from $X.$ 
The equivalence between D-achievability and E-D-achievability in the setting of Proposition \ref{prop:gaussian_infinite_common_randomness} breaks down when common randomness is limited. Indeed, consider the case where the source is independent from the side information ($\eta=0$). On the one hand, the latter is not useful when only available at the decoder, and the corresponding trade-off is given by Proposition \ref{prop:standard_normal_source}. On the other hand, when the side information is available at both terminals, it acts as
a
common randomness
with infinite rate,
which corresponds to the scenario $R_c \to \infty$ in Proposition \ref{prop:standard_normal_source}.

\section{Proof of Theorem \ref{theorem:E_D_rates_region_marginal_realism}}\label{sec:achievability_two_sided}

\subsection{Converse}\label{sec:converse_E_D_marginal}

We prove that $\overline{\mathcal{A}}^{(m)}_{E\text{-}D} \subset \overline{\mathcal{S}}^{(m)}_{E\text{-}D}$ by proving that $\mathcal{A}^{(m)}_{E\text{-}D} \subset \overline{\mathcal{S}}^{(m)}_{E\text{-}D}.$ 
This --- self-contained --- proof builds on the approach of \cite{2015TITYassaeeEtAlChannelSimulationInteractiveComm}, where the same Markov chain as in \eqref{eq:def_D_E_D} is studied. Let $(R, R_c,\Delta)$ be E-D-achievable with near-perfect marginal realism. Then, by Theorem \ref{theorem:equivalence_perfect_realism}, it is E-D-achievable with perfect marginal realism. Fix $\varepsilon$\hspace{2pt}$>$\hspace{2pt}$0.$
Then, there exists a $(n,R, R_c)$ E-D-code inducing a joint distribution $P$ such that
$\mathbb{E}_P[d(X^n, Y^n)] \, {\leq} \, \Delta \, + \, \varepsilon$ and $P_{Y^n} \, {\equiv} \, p_X^{\otimes n}.$
Let $T$ be a uniform random variable over $[n],$
independent of all other variables.
Define $V=(M, J, Z_{[n] \backslash T}, T),$ 
which has a finite alphabet.
Since $(X^n,Z^n) \sim p_{X,Z}^{\otimes n},$ the distribution of $(X_T,Z_T)$ is $p_{X,Z}.$ 
Similarly, 
we have $P_{Y_T} \equiv p_X.$
Distribution $P_{X_T,Z_T,V,Y_T}$ satisfies the
Markov chain in $\mathcal{D}^{(m)}_{E\text{-}D}$ (Eq. \ref{eq:def_D_E_D}) by Lemma \ref{lemma:Markov_chain_Y_T_X_T} (Appendix \ref{app:subsec:Markov_chains_precise_reference_for_converse}).
Thus, $P_{X_T, Y_T, Z_T, V} \in \mathcal{D}^{(m)}_{E\text{-}D}.$ It remains to check that the inequalities in the definition of $\mathcal{S}^{(m)}_{E\text{-}D}$ are satisfied.
Since $P_{X_T,Y_T} \equiv \hat{P}_{X^n,Y^n},$ we have $\mathbb{E}[\scalebox{1.0}{$d(X_T, Y_T)$}] = \mathbb{E}[\scalebox{1.0}{$d(X^n, Y^n)$}] \leq \Delta + \varepsilon.$
We first lower bound the rates using quantized variables before taking a limit.
Consider any (finite-valued) quantizer on $\mathcal{X}.$ The image of a letter $x$ by the latter is denoted $[x].$
We have
\begin{IEEEeqnarray}{rCl}
    nR \geq H(M)
    \geq
    \scalebox{1.0}{$I(M;[X]^n|Z^n, J)$}
    &=& \scalebox{1.0}{$I(M, J ;[X]^n|Z^n)$} \label{eq:converse_E_D_marginal_R_using_J_indep} \\
    &=& \scalebox{1.0}{$\sum_{t=1}^n I(M, J ; [X]_t | Z^n, [X]_{t+1:n})$} \nonumber \\ 
    &=& \scalebox{1.0}{$\sum_{t=1}^n I(M, J, [X]_{t+1:n}, Z_{[n] \backslash t} ; [X]_t | Z_t)$} \nonumber \\
    &\geq& \scalebox{1.0}{$\sum_{t=1}^n I(M, J, Z_{[n] \backslash t} ; [X]_t | Z_t)$} \nonumber \\
    &=& \scalebox{1.0}{$n I(M, J, Z_{[n] \backslash T} ; [X]_T | Z_T, T)$}\IEEEeqnarraynumspace\label{eq:converse_E_D_marginal_lower_bound_R_t_to_T} \\
    &=& \scalebox{1.0}{$n I(V ; [X]_T | Z_T),$}\label{eq:converse_E_D_marginal_lower_bound_R_T_indep}
\end{IEEEeqnarray}
where \eqref{eq:converse_E_D_marginal_R_using_J_indep} follows from the independence between the common randomness and the sources; and \eqref{eq:converse_E_D_marginal_lower_bound_R_t_to_T} and \eqref{eq:converse_E_D_marginal_lower_bound_R_T_indep} follow from the independence of $T$ and all other variables and from the fact that couples in $\{([X]_t,Z_t)\}_{t \in [n]}$ are i.i.d. Moreover, we have
\begin{IEEEeqnarray}{rCl}
    n[R+R_c+H(p_Z)]
    \geq
    I(M,J,Z^n;[Y]^n)
    &=&
    \sum_{t=1}^n I(M,J,Z^n ;[Y]_t|[Y]_{1:t-1})
    \nonumber \\*
    &=& \sum_{t=1}^n I(M,J,Z^n,[Y]_{1:t-1} ;[Y]_t)
    \label{eq:in_converse_marginal_E_D_perfect_realism1} \\*
    &\geq& \sum_{t=1}^n I(M,J,Z^n ;[Y]_t)
    \nonumber\\*
    &=& nI(M,J, Z_{\scalebox{0.64}{$[n]{\setminus}T$}}, Z_T ;[Y]_T|T)
    \nonumber \\*
    &=& nI(V,Z_T;[Y]_T)
    ,\label{eq:in_converse_marginal_E_D_perfect_realism2}
\end{IEEEeqnarray}where \eqref{eq:in_converse_marginal_E_D_perfect_realism1} and \eqref{eq:in_converse_marginal_E_D_perfect_realism2} follow from the fact that variables in $\{[Y]_t\}_{t \in [n]}$ are i.i.d.
This is true for any quantizer on $\mathcal{X}.$ Moreover, by Lemma \ref{lemma:entropy_geq_mutual_info} (Appendix \ref{app:subsubsec:info_theory_for_general_spaces}) we have $I(Z_T;V) \leq H(Z_T)<\infty.$
Then, by Proposition \ref{prop:limit_on_sequence_of_quantizers} (Appendix \ref{app:subsubsec:info_theory_for_general_spaces}) we have
\begin{IEEEeqnarray}{rCl}
    R \geq I(X_T;V|Z_T)
    \ \text{and} \
    R+R_c+H(Z_T) \geq I(Y_T;V,Z_T).\nonumber
\end{IEEEeqnarray}
By Lemma \ref{lemma:entropy_geq_mutual_info} we have that $I(Z_T;Y_T) \leq H(Z_T)<\infty,$ and that consequently $H(Z_T|Y_T)$ is well-defined and equal to $H(Z_T)-I(Z_T;Y_T).$ Hence, by the chain rule for general alphabets (Proposition \ref{prop:chain_rule}) we have
\begin{IEEEeqnarray}{c}
    I(Y_T;V,Z_T)-H(Z_T) = I(Y_T;V|Z_T) - H(Z_T|Y_T).
\end{IEEEeqnarray}
Therefore, $(R,R_c, \Delta+\varepsilon) \in \mathcal{S}^{(m)}_{E\text{-}D}.$ Thus, $(R,R_c, \Delta) \in \overline{\mathcal{S}}^{(m)}_{E\text{-}D},$ which concludes the converse proof.

\subsection{Informal outline of the achievability proof}

We first construct a distribution $Q^{(1)}$ (Subsection \ref{sec:Q_1_two_sided}), analyzed using the soft covering lemma with side information \cite[Corollary~VII.5]{2013PaulCuffDistributedChannelSynthesis}. The codebook used consists of one sub-codebook per possible realization $z^n$ in $\mathcal{Z}^n$ (Subsection \ref{subsec:introducing_R_R_c_Delta_achievability_marginal_two_sided}). Similarly to the approach of \cite[Section~V]{2013PaulCuffDistributedChannelSynthesis} --- for the channel synthesis problem without side information ---, we modify $Q^{(1)}$ to obtain a distribution $P^{(1)}$ corresponding to a coding scheme
(Subsection \ref{subsec:E_D_MR_from_Q_to_P}).
We show that the distance between these two distributions is negligible (Subsection \ref{subsec:distance_P_1_Q_1_E_D_MR}). While most results in \cite{2013PaulCuffDistributedChannelSynthesis} are only proved for finite alphabets, we propose an approach which allows to handle general alphabets. In particular, in order to obtain a final distortion bound (Subsection \ref{sec:conclusion_of_achievability_two_sided}), we use the uniform integrability in conjunction with the equivalence between achievability with near-perfect and perfect realism.

\subsection{Random codebook}\label{subsec:introducing_R_R_c_Delta_achievability_marginal_two_sided}
Here, we prove that $\overline{\mathcal{S
}}^{(m)}_{E\text{-}D} \subseteq \overline{\mathcal{A}}^{(m)}_{E\text{-}D}$ by proving that $\mathcal{S}^{(m)}_{E\text{-}D} \subseteq \overline{\mathcal{A}}^{(m)}_{E\text{-}D}.$ 
Let $(R,R_c,\Delta)$ be a triplet in $\mathcal{S}^{(m)}_{E\text{-}D}.$ 
Let $p_{X,Y,Z,V}$ be a corresponding distribution from the definition of $\mathcal{S}^{(m)}_{E\text{-}D}.$ Then,
\begin{IEEEeqnarray}{rCl}
R &\geq& I_p(X;V|Z)
\label{eq:introducing_R_two_sided}\\
R+R_c &\geq& I_p(Y;V|Z) - H_p(Z|Y)
\label{eq:introducing_R_c_two_sided}\\
\Delta &\geq& \mathbb{E}_p[d(X,Y)].
\label{eq:introducing_Delta_two_sided}
\end{IEEEeqnarray}
We start by rewriting these expressions. By Lemma \ref{lemma:entropy_geq_mutual_info} (Appendix \ref{app:subsubsec:info_theory_for_general_spaces}) we have that $I_p(Z;Y) \leq H_p(Z)<\infty,$ and that consequently $H_p(Z|Y)$ is well-defined and equal to $H_p(Z)-I_p(Z;Y).$ Hence, by the chain rule for general alphabets (Proposition \ref{prop:chain_rule}) we have
\begin{IEEEeqnarray}{c}
I_p(Y;V|Z) - H_p(Z|Y) = I_p(Y;V,Z)-H_p(Z).\label{eq:replace_conditional_entropy_by_entropy_in_proof}
\end{IEEEeqnarray}
Moreover, the right hand side of \eqref{eq:introducing_R_two_sided} can be rewritten. For any quantized version $[X]$ of $X$ we have
\begin{IEEEeqnarray}{c}
I_p([X];V|Z) = I_p([X],Z;V,Z) - H_p(Z).\label{eq:rewriting_with_quantized_X}
\end{IEEEeqnarray}Since
$Z$
is
finite-valued, we have $I(Z;V)<\infty.$ Thus, from \eqref{eq:rewriting_with_quantized_X} and
Proposition \ref{prop:limit_on_sequence_of_quantizers} (Appendix \ref{app:subsubsec:info_theory_for_general_spaces}),
we have
\begin{IEEEeqnarray}{c}
I_p(X;V|Z) = I_p(X,Z;V,Z) - H_p(Z).\label{eq:rewrite_conditional_info_with_entropy_in_proof}
\end{IEEEeqnarray}
Fix some $\varepsilon>0.$ For every $n \geq 1,$ let $(\mathcal{C}^{(n)}_{z^n})_{z^n \in \mathcal{Z}^n}$ be a random family of mutually independent codebooks, with the codebook of index $z^n$ having $\lfloor 2^{n(R+\varepsilon)}\rfloor \times \lfloor 2^{nR_c}\rfloor$ i.i.d. codewords sampled from $\prod_{t=1}^n p_{V|Z{=}z_t}.$ The codewords are indexed by couples $(m,j).$ We also denote this random family by $\mathcal{C}^{(n)}_{\mathcal{Z}^n},$ and its distribution by $\mathbb{Q}_{\mathcal{C}^{(n)}_{\mathcal{Z}^n}}.$

\subsection{Distribution $Q^{(1)}$ and soft-covering lemma}\label{sec:Q_1_two_sided}
\begin{figure}[t!]
\centering\includegraphics[width=0.7\textwidth]{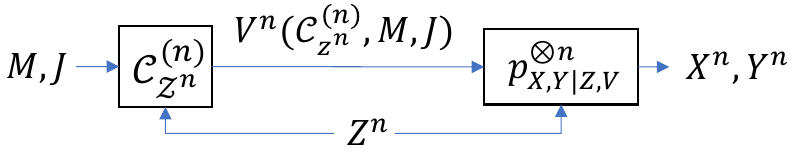}
\caption{
Graphical model for the distribution $Q^{(1)}$ appearing in the proof of Theorem \ref{theorem:E_D_rates_region_marginal_realism} (E-D-achievable tuples with marginal realism).
This is an instance of the graphical model involved in the soft covering lemma with side information \cite[Lemma~VII.5]{2013PaulCuffDistributedChannelSynthesis}.
}
\label{fig:soft_covering_setup_two_sided}
\end{figure}
For every positive integer $n,$ we define the distribution $Q^{(1)},$ described in Figure \ref{fig:soft_covering_setup_two_sided}, as follows:
\begin{IEEEeqnarray}{rCl}
Q^{(1)}_{\mathcal{C}^{(n)}_{\mathcal{Z}^n}, M,J,Z^n,V^n,X^n,Y^n}
&:=& \ \mathbb{Q}_{\mathcal{C}^{(n)}_{\mathcal{Z}^n}}
\cdot p^{\mathcal{U}}_{[2^{\scalebox{0.45}{$n(R+\varepsilon)$}}]} \cdot p^{\mathcal{U}}_{[2^{nR_c}]}
\cdot p_{Z}^{\otimes n}
\cdot \mathbf{1}_{V^n{=}v^n(c^{(n)}_{z^n},m,j)}
\nonumber\\*
&&\cdot \prod_{t=1}^n 
p_{\scalebox{0.55}{$X|Z\text{=}z_t, V\text{=}v_t$}}
\cdot \prod_{t=1}^n 
p_{\scalebox{0.55}{$Y|Z\text{=}z_t, V\text{=}v_t$}}.
\IEEEeqnarraynumspace\label{eq:def_Q_1_two_sided}
\end{IEEEeqnarray}
Hereafter, when referring to a conditional probability kernel of $Q^{(1)},$ it will be the corresponding one appearing in \eqref{eq:def_Q_1_two_sided} unless stated otherwise.
By the definition of $\mathbb{Q}_{\mathcal{C}^{(n)}_{\mathcal{Z}^n}},$ we have $Q^{(1)}_{Z^n,V^n} \equiv p_{Z,V}^{\otimes n},$ and therefore, from \eqref{eq:def_Q_1_two_sided} and the Markov chain in \eqref{eq:def_D_E_D} we have 
\begin{IEEEeqnarray}{c}
Q^{(1)}_{X^n, Y^n, Z^n,V^n} \equiv p_{X,Y,Z,V}^{\otimes n}.\label{eq:Q_1_two_sided_averaged_over_codebook_gives_p}
\end{IEEEeqnarray}
Hence, 
from \eqref{eq:introducing_Delta_two_sided} and the additivity of $d,$ we have
\begin{equation}\label{eq:Distortion_Q_1_two_sided}
    \mathbb{E}_{Q^{(1)}}[d(X^n,Y^n)] \leq \Delta.
\end{equation}

\subsubsection{Marginals of $Y^n$ and $(X^n,Z^n)$ knowing a codebook}
\hfill\\
\indent We start by stating a standard lemma regarding the total variation distance. See Appendix \ref{app:subsec:proofs_of_measure_theory_lemmas} for a proof for general alphabets.
\begin{lemma}\label{lemma:get_expectation_out_of_TV}
    Let $\Pi$ and $\Gamma$ be two distributions on the product of two Polish spaces $\mathcal{W} $ and $\mathcal{L},$ and let $\Pi_{L|W}, \Gamma_{L|W}$ be two channels. Then, we have \begin{equation*}
        \| \Pi_W \Pi_{L|W} - \Pi_W \Gamma_{L|W} \|_{TV} = \mathbb{E}_{\Pi_W} \big[ \| \Pi_{L|W} - \Gamma_{L|W} \|_{TV} \big].
    \end{equation*}
\end{lemma}

We use the \textit{soft covering lemma} with side information
\cite[Corollary~VII.5]{2013PaulCuffDistributedChannelSynthesis} in its memoryless case. A nearly direct application of these two lemmas gives --- see Appendix \ref{app:subsec:use_of_soft_covering_marginal_two_sided} for details ---, from \eqref{eq:introducing_R_two_sided}, \eqref{eq:introducing_R_c_two_sided}, \eqref{eq:replace_conditional_entropy_by_entropy_in_proof} and \eqref{eq:rewrite_conditional_info_with_entropy_in_proof}:
\begin{claim}\label{claim:use_of_soft_covering_marginal_two_sided}
\begin{IEEEeqnarray}{rCl}
\mathbb{E}_{\mathcal{C}^{(n)}_{\mathcal{Z}^n}}\big[\|Q^{(1)}_{Y^n|\mathcal{C}^{(n)}_{\mathcal{Z}^n}} - p_{X}^{\otimes n}\|_{TV}\big] &\underset{n \to \infty}{\longrightarrow}& 0 \label{eq:TV_Q_1_Y_two_sided}\\
\mathbb{E}_{\mathcal{C}^{(n)}_{\mathcal{Z}^n}}\big[\|Q^{(1)}_{J, X^n, Z^n|\mathcal{C}^{(n)}_{\mathcal{Z}^n}} - p^{\mathcal{U}}_{[2^{nR_c}]}p_{X, Z}^{\otimes n}\|_{TV}\big] &\underset{n \to \infty}{\longrightarrow}& 0.\IEEEeqnarraynumspace\label{eq:TV_Q_1_J_X_Z_two_sided}
\end{IEEEeqnarray}
\end{claim}

\vspace{10pt}
\subsubsection{Choosing a codebook}
\hfill\\
\indent From \eqref{eq:Distortion_Q_1_two_sided},
Lemma \ref{lemma:get_expectation_out_of_TV}, and
the Markov inequality
we obtain
\begin{IEEEeqnarray}{c}\label{eq:Markov_ineq_applied_two_sided}
    \scalebox{0.87}{$\mathbb{Q}_{\mathcal{C}^{(n)}_{\mathcal{Z}^n}}\Big(\mathbb{E}_{Q^{(1)}}[d(X^n,Y^n)|\mathcal{C}^{(n)}_{\mathcal{Z}^n}] \leq \Delta + \varepsilon \Big)
    \geq \varepsilon/(\Delta+\varepsilon).$}\IEEEeqnarraynumspace
\end{IEEEeqnarray}
In addition, 
we obtain convergence in probability from \eqref{eq:TV_Q_1_Y_two_sided} and \eqref{eq:TV_Q_1_J_X_Z_two_sided}. Combining this with \eqref{eq:Markov_ineq_applied_two_sided} gives that for a certain $N_0,$ $\forall n\geq N_0$ there is a codebook $c^{(n)}_{\mathcal{Z}^n,*}$ such that
\begin{IEEEeqnarray}{rCl}
   \scalebox{0.9}{$\mathbb{E}$}_{Q^{(1)}_{X^n,Y^n|\mathcal{C}^{(n)}_{\mathcal{Z}^n}{=}c^{(n)}_{\mathcal{Z}^n,*}}} \scalebox{0.9}{$[d(X^n,Y^n)]$} &\leq& \scalebox{0.9}{$\Delta+\varepsilon, $}\label{eq:Distortion_Q_1_fixed_codebook_two_sided}
   \IEEEeqnarraynumspace\\
    \scalebox{0.9}{$\|Q^{(1)}_{Y^n|\mathcal{C}^{(n)}_{\mathcal{Z}^n}{=}c^{(n)}_{\mathcal{Z}^n,*}} - p_{X}^{\otimes n}\|_{TV}$} &\leq&
    \scalebox{0.9}{$\varepsilon, $}\label{eq:TV_Q_1_Y_fixed_codebook_two_sided}
    \IEEEeqnarraynumspace\\
    \scalebox{0.9}{$\|Q^{(1)}_{J, X^n, Z^n|\mathcal{C}^{(n)}_{\mathcal{Z}^n}{=}c^{(n)}_{\mathcal{Z}^n,*}} - p^{\mathcal{U}}_{[2^{nR_c}]}p_{X, Z}^{\otimes n}\|_{TV}$} &\leq& \scalebox{0.9}{$\varepsilon. $}\label{eq:TV_Q_1_J_X_Z_fixed_codebook_two_sided}
    \IEEEeqnarraynumspace
\end{IEEEeqnarray}

\subsection{Constructing a code}\label{subsec:E_D_MR_from_Q_to_P}
In this subsection, we show how the distribution \scalebox{0.9}{$Q^{(1)}_{|\mathcal{C}^{(n)}_{\mathcal{Z}^n}{=}c^{(n)}_{\mathcal{Z}^n,*}}$} can be modified
to lead to a code,
with a negligible effect on the two distributions of interest: the marginal of $Y^n$ and the average empirical distribution of $(X^n,Y^n).$
The analysis of the expected distortion is carried out in Section \ref{sec:conclusion_of_achievability_two_sided}, where we use the uniform integrability in conjunction with the equivalence between near-perfect and perfect realism.
By definition \eqref{eq:def_Q_1_two_sided} of \scalebox{0.9}{$Q^{(1)}_{|\mathcal{C}^{(n)}_{\mathcal{Z}^n}{=}c^{(n)}_{\mathcal{Z}^n,*}},$} we have
\begin{IEEEeqnarray}{rCl}
Q^{(1)}_{\scalebox{0.7}{$J, Z^n,M,V^n,X^n,Y^n|\mathcal{C}^{(n)}_{\mathcal{Z}^n}{=}c^{(n)}_{\mathcal{Z}^n,*}$}} 
&\equiv& 
Q^{(1)}_{\scalebox{0.7}{$J, Z^n,M,V^n,X^n|\mathcal{C}^{(n)}_{\mathcal{Z}^n}{=}c^{(n)}_{\mathcal{Z}^n,*}$}} 
\cdot \prod_{t=1}^n p_{\scalebox{0.55}{$Y|Z\text{=}z_t, V\text{=}v_t$} } \nonumber\\*
&\equiv&
Q^{(1)}_{\scalebox{0.7}{$J, X^n, Z^n|\mathcal{C}^{(n)}_{\mathcal{Z}^n}{=}c^{(n)}_{\mathcal{Z}^n,*}$}} 
\cdot Q^{(1)}_{\scalebox{0.7}{$M,V^n| J\text{=}j, X^n\text{=}x^n, Z^n\text{=}z^n,\mathcal{C}^{(n)}_{\mathcal{Z}^n}{=}c^{(n)}_{\mathcal{Z}^n,*}$}}
\cdot \prod_{t=1}^n p_{\scalebox{0.55}{$Y|Z\text{=}z_t, V\text{=}v_t$} }.\label{eq:rewriting_of_Q_1_two_sided}\IEEEeqnarraynumspace
\end{IEEEeqnarray}
We construct a distribution $P^{(1)}$ which differs from \scalebox{0.9}{$Q^{(1)}_{|\mathcal{C}^{(n)}_{\mathcal{Z}^n}{=}c^{(n)}_{\mathcal{Z}^n,*}}$} in having the correct marginal for $(X^n,Z^n,J)$ as follows:
\begin{IEEEeqnarray}{rCl}
P^{(1)}_{J,X^n,Z^n,M,V^n,Y^n}
&:=& p^{\mathcal{U}}_{[2^{nR_c}]}
\cdot p_{\scalebox{0.7}{$X, Z$}}^{\otimes n}
\cdot Q^{(1)}_{\scalebox{0.7}{$M,V^n| J\text{=}j, X^n\text{=}x^n, Z^n\text{=}z^n,\mathcal{C}^{(n)}_{\mathcal{Z}^n}{=}c^{(n)}_{\mathcal{Z}^n,*}$}}
\cdot \prod_{t=1}^n p_{\scalebox{0.55}{$Y|Z\text{=}z_t, V\text{=}v_t$} },\label{eq:long_def_P_1_two_sided}
\end{IEEEeqnarray}
which from \eqref{eq:rewriting_of_Q_1_two_sided}
can be written as
\begin{IEEEeqnarray}{rCl}
P^{(1)}_{J,X^n,Z^n,M,V^n,Y^n}
&\equiv&
p^{\mathcal{U}}_{[2^{nR_c}]}
\cdot p_{\scalebox{0.7}{$X, Z$}}^{\otimes n}
\cdot Q^{(1)}_{\scalebox{0.7}{$M,V^n,Y^n| J\text{=}j, X^n\text{=}x^n, Z^n\text{=}z^n,\mathcal{C}^{(n)}_{\mathcal{Z}^n}{=}c^{(n)}_{\mathcal{Z}^n,*}$}}
.\label{eq:short_def_P_1_two_sided} 
\end{IEEEeqnarray}

From \eqref{eq:long_def_P_1_two_sided},
distribution $P^{(1)}$ satisfies
\begin{equation}
    (X^n,M,J)-(Z^n,V^n)-Y^n \ \text{, and hence } \ X^n-(Z^n,M,J,V^n)-Y^n.\label{eq:Markov_for_P_1_with_quantizations_two_sided}
\end{equation}
Hence, by Lemma \ref{lemma:conditions_for_P_to_define_a_code}, $P^{(1)}$ defines a $(n,R+\varepsilon,R_c)$ E-D-code.

\subsection{Distance between $P^{(1)}$ and $Q^{(1)}$}
\label{subsec:distance_P_1_Q_1_E_D_MR}
\indent We start by citing some lemmas appearing in \cite{2013PaulCuffDistributedChannelSynthesis} with finite alphabets.
See Appendix \ref{app:subsec:proofs_of_measure_theory_lemmas} for proofs for general alphabets.
\begin{lemma}\label{lemma:TV_joint_to_TV_marginal}
    Let $\Pi$ and $\Gamma$ be two distributions on an alphabet $\mathcal{W} \times \mathcal{L}.$ Then,
    \begin{equation*}
        \| \Pi_W - \Gamma_W \|_{TV} \leq \| \Pi_{W,L} - \Gamma_{W,L} \|_{TV}.
    \end{equation*}
\end{lemma}
\begin{lemma}\label{lemma:TV_same_channel}
    Let $\Pi$ and $\Gamma$ be two distributions on an alphabet $\mathcal{W} \times \mathcal{L}.$ Then, when using the same channel $\Pi_{L|W}$ we have \begin{equation*}
        \| \Pi_W \Pi_{L|W} - \Gamma_W \Pi_{L|W} \|_{TV} = \| \Pi_W - \Gamma_W \|_{TV}.
    \end{equation*}
\end{lemma}

From \eqref{eq:short_def_P_1_two_sided} and Lemma \ref{lemma:TV_same_channel}, comparing $P^{(1)}$ with $Q^{(1)}$ reduces to comparing marginals, i.e., to \eqref{eq:TV_Q_1_J_X_Z_fixed_codebook_two_sided}:
\begin{IEEEeqnarray}{c}
\Big\|P^{(1)}_{\scalebox{0.6}{$M, J, V^n, X^n, Z^n, Y^n$}} - 
\ Q^{(1)}_{\scalebox{0.6}{$M, J, V^n, X^n, Z^n, Y^n|\mathcal{C}^{(n)}_{\mathcal{Z}^n}{=}c^{(n)}_{\mathcal{Z}^n,*}$}}\Big\|_{TV}
\nonumber
=
\Big\|P^{(1)}_{\scalebox{0.6}{$J, X^n, Z^n$}} - Q^{(1)}_{\scalebox{0.6}{$J, X^n, Z^n|\mathcal{C}^{(n)}_{\mathcal{Z}^n}{=}c^{(n)}_{\mathcal{Z}^n,*}$}} \Big\|_{TV}\leq \varepsilon. \label{eq:TV_P_1_Q_1_two_sided}
\end{IEEEeqnarray} 
Therefore, by Lemma \ref{lemma:TV_joint_to_TV_marginal} with $W=(X^n, Y^n)$ and again $W=Y^n,$
we obtain
\begin{IEEEeqnarray}{c}
    \Big\|P^{(1)}_{Y^n} - Q^{(1)}_{Y^n|\mathcal{C}^{(n)}_{\mathcal{Z}^n}{=}c^{(n)}_{\mathcal{Z}^n,*}}\Big\|_{TV} \leq \Big\|P^{(1)}_{X^n, Y^n} - Q^{(1)}_{X^n, Y^n|\mathcal{C}^{(n)}_{\mathcal{Z}^n}{=}c^{(n)}_{\mathcal{Z}^n,*}}\Big\|_{TV} \leq \varepsilon. \label{eq:TV_sur_X_Y_P_1_Q_1_two_sided}
\end{IEEEeqnarray}
By the triangle inequality and Lemma \ref{lemma:TV_joint_to_TV_marginal} with $W=(X_t,Y_t)$ for $t\in [n],$ we obtain
\begin{IEEEeqnarray}{c}
    \Big\|\hat{P}^{(1)}_{X^n, Y^n} - \hat{Q}^{(1)}_{X^n, Y^n|\mathcal{C}^{(n)}_{\mathcal{Z}^n}{=}c^{(n)}_{\mathcal{Z}^n,*}}\Big\|_{TV} \leq \Big\|P^{(1)}_{X^n, Y^n} - Q^{(1)}_{X^n, Y^n|\mathcal{C}^{(n)}_{\mathcal{Z}^n}{=}c^{(n)}_{\mathcal{Z}^n,*}}\Big\|_{TV} \leq \varepsilon. \label{eq:TV_empirical_sur_X_Y_P_1_Q_1_two_sided}
\end{IEEEeqnarray}

\subsection{Conclusion}\label{sec:conclusion_of_achievability_two_sided}
Since $(d,p_X)$ is uniformly integrable, then from \eqref{eq:Distortion_Q_1_fixed_codebook_two_sided}, \eqref{eq:TV_Q_1_Y_fixed_codebook_two_sided}, \eqref{eq:TV_sur_X_Y_P_1_Q_1_two_sided} and \eqref{eq:TV_empirical_sur_X_Y_P_1_Q_1_two_sided}, for every $n\geq N_0,$ we can use Proposition \ref{proposition:to_perfect_realism} with $Q = Q^{(1)}_{|\mathcal{C}^{(n)}_{\mathcal{Z}^n}{=}c^{(n)}_{\mathcal{Z}^n,*}},$ $P=P^{(1)},$ $\varepsilon_1 = \varepsilon_2 = \varepsilon_3 = \varepsilon_4 = \varepsilon.$
Then, there is a sequence of $(n,R+\varepsilon,R_c)$ E-D-codes with the same encoder as $P^{(1)},$ and inducing a distribution $P^{(2)}$
such that for large enough $n$ we have
\begin{IEEEeqnarray}{rCl}
\mathbb{E}_{P^{(2)}}[d(X^n, Y^n)] \leq \Delta + \varepsilon + \sup_{\mathbb{P}_{X,Y,\xi}} \mathbb{E}[d(X,Y) \cdot \xi]
\label{eq:distortion_P_3_two_sided}
\
&\text{ and } P^{(2)}_{Y^n} \equiv p_X^{\otimes n},&
\end{IEEEeqnarray}
where the supremum is taken over all distributions $\mathbb{P}_{X,Y,\xi}$ on $\mathcal{X}^2 \times \{0,1\}$ satisfying $\mathbb{P}_X \equiv \mathbb{P}_Y \equiv p_X$ and $\mathbb{P}(\{\xi=1\}) \leq 3\varepsilon.$
This proves the E-D-achievability of $(R+\varepsilon, R_c, \Delta+\varepsilon +s(\varepsilon)),$ 
where $s(\varepsilon)$ is the supremum in \eqref{eq:distortion_P_3_two_sided} --- which does not depend on $n.$ This is true for every $\varepsilon>0.$ By
the uniform integrability assumption, we have $s(\varepsilon) \to 0$
if $\varepsilon \to 0.$ Therefore, we obtain $(R, R_c, \Delta) \in \overline{\mathcal{A}}^{(m)}_{E\text{-}D},$ as desired.

\section{Proof of Theorem \ref{theorem:D_rates_region_marginal_realism}}\label{sec:achievability}

\subsection{Informal outline} 
Similarly to the approach of \cite[Section~IV]{2016SongCuffLikelihoodEncoder}, we introduce a virtual message $M'$ with rate $R'$ generated and used by the encoder, but not transmitted (Subsection \ref{sec:where_we_fix_epsilon}). The decoder will then guess it.
We start with a distribution $Q^{(1)}$
(Subsection \ref{sec:Q_1}), analyzed using the soft covering lemma of \cite{2013PaulCuffDistributedChannelSynthesis}, 
then change it little by little to obtain an intermediate distribution $Q^{(2)}$
(Subsection \ref{subsubsec:def_Q_2}) and a distribution $P^{(1)}$ corresponding to a coding scheme (Subsection \ref{subsec:defining_P_1_one_sided}). In order to obtain a final distortion bound, we use uniform integrability in conjunction with the equivalence between achievability with near-perfect and perfect realism. The latter argument is needed if the source alphabet is not finite.

\subsection{Random codebook indexed by a virtual message}\label{sec:where_we_fix_epsilon}
Here, we prove that $\overline{\mathcal{S
}}^{(m)}_{D} \subseteq \overline{\mathcal{A}}^{(m)}_{D}$ by proving that $\mathcal{S}^{(m)}_{D} \subseteq \overline{\mathcal{A}}^{(m)}_{D}.$
Let $(R,R_c,\Delta)$ be a triplet in $\mathcal{S}^{(m)}_{D}.$
Let $p_{X,Y,Z,V}$ be a corresponding distribution from the definition of $\mathcal{S
}^{(m)}_{D}.$ Then, $I_p(Z;V) < \infty,$ and
\begin{IEEEeqnarray}{rCl}
R &\geq& I_p(X;V) - I_p(Z;V)
\label{eq:introducing_R}
\\
R+R_c &\geq& I_p(Y;V) - I_p(Z;V)
\label{eq:introducing_R_c}
\\
\Delta &\geq& \mathbb{E}_p[d(X,Y)].
\label{eq:introducing_Delta}
\end{IEEEeqnarray}
Fix some $\varepsilon>0.$ We introduce a rate $R'$ corresponding to a virtual message $M',$ as follows. If $I_p(Z;V)=0$ then set $R'=0.$ Otherwise, fix $R'$ in $ \big(I_p(Z;V) - \varepsilon, I_p(Z;V)\big) \cap \big(0,+\infty \big).$ 
Then, from \eqref{eq:introducing_R}, 
\eqref{eq:introducing_R_c}, the Markov chain property $Z-X-V,$ and the chain rule for general alphabets (Proposition \ref{prop:chain_rule}, Appendix \ref{app:subsubsec:info_theory_for_general_spaces}) we have:
\begin{IEEEeqnarray}{rCl}\label{eq:sum_rate_large_enough_for_X_and_Z}
    R+ R' +\varepsilon &>& I_p(X;V) = I_p(X,Z;V),\\
\label{eq:sum_rate_large_enough_for_Y}
    R + R' + R_c +\varepsilon &>& I_p(Y;V),
\end{IEEEeqnarray}where the equality in \eqref{eq:sum_rate_large_enough_for_X_and_Z} follows from the Markov chain $Z-X-V$ of \eqref{eq:def_D_D} and the chain rule for general alphabets (Proposition \ref{prop:chain_rule}).
For every $n \geq 1,$ let $\mathcal{C}^{(n)}$ be a random codebook with $\lfloor 2^{n(R+\varepsilon)}\rfloor \times \lfloor 2^{nR'}\rfloor \times \lfloor 2^{nR_c}\rfloor$ i.i.d. codewords sampled from $p_{V}^{\otimes n}.$ The codewords are indexed by triples $(m,m',j).$ We denote this random codebook distribution by $\mathbb{Q}_{\mathcal{C}^{(n)}}.$

\subsection{Distribution $Q^{(1)}$ and soft-covering lemma}\label{sec:Q_1}
\begin{figure}[t!]
\centering\includegraphics[width=0.7\textwidth]{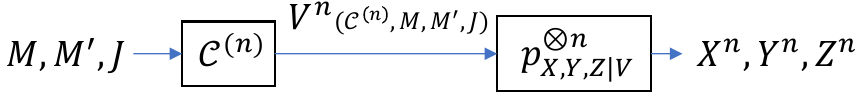}
\caption{
Graphical model for the distribution $Q^{(1)}$ appearing in the proof of Theorem \ref{theorem:D_rates_region_marginal_realism} (D-achievable tuples with marginal realism).
This is an instance of the graphical model involved in the soft covering lemma \cite[Lemma~VII.4]{2013PaulCuffDistributedChannelSynthesis}.
}
\label{fig:soft_covering_setup}
\end{figure}
For every positive integer $n,$ we define the distribution $Q^{(1)},$ described in Figure \ref{fig:soft_covering_setup}, as follows:
\begin{IEEEeqnarray}{rCl}
Q^{(1)}_{\mathcal{C}^{(n)}, M,M',J,Z^n,V^n,X^n,Y^n}
&:=& \ \mathbb{Q}_{\mathcal{C}^{(n)}}
\cdot p^{\mathcal{U}}_{[2^{\scalebox{0.45}{$n(R+\varepsilon)$}}]}
\cdot p^{\mathcal{U}}_{[2^{\scalebox{0.45}{$nR'$}}]} 
\cdot p^{\mathcal{U}}_{[2^{nR_c}]}
\cdot \mathbf{1}_{V^n{=}v^n(c^{(n)},m,m',j)}
\nonumber\\*
&&\cdot \prod_{t=1}^n 
p_{\scalebox{0.55}{$X|V\text{=}v_t$}}
\cdot \prod_{t=1}^n 
p_{\scalebox{0.55}{$Z|X\text{=}x_t$}}
\cdot \prod_{t=1}^n 
p_{\scalebox{0.55}{$Y|Z\text{=}z_t, V\text{=}v_t$}}.
\IEEEeqnarraynumspace\label{eq:def_Q_1_one_sided}
\end{IEEEeqnarray}
Hereafter, when referring to a conditional probability kernel of $Q^{(1)},$ it will be the corresponding one appearing in \eqref{eq:def_Q_1_one_sided} unless stated otherwise. By the definition of $\mathbb{Q}_{\mathcal{C}^{(n)}},$ we have $Q^{(1)}_{V^n} \equiv p_V^{\otimes n},$ and therefore, from \eqref{eq:def_Q_1_one_sided} and the Markov chains in \eqref{eq:def_D_D}, we have
\begin{IEEEeqnarray}{c}
Q^{(1)}_{V^n,X^n, Z^n, Y^n} \equiv p_{V,X,Z,Y}^{\otimes n}.\label{eq:Q_1_averaged_over_codebook_gives_p}
\end{IEEEeqnarray}
Hence, 
from \eqref{eq:introducing_Delta} and the additivity of $d,$ we have
\begin{equation}\label{eq:Distortion_Q_1}
    \mathbb{E}_{Q^{(1)}}[d(X^n,Y^n)] \leq \Delta.
\end{equation}

\vspace{10pt}
\subsubsection{Marginals of $Y^n$ and $(X^n,Z^n)$ knowing a codebook}
\hfill\\
\indent We use the \textit{soft covering lemma} for general alphabets \cite[Corollary~VII.4]{2013PaulCuffDistributedChannelSynthesis} in its memoryless case. A nearly direct application gives --- see Appendix \ref{app:subsec:use_of_soft_covering_marginal_one_sided} for details ---, from \eqref{eq:sum_rate_large_enough_for_X_and_Z} and \eqref{eq:sum_rate_large_enough_for_Y}:
\begin{claim}\label{claim:use_of_soft_covering_marginal_one_sided}
\begin{IEEEeqnarray}{rCl}
\mathbb{E}_{\mathcal{C}^{(n)}}\big[\|Q^{(1)}_{Y^n|\mathcal{C}^{(n)}} - p_{X}^{\otimes n}\|_{TV}\big] &\underset{n \to \infty}{\longrightarrow}& 0 \label{eq:TV_Q_1_Y_one_sided}\\
\mathbb{E}_{\mathcal{C}^{(n)}}\big[\|Q^{(1)}_{J, X^n, Z^n|\mathcal{C}^{(n)}} - p^{\mathcal{U}}_{[2^{nR_c}]}p_{X, Z}^{\otimes n}\|_{TV}\big] &\underset{n \to \infty}{\longrightarrow}& 0.\IEEEeqnarraynumspace\label{eq:TV_Q_1_J_X_Z_one_sided}
\end{IEEEeqnarray}
\end{claim}

\subsubsection{Decoding of $M'$}\label{subsec:defining_M_hat_prime}
\hfill\\
\indent The virtual message $M'$ is to be decoded as $\hat{M'},$ conditionally independent from $M',X^n,Y^n$ knowing $M,J,Z^n,\mathcal{C}^{(n)}:$ in the case of finite alphabets,
\begin{equation*}
Q^{(1)}_{\scalebox{0.7}{$\hat{M}' | \mathcal{C}^{(n)}{=}c^{(n)}, M{=}m, J{=}j, Z^n{=}z^n$}}
\end{equation*}
is a joint $p_{V,Z}$-typicality decoder for sub-codebook $(v^n(c^{(n)},m,a,j))_a$ (see, e.g., \cite[page~201]{Cover&Thomas2006}).
By the standard random coding argument for channel coding (see, e.g., \cite[Section~7.7]{Cover&Thomas2006}) and using the fact that $R'<I_p(Z;V)$ (by construction), we can obtain
\begin{IEEEeqnarray}{c}
\scalebox{0.9}{$\substack{\scalebox{1.0}{$Q^{(1)} \; (\hat{M}'\neq M') $} \\ \hat{M}', M' \qquad \qquad} \underset{n \to \infty}{\longrightarrow} 0$}\label{eq:Q_1_decoding_error_fixed_codebook_finite_valued_sources} \IEEEeqnarraynumspace
\end{IEEEeqnarray}
in the case of finite alphabets.
\begin{claim}\label{claim:decoding_virtual_message_general_alphabets}
For general sources, there exists a decoder satisfying \eqref{eq:Q_1_decoding_error_fixed_codebook_finite_valued_sources}.
\end{claim}

See Appendix \ref{app:decoder_M_prime} for
the
details and
proof. 
The probability in \eqref{eq:Q_1_decoding_error_fixed_codebook_finite_valued_sources} can be rewritten as an expectation over $\mathcal{C}^{(n)}.$
Since the convergence towards zero in expectation implies a convergence in probability for a non-negative variable, we obtain:
\begin{IEEEeqnarray}{c}\label{eq:decoding_error_Q_1}
     \scalebox{0.9}{$Q^{(1)} \big( M' \neq \hat{M}' \big| \ \mathcal{C}^{(n)} \big) \overset{\mathcal{P}}{\underset{n \to \infty}{\longrightarrow}} 0.$}\IEEEeqnarraynumspace
\end{IEEEeqnarray}

\subsubsection{Choosing a codebook}
\hfill\\
\indent From \eqref{eq:Distortion_Q_1}
and
the Markov inequality
we obtain
\begin{IEEEeqnarray}{c}\label{eq:Markov_ineq_applied}
    \scalebox{0.87}{$\mathbb{Q}_{\mathcal{C}^{(n)}}\Big(\mathbb{E}_{Q^{(1)}}[d(X^n,Y^n)|\mathcal{C}^{(n)}] \leq \Delta + \varepsilon \Big)
    \geq \varepsilon/(\Delta+\varepsilon).$}\IEEEeqnarraynumspace
\end{IEEEeqnarray}
In addition, similarly to \eqref{eq:decoding_error_Q_1}, we obtain convergence in probability from \eqref{eq:TV_Q_1_Y_one_sided} and \eqref{eq:TV_Q_1_J_X_Z_one_sided}. Combining this with
\eqref{eq:Markov_ineq_applied} gives that for a certain $N_0,$ $\forall n\geq N_0$ there is a codebook $c_*^{(n)}$ such that
\begin{IEEEeqnarray}{rCl}
   \scalebox{0.9}{$\mathbb{E}$}_{Q^{(1)}_{X^n,Y^n|\mathcal{C}^{(n)}{=}c_*^{(n)}}} \scalebox{0.9}{$[d(X^n,Y^n)]$} &\leq& \scalebox{0.9}{$\Delta+\varepsilon, $}
   \nonumber\\
    \scalebox{0.9}{$\|Q^{(1)}_{Y^n|\mathcal{C}^{(n)}{=}c_*^{(n)}} - p_{X}^{\otimes n}\|_{TV}$} &\leq& \scalebox{0.9}{$\varepsilon, $}
    \nonumber\\
    \scalebox{0.9}{$\|Q^{(1)}_{J, X^n, Z^n|\mathcal{C}^{(n)}{=}c_*^{(n)}} - p^{\mathcal{U}}_{[2^{nR_c}]}p_{X, Z}^{\otimes n}\|_{TV}$} &\leq& \scalebox{0.9}{$\varepsilon, $}
    \label{eq:TV_Q_1_J_X_Z_fixed_codebook}
    \IEEEeqnarraynumspace\\
    \scalebox{0.9}{$\substack{\scalebox{1.0}{$Q^{(1)} \; (\hat{M}'\neq M') $} \\ M', \hat{M}' |\mathcal{C}^{(n)}{=}c_*^{(n)}}$} &\leq& \scalebox{0.9}{$\varepsilon .$}
    \label{eq:Q_1_decoding_error_fixed_codebook}
    \IEEEeqnarraynumspace
\end{IEEEeqnarray}

\subsection{Construction of a code}
In this subsection, we show how the distribution 
\scalebox{0.9}{$Q^{(1)}_{|\mathcal{C}^{(n)}{=}c^{(n)}_*}$} can be modified
to lead to a code,
with a negligible effect on the error for decoding $M'$ and on the two distributions of interest: the marginal of $Y^n$ and the average empirical distribution of $(X^n,Y^n).$
The analysis of the expected distortion is carried out in Section \ref{sec:conclusion_of_achievability_two_sided}, where we use the uniform integrability in conjunction with the equivalence between near-perfect and perfect realism.

\subsubsection{Effect of using the decoded virtual message $\hat{M}'$}\label{subsubsec:def_Q_2}
\hfill\\
\indent For every positive integer $n$ we define the following distribution which differs from
\scalebox{0.9}{$Q^{(1)}_{|\mathcal{C}^{(n)}{=}c^{(n)}_*}$}
in that $Y^n$ is sampled using $\hat{M}'$ instead of $M':$
\begin{IEEEeqnarray}{rCl}
Q^{(2)}_{M,M',J,V^n,X^n,Z^n,\hat{M}',Y^n}
&:=& \ p^{\mathcal{U}}_{[2^{\scalebox{0.45}{$n(R+\varepsilon)$}}]}
\cdot p^{\mathcal{U}}_{[2^{\scalebox{0.45}{$nR'$}}]} 
\cdot p^{\mathcal{U}}_{[2^{nR_c}]}
\cdot \mathbf{1}_{\scalebox{0.65}{$V^n{=}v^n(c_*^{(n)},m,m',j)$}}
\cdot \prod_{t=1}^n 
p_{\scalebox{0.6}{$X|V\text{=}v_t$}}
\cdot \prod_{t=1}^n 
p_{\scalebox{0.6}{$Z|X\text{=}x_t$}}
\nonumber\\*
&&\cdot \;
Q^{(1)}_{\scalebox{0.65}{$\hat{M}' | M{=}m, J{=}j, Z^n{=}z^n,\mathcal{C}^{(n)}{=}c^{(n)}_*$}}
\cdot \prod_{t=1}^n p_{\scalebox{0.65}{$Y|Z\text{=}z_t, V\text{=}v_t(c_*^{(n)},m,\hat{m}',j)$}}
\IEEEeqnarraynumspace\label{eq:long_def_Q_2}
\end{IEEEeqnarray}
By definition \eqref{eq:def_Q_1_one_sided} of
\scalebox{0.9}{$Q^{(1)}_{|\mathcal{C}^{(n)}}$}
this can be rewritten as
\begin{IEEEeqnarray}{rCl}
Q^{(2)}_{M,M',J,V^n,X^n,Z^n,\hat{M}',Y^n}
&\equiv& 
Q^{(1)}_{\scalebox{0.7}{$M,M',J,V^n,X^n,Z^n|\mathcal{C}^{(n)}{=}c^{(n)}_*$}}
\cdot Q^{(1)}_{\scalebox{0.7}{$\hat{M}' | M{=}m, J{=}j, Z^n{=}z^n,\mathcal{C}^{(n)}{=}c^{(n)}_*$}}
\nonumber\\*
&&\cdot 
\prod_{t=1}^n p_{\scalebox{0.7}{$Y|Z\text{=}z_t, V\text{=}v_t(c_*^{(n)},m,\hat{m}',j)$}},\IEEEeqnarraynumspace\label{eq:intermediary_rewriting_Q_2}
\end{IEEEeqnarray}
which can be further rewritten --- e.g. by Corollary \ref{cor:transition_kernel_with_few_active_variables} --- as
\begin{IEEEeqnarray}{c}
Q^{(2)}_{M,M',J,Z^n,V^n,X^n,Y^n,\hat{M}'} 
\equiv Q^{(1)}_{M,M',J,Z^n,V^n,X^n,\hat{M}'|\mathcal{C}^{(n)}{=}c^{(n)}_*} 
\cdot \prod_{t=1}^n p_{\scalebox{0.7}{$Y|Z\text{=}z_t, V\text{=}v_t(c_*^{(n)},m,\hat{m}',j)$}}.\label{eq:def_Q_2}
\end{IEEEeqnarray}
The following lemma appears in \cite{2016SongCuffLikelihoodEncoder} with finite alphabets,
and also holds
for general
ones: it is a direct consequence of the standard relation between coupling and the total variation distance (see, e.g., \cite[Chapter~I, Theorem~5.2]{1992Coupling_Book}).
\begin{lemma}\label{lemma:TV_due_to_M_and_hatM}
    Let $P_{UWL}$ be a distribution on an alphabet of the form $\mathcal{U}\times\mathcal{U}\times\mathcal{L}$ and let $\eta \in (0,1).$ If $P(U \neq W) \leq \eta$ we have $\| P_{UL} - P_{WL}\|_{TV} \leq \eta.$
\end{lemma}
\vspace{5pt}

Using Lemma \ref{lemma:TV_due_to_M_and_hatM} on $Q^{(1)}$ with $U=M', W=\hat{M}',$
$L=(M, J, Z^n, X^n)$ 
we have from \eqref{eq:Q_1_decoding_error_fixed_codebook}
\begin{IEEEeqnarray}{c}
    \big\|Q^{(1)}_{M', M, J, Z^n, X^n|\mathcal{C}^{(n)}{=}c^{(n)}_*} - Q^{(1)}_{\hat{M}', M, J, Z^n, X^n|\mathcal{C}^{(n)}{=}c^{(n)}_*}\big\|_{TV} \leq \varepsilon. \nonumber
\end{IEEEeqnarray}
As a consequence, and by construction of $Q^{(2)}$ and its similarity to that of $Q^{(1)},$ we have by Lemma \ref{lemma:TV_same_channel} with $L=Y^n:$
\begin{IEEEeqnarray}{c}
    \big\|Q^{(1)}_{\scalebox{0.7}{$M', M, J, Z^n, X^n, Y^n|\mathcal{C}^{(n)}{=}c^{(n)}_*$}} - Q^{(2)}_{\scalebox{0.7}{$\hat{M}', M, J, Z^n, X^n, Y^n$}} \big\|_{TV} \leq \varepsilon.
    \nonumber
\end{IEEEeqnarray}Hence, by Lemma \ref{lemma:TV_joint_to_TV_marginal} with $W=(X^n,Y^n)$ we have
\begin{IEEEeqnarray}{c}\label{eq:TV_Q_1_Q_2}
    \big\|Q^{(1)}_{\scalebox{0.7}{$X^n, Y^n|\mathcal{C}^{(n)}{=}c^{(n)}_*$}} - Q^{(2)}_{\scalebox{0.7}{$X^n, Y^n$}} \big\|_{TV} \leq \varepsilon. \IEEEeqnarraynumspace
\end{IEEEeqnarray}
Since $Q^{(2)}_{J, X^n, Z^n} \equiv Q^{(1)}_{J, X^n, Z^n|\mathcal{C}^{(n)}{=}c^{(n)}_*}$ we also have from \eqref{eq:TV_Q_1_J_X_Z_fixed_codebook}:
\begin{equation}\label{eq:TV_Q_2_J_X_Z}
    \|Q^{(2)}_{J, X^n, Z^n} - p^{\mathcal{U}}_{[2^{nR_c}]}p_{X, Z}^{\otimes n}\|_{TV} \leq \varepsilon.
\end{equation}

\subsubsection{Finalizing the code construction}\label{subsec:defining_P_1_one_sided}
\hfill\\
\indent From \eqref{eq:long_def_Q_2}
distribution $Q^{(2)}$ satisfies
\begin{equation}
    Z^n-X^n-(M,M',J,V^n), \text{ and hence } Z^n-(X^n,J)-M.\label{eq:short_Markov_for_Q_2_one_sided}
\end{equation}
From \eqref{eq:intermediary_rewriting_Q_2}, when considering variables other than $M',V^n$ we obtain --- e.g., Corollary \ref{cor:transition_kernel_with_few_active_variables}:
\begin{IEEEeqnarray}{c}
Q^{(2)}_{\scalebox{0.7}{$J,X^n,Z^n,M,\hat{M}',Y^n$}}
\equiv
Q^{(2)}_{\scalebox{0.7}{$J, X^n, Z^n,M$}}
\cdot
Q^{(2)}_{\scalebox{0.7}{$\hat{M}',Y^n| M{=}m, J{=}j, Z^n{=}z^n$}}
.\nonumber
\end{IEEEeqnarray} 
From Markov chain \eqref{eq:short_Markov_for_Q_2_one_sided}, this can be rewritten as
\begin{IEEEeqnarray}{c}
Q^{(2)}_{\scalebox{0.7}{$J,X^n,Z^n,M,\hat{M}',Y^n$}}
\equiv
Q^{(2)}_{\scalebox{0.7}{$J, X^n, Z^n$}}
\cdot
Q^{(2)}_{\scalebox{0.7}{$M| J\text{=}j, X^n\text{=}x^n$}}
\cdot
Q^{(2)}_{\scalebox{0.7}{$\hat{M}',Y^n| M{=}m, J{=}j, Z^n{=}z^n$}}
.\label{eq:rewriting_of_Q_2_for_P_1}
\end{IEEEeqnarray}
We define the distribution $P^{(1)}$ achieving near-perfect marginal realism. It differs from $Q^{(2)}$ in having the correct marginal for $(X^n,Z^n,J){:}$ fix a transition kernel of $Q^{(2)}$ for $M$ knowing $(X^n,J),$ and let
\begin{IEEEeqnarray}{c}
P^{(1)}_{J,X^n,Z^n,M,\hat{M}',Y^n}
:=
p^{\mathcal{U}}_{[2^{nR_c}]}
\cdot
p_{\scalebox{0.7}{$X, Z$}}^{\otimes n}
\cdot
Q^{(2)}_{\scalebox{0.7}{$M| J\text{=}j, X^n\text{=}x^n$}}
\cdot
Q^{(2)}_{\scalebox{0.7}{$\hat{M}',Y^n| M{=}m, J{=}j, Z^n{=}z^n$}}
,\label{eq:long_def_P_1_one_sided}    
\end{IEEEeqnarray}
which from \eqref{eq:rewriting_of_Q_2_for_P_1} can be written as
\begin{IEEEeqnarray}{rCl}
P^{(1)}_{J,X^n,Z^n,M,\hat{M}',Y^n}
&=&
p^{\mathcal{U}}_{[2^{nR_c}]}
\cdot p_{\scalebox{0.7}{$X, Z$}}^{\otimes n}
\cdot Q^{(2)}_{\scalebox{0.7}{$M,\hat{M}',Y^n| J\text{=}j, X^n\text{=}x^n, Z^n\text{=}z^n$}}
.\label{eq:short_def_P_1_one_sided} 
\end{IEEEeqnarray}
From \eqref{eq:long_def_P_1_one_sided} distribution $P^{(1)}$ satisfies
\begin{IEEEeqnarray}{c}
    Z^n-(X^n,J)-M
\ \ \text{and} \ \
X^n-(M,J,Z^n)-(\hat{M}',Y^n).\label{eq:long_Markov_for_P_1_one_sided}
\end{IEEEeqnarray}
The alphabet of $M$ is $[2^{n(R+\varepsilon)}]$ throughout this proof by construction of the random codebook $\mathcal{C}^{(n)}.$
Therefore, from \eqref{eq:short_def_P_1_one_sided}-\eqref{eq:long_Markov_for_P_1_one_sided}
and Lemma \ref{lemma:conditions_for_P_to_define_a_code},
distribution $P^{(1)}$ defines a $(n,R+\varepsilon,R_c)$ D-code.
From \eqref{eq:short_def_P_1_one_sided} and Lemma \ref{lemma:TV_same_channel}, comparing $P^{(1)}$ with $Q^{(2)}$ reduces to comparing marginals, i.e., to \eqref{eq:TV_Q_2_J_X_Z}:
\begin{IEEEeqnarray}{rCl}
\big\|P^{(1)}_{\scalebox{0.6}{$M, M', J, X^n, Z^n, \hat{M}', Y^n$}} - Q^{(2)}_{\scalebox{0.6}{$M, M', J, X^n, Z^n, \hat{M}', Y^n$}}\big\|_{TV}
=
\big\|P^{(1)}_{J, X^n, Z^n} - Q^{(2)}_{J, X^n, Z^n}\big\|_{TV}\leq \varepsilon. \label{eq:TV_P_1_Q_2}
\end{IEEEeqnarray} 
Therefore, by Lemma \ref{lemma:TV_joint_to_TV_marginal} with $W=Y^n$ and $W=(X^n, Y^n),$ the triangle inequality and \eqref{eq:TV_Q_1_Q_2}, we obtain
\begin{IEEEeqnarray}{c}
    \big\|P^{(1)}_{Y^n} - Q^{(1)}_{Y^n|\mathcal{C}^{(n)}{=}c^{(n)}_*}\big\|_{TV} \leq \big\|P^{(1)}_{X^n, Y^n} - Q^{(1)}_{X^n, Y^n|\mathcal{C}^{(n)}{=}c^{(n)}_*}\big\|_{TV} \leq 2\varepsilon. \label{eq:TV_sur_X_Y_P_1_Q_1}
\end{IEEEeqnarray}
By the triangle inequality and Lemma \ref{lemma:TV_joint_to_TV_marginal} with $W=(X_t,Y_t)$ for $t\in [n],$ we obtain
\begin{IEEEeqnarray}{c}
    \big\|\hat{P}^{(1)}_{X^n, Y^n} - \hat{Q}^{(1)}_{X^n, Y^n|\mathcal{C}^{(n)}{=}c^{(n)}_*}\big\|_{TV} \leq \big\|P^{(1)}_{X^n, Y^n} - Q^{(1)}_{X^n, Y^n|\mathcal{C}^{(n)}{=}c^{(n)}_*}\big\|_{TV} \leq 2\varepsilon. \label{eq:TV_empirical_sur_X_Y_P_1_Q_1_one_sided}
\end{IEEEeqnarray}
The end of the proof is identical to Subsection \ref{sec:conclusion_of_achievability_two_sided} and is omitted for the purpose of conciseness.

\section{Proof of Theorem \ref{theorem:E_D_rates_region_joint_realism}}\label{sec:achievability_joint_two_sided}

\subsection{Converse}\label{sec:converse_E_D_joint}

We prove that $\overline{\mathcal{A}}^{(j)}_{E\text{-}D} \subset \overline{\mathcal{S}}^{(j)}_{E\text{-}D}$ by proving that $\mathcal{A}^{(j)}_{E\text{-}D} \subset \overline{\mathcal{S}}^{(j)}_{E\text{-}D}.$
We proceed as in the case of marginal realism (Section \ref{sec:converse_E_D_marginal}).
Let $(R, R_c,\Delta)$ be E-D-achievable with near-perfect joint realism. Then, by Theorem \ref{theorem:equivalence_perfect_realism}, it is E-D-achievable with perfect joint realism. Fix $\varepsilon$\hspace{2pt}$>$\hspace{2pt}$0.$
Then, there exists a $(n,R, R_c)$ E-D-code inducing a joint distribution $P$ such that
$\mathbb{E}_P[d(X^n, Y^n)] \, {\leq} \, \Delta \, + \, \varepsilon$ and $P_{Y^n,Z^n} \, {\equiv} \, p_{X,Z}^{\otimes n}.$
Let $T$ be a uniform random variable over $[n].$ 
Define $V=(M, J, Z_{[n] \backslash T}, T),$ 
which has a finite alphabet.
Since $(X^n,Z^n) \sim p_{X,Z}^{\otimes n},$ the distribution of $(X_T,Z_T)$ is $p_{X,Z}.$ 
Similarly, 
we have $P_{Y_T,Z_T} \equiv p_{X,Z}.$
Distribution $P_{X_T,Z_T,V,Y_T}$ satisfies the
Markov chain in $\mathcal{D}^{(j)}_{E\text{-}D}$ (Eq. \ref{eq:def_D_E_D_joint}) by Lemma \ref{lemma:Markov_chain_Y_T_X_T} (Appendix \ref{app:subsec:Markov_chains_precise_reference_for_converse}).
Thus, $P_{X_T, Y_T, Z_T, V} \in \mathcal{D}^{(j)}_{E\text{-}D}.$ It remains to check that the inequalities in the definition of $\mathcal{S}^{(j)}_{E\text{-}D}$ are satisfied.
Since $P_{X_T,Y_T} \equiv \hat{P}_{X^n,Y^n},$ we have $\mathbb{E}[\scalebox{1.0}{$d(X_T, Y_T)$}] = \mathbb{E}[\scalebox{1.0}{$d(X^n, Y^n)$}] \leq \Delta + \varepsilon.$
We first lower bound the rates using quantized variables before taking the limit. 
Consider any (finite-valued) quantizer on $\mathcal{X}.$ The image of a letter $x$ by the latter is denoted $[x].$ We know that independence is preserved by deterministic mappings.
Hence,
\begin{IEEEeqnarray}{rCl}
    nR \geq H(M)
    \geq
    \scalebox{1.0}{$I(M;[X]^n|Z^n, J)$}
    &=& \scalebox{1.0}{$I(M, J ;[X]^n|Z^n)$} \label{eq:converse_E_D_joint_R_using_J_indep} \\
    &=& \scalebox{1.0}{$\sum_{t=1}^n I(M, J ; [X]_t | Z^n, [X]_{t+1:n})$} \nonumber \\ 
    &=& \scalebox{1.0}{$\sum_{t=1}^n I(M, J, [X]_{t+1:n}, Z_{[n] \backslash t} ; [X]_t | Z_t)$} \nonumber \\
    &\geq& \scalebox{1.0}{$\sum_{t=1}^n I(M, J, Z_{[n] \backslash t} ; [X]_t | Z_t)$} \nonumber \\
    &=& \scalebox{1.0}{$n I(M, J, Z_{[n] \backslash T} ; [X]_T | Z_T, T)$}\IEEEeqnarraynumspace\label{eq:converse_E_D_joint_lower_bound_R_t_to_T} \\
    &=& \scalebox{1.0}{$n I(V ; [X]_T | Z_T),$}\label{eq:converse_E_D_joint_lower_bound_R_T_indep}
\end{IEEEeqnarray}
where \eqref{eq:converse_E_D_joint_R_using_J_indep} follows from the independence between the common randomness and the sources; and \eqref{eq:converse_E_D_joint_lower_bound_R_t_to_T} and \eqref{eq:converse_E_D_joint_lower_bound_R_T_indep} follow from the independence of $T$ and all other variables and from the fact that couples in $\{([X]_t,Z_t)\}_{t \in [n]}$ are i.i.d. Moreover, we have
\begin{IEEEeqnarray}{rCl}
    n(R+R_c) \geq H(M,J)
    \geq
    I(M,J;[Y]^n|Z^n)
    &=& \sum_{t=1}^n I(M,J ;[Y]_t|Z^n, [Y]_{1:t-1}) \nonumber \\
    &=& \sum_{t=1}^n I(M,J,[Y]_{1:t-1} ;[Y]_t|Z^n)
    \label{eq:in_converse_joint_E_D_perfect_realism1} \\*
    &\geq& \sum_{t=1}^n I(M,J ;[Y]_t|Z^n)
    \nonumber\IEEEeqnarraynumspace \\
    &=& \sum_{t=1}^n
    I(M,J, Z_{[n] \backslash t} ;[Y]_t|Z_t) 
    \label{eq:in_converse_joint_E_D_perfect_realism2} \\
    &=& nI(M,J, Z_{\scalebox{0.64}{$[n]{\setminus}T$}} ;[Y]_T|Z_T,T)
    \nonumber \\
    &=& nI(V;[Y]_T|Z_T)
    ,\label{eq:in_converse_joint_E_D_perfect_realism3}
\end{IEEEeqnarray}
where \eqref{eq:in_converse_joint_E_D_perfect_realism1}, \eqref{eq:in_converse_joint_E_D_perfect_realism2} and \eqref{eq:in_converse_joint_E_D_perfect_realism3} follow from the fact that couples in $\{([Y]_t,Z_t)\}_{t \in [n]}$ are i.i.d. and that $T$ is independent from $(Y^n,Z^n).$
This is true for any quantizer on $\mathcal{X}.$ Moreover, 
since $Z^n$
is
finite-valued, we have $I(Z_T;V)<\infty.$
Then, by Proposition \ref{prop:limit_on_sequence_of_quantizers} we have
\begin{IEEEeqnarray}{c}
R \geq I(X_T;V|Z_T)
\ \ \text{and} \ \
R+R_c \geq I(Y_T;V|Z_T).\nonumber
\end{IEEEeqnarray}
Therefore, $(R,R_c, \Delta+\varepsilon) \in \mathcal{S}^{(j)}_{E\text{-}D}.$ Hence, $(R,R_c, \Delta) \in \overline{\mathcal{S}}^{(j)}_{E\text{-}D},$ which concludes the converse proof.

\subsection{Achievability}
We prove that $\overline{\mathcal{S
}}^{(j)}_{E\text{-}D} \subseteq \overline{\mathcal{A}}^{(j)}_{E\text{-}D}$ by proving that $\mathcal{S}^{(j)}_{E\text{-}D} \subseteq \overline{\mathcal{A}}^{(j)}_{E\text{-}D}.$ 
We proceed exactly as in Section \ref{sec:achievability_two_sided}.
Let $(R,R_c,\Delta)\in\mathcal{S}^{(j)}_{E\text{-}D}.$ 
Let $p_{X,Y,Z,V}$ be a corresponding distribution from the definition of $\mathcal{S}^{(j)}_{E\text{-}D}.$ Then,
\begin{equation}\label{eq:introducing_R_joint_two_sided}
    R \geq I_p(X;V|Z) 
\end{equation}
\begin{equation}\label{eq:introducing_R_c_joint_two_sided}
    R+R_c \geq I_p(Y;V|Z)
\end{equation}
\begin{equation}\label{eq:introducing_Delta_joint_two_sided}
    \Delta \geq \mathbb{E}_p[d(X,Y)].
\end{equation}
For every $n \geq 1,$ we define codebook $(\mathcal{C}^{(n)}_{z^n})_{z^n \in \mathcal{Z}^n}$ and distribution $Q^{(1)}$ exactly as in Section \ref{sec:achievability_two_sided}. Similarly to \eqref{eq:rewrite_conditional_info_with_entropy_in_proof} therein, Inequalities \eqref{eq:introducing_R_joint_two_sided} and \eqref{eq:introducing_R_c_joint_two_sided} rewrite as
\begin{IEEEeqnarray}{c}
R \geq I_p(X,Z;V,Z) - H_p(Z)
\ \ \text{and} \ \
R+R_c \geq I_p(Y,Z;V,Z) - H_p(Z).\nonumber    
\end{IEEEeqnarray}
Therefore, in the same manner as in Section \ref{sec:achievability_two_sided}, the soft covering lemma with side information gives
\begin{IEEEeqnarray}{rCl}
\mathbb{E}_{\mathcal{C}^{(n)}_{\mathcal{Z}^n}}\big[\|Q^{(1)}_{Y^n,Z^n|\mathcal{C}^{(n)}_{\mathcal{Z}^n}} - p_{X,Z}^{\otimes n}\|_{TV}\big] &\underset{n \to \infty}{\longrightarrow}& 0 
\nonumber\\
\mathbb{E}_{\mathcal{C}^{(n)}_{\mathcal{Z}^n}}\big[\|Q^{(1)}_{J, X^n, Z^n|\mathcal{C}^{(n)}_{\mathcal{Z}^n}} - p^{\mathcal{U}}_{[2^{nR_c}]}p_{X, Z}^{\otimes n}\|_{TV}\big] &\underset{n \to \infty}{\longrightarrow}& 0.\nonumber
\end{IEEEeqnarray}Similarly to \eqref{eq:Distortion_Q_1_fixed_codebook_two_sided}, \eqref{eq:TV_Q_1_Y_fixed_codebook_two_sided}, \eqref{eq:TV_Q_1_J_X_Z_fixed_codebook_two_sided}, there exists $N_0$ such that for $n\geq N_0$ there is a codebook $c^{(n)}_{\mathcal{Z}^n,*}$ such that
\begin{IEEEeqnarray}{rCl}
   \scalebox{0.99}{$\mathbb{E}$}_{Q^{(1)}_{X^n,Y^n|\mathcal{C}^{(n)}_{\mathcal{Z}^n}{=}c^{(n)}_{\mathcal{Z}^n,*}}} \scalebox{0.94}{$[d(X^n,Y^n)]$} &\leq& \scalebox{0.94}{$\Delta+\varepsilon, $}
    \label{eq:Distortion_Q_1_fixed_codebook_joint_two_sided}
    \IEEEeqnarraynumspace\\
    \scalebox{0.99}{$\|Q^{(1)}_{Y^n,Z^n|\mathcal{C}^{(n)}_{\mathcal{Z}^n}{=}c^{(n)}_{\mathcal{Z}^n,*}} - p_{X,Z}^{\otimes n}\|_{TV}$} &\leq& \scalebox{0.99}{$\varepsilon, $}
    \label{eq:TV_Q_1_Y_Z_fixed_codebook_joint_two_sided}
    \IEEEeqnarraynumspace\\
    \scalebox{0.99}{$\|Q^{(1)}_{J, X^n, Z^n|\mathcal{C}^{(n)}_{\mathcal{Z}^n}{=}c^{(n)}_{\mathcal{Z}^n,*}} - p^{\mathcal{U}}_{[2^{nR_c}]}p_{X, Z}^{\otimes n}\|_{TV}$} &\leq& \scalebox{0.99}{$\varepsilon. $}
    \label{eq:TV_Q_1_J_X_Z_fixed_codebook_joint_two_sided}
    \IEEEeqnarraynumspace
\end{IEEEeqnarray}
We construct the E-D-code $P^{(1)}$ exactly as in Section \ref{sec:achievability_two_sided}, and by the same arguments that brought inequality \eqref{eq:TV_P_1_Q_1_two_sided}, we obtain from \eqref{eq:TV_Q_1_J_X_Z_fixed_codebook_joint_two_sided}:
\begin{IEEEeqnarray}{rCl}
\big\|P^{(1)}_{\scalebox{0.6}{$M, J, V^n, X^n, Z^n, Y^n$}} - 
\ Q^{(1)}_{\scalebox{0.6}{$M, J, V^n, X^n, Z^n, Y^n$}}\big\|_{TV}
=
\big\|P^{(1)}_{J, X^n, Z^n} - Q^{(1)}_{J, X^n, Z^n} \big\|_{TV}\leq \varepsilon.\label{eq:TV_P_1_Q_1_joint_two_sided}
\end{IEEEeqnarray}This and \eqref{eq:TV_Q_1_Y_Z_fixed_codebook_joint_two_sided} imply that $P^{(1)}$ achieves near-perfect joint realism. We conclude as in Section \ref{sec:conclusion_of_achievability_two_sided}, but using Proposition \ref{proposition:to_perfect_joint_realism} instead of Proposition \ref{proposition:to_perfect_realism}, with \eqref{eq:Distortion_Q_1_fixed_codebook_joint_two_sided}, \eqref{eq:TV_Q_1_Y_Z_fixed_codebook_joint_two_sided} and \eqref{eq:TV_P_1_Q_1_joint_two_sided}.

\section{Proof of Theorem \ref{theorem:D_rates_region_joint_realism}}\label{sec:achievability_joint_one_sided}

\subsection{Converse}
The converse proof is identical to that of Theorem \ref{theorem:E_D_rates_region_joint_realism} (E-D-codes with (near)-perfect joint realism), where bounds on rates and distortion do not rely on any Markov chain properties,
and only rely on independence properties, which are preserved by quantization.
We only add the following arguments, related to the difference between
$\mathcal{D}^{(j)}_{E\text{-}D}$
and
$\mathcal{D}^{(j)}_D$
(Eq.
\ref{eq:def_D_E_D_joint}
and
\ref{eq:def_D_D_joint})
regarding the single-letter distributions.
First, Markov chain $Z_T-X_T-V$ is satisfied by Corollary \ref{lemma:Markov_chain_X_T_Z_T} (Appendix \ref{app:subsec:Markov_chains_precise_reference_for_converse}).
Second,
the quantization argument
in bounds on rates
is done with
three quantizers, on $\mathcal{X},\mathcal{Z},$ and $\mathcal{Z}^{n-1},$
rather than just on $\mathcal{X}.$
Lastly,
we have $I(Z_T;V)<\infty.$
This is proved in
the remainder of this converse argument.
Fix $n\in\mathbb{N},$ and quantizers $\kappa_1$ and $\kappa_2,$ on $\mathcal{Z}$ and $\mathcal{Z}^{n-1},$ respectively.
Then,
we have
\begin{IEEEeqnarray}{rCl}
    \scalebox{1.0}{$I(M, J, \kappa_2(Z_{[n] \backslash T}), T ; \kappa_1(Z_T)) $}
    =
    \scalebox{1.0}{$I(M, J, \kappa_2(Z_{[n] \backslash T}) ; \kappa_1(Z_T)  | T) $}
    &=& \tfrac{1}{n} \scalebox{1.0}{$\sum_{t=1}^n I(M, J, \kappa_2(Z_{[n] \backslash t}) ; \kappa_1(Z_t) ) $}\nonumber \\*
    &=& \tfrac{1}{n} \scalebox{1.0}{$\sum_{t=1}^n I(M ; \kappa_1(Z_t) | J, \kappa_2(Z_{[n] \backslash t}))
    $}\nonumber\\*
    &\leq& H(M), \nonumber
\end{IEEEeqnarray}
where the first two equalities use the independence of $T$ from all other variables;
the third uses the independence of $J$ from $[Z]^n$ and the fact that variables in $\{[Z]_t\}_{t \in [n]}$ are i.i.d.; and the final inequality follows from known properties of finite alphabets. Indeed, independence is preserved by deterministic quantization.
This equality is true for any
quantizers
on
$\mathcal{Z}$ and $\mathcal{Z}^{n-1}.$
Then, from Proposition \ref{prop:limit_on_sequence_of_quantizers},
\begin{IEEEeqnarray}{c}
I(V;Z_T)
\ \leq \
\scalebox{1.0}{$
H(M),
$}\nonumber 
\end{IEEEeqnarray}which is finite because $M$ is finite-valued (Lemma \ref{lemma:entropy_geq_mutual_info}).

\subsection{Achievability}

We prove that $\overline{\mathcal{S
}}^{(j)}_{D} \subseteq \overline{\mathcal{A}}^{(j)}_{D}$ by proving that $\mathcal{S}^{(j)}_{D} \subseteq \overline{\mathcal{A}}^{(j)}_{D}.$ 
We proceed exactly as in the marginal realism case (Section \ref{sec:achievability}).
Let $(R,R_c,\Delta)\in\mathcal{S}^{(j)}_{D}$ 
and $p$ be a corresponding distribution from
\eqref{eq:def_S_D_joint}.
\begin{equation}\label{eq:introducing_R_joint_one_sided}
    R \geq I_p(X;V|Z) 
\end{equation}
\begin{equation}\label{eq:introducing_R_c_joint_one_sided}
    R+R_c \geq I_p(Y;V|Z)
\end{equation}
\begin{equation}\label{eq:introducing_Delta_joint_one_sided}
    \Delta \geq \mathbb{E}_p[d(X,Y)].
\end{equation}
By assumption, $p$ satisfies the Markov chain $Z-X-V$ of \eqref{eq:def_D_D_joint} and $I_p(Z;V)<\infty.$ Therefore, by the chain rule for general alphabets (Proposition \ref{prop:chain_rule}), inequalities \eqref{eq:introducing_R_joint_one_sided} and \eqref{eq:introducing_R_c_joint_one_sided} rewrite as
\begin{IEEEeqnarray}{c}
R \geq I_p(X,Z;V) - I_p(Z;V)
\ \ \text{and} \ \
R+R_c \geq I_p(Y,Z;V) - I_p(Z;V).\label{eq:rewritten_ineqs_on_R_and_R_c_one_sided_joint}    
\end{IEEEeqnarray}
Hence, we can define rate R', codebook $\mathcal{C}^{(n)}$ and distribution $Q^{(1)}$ exactly as in Section \ref{sec:achievability} and obtain, from
\eqref{eq:rewritten_ineqs_on_R_and_R_c_one_sided_joint} and the soft covering lemma:
\begin{IEEEeqnarray}{rCl}
\mathbb{E}_{\mathcal{C}^{(n)}}\big[\|Q^{(1)}_{J, X^n, Z^n|\mathcal{C}^{(n)}} - p^{\mathcal{U}}_{[2^{nR_c}]}p_{X, Z}^{\otimes n}\|_{TV}\big] &\underset{n \to \infty}{\longrightarrow}& 0\IEEEeqnarraynumspace\label{eq:TV_Q_1_J_X_Z_one_sided_joint}\\*
\mathbb{E}_{\mathcal{C}^{(n)}}\big[\|Q^{(1)}_{Y^n,Z^n|\mathcal{C}^{(n)}} - p_{X,Z}^{\otimes n}\|_{TV}\big] &\underset{n \to \infty}{\longrightarrow}& 0. \label{eq:TV_Q_1_Y_one_sided_joint}
\end{IEEEeqnarray}In the same manner as in Section \ref{sec:achievability}, for a certain $N_0,$ $\forall n\geq N_0$ there is a codebook $c^{(n)}_{*}$ such that
\begin{IEEEeqnarray}{rCl}
   \scalebox{0.99}{$\mathbb{E}$}_{Q^{(1)}_{X^n,Y^n|\mathcal{C}^{(n)}{=}c^{(n)}_{*}}} \scalebox{0.94}{$[d(X^n,Y^n)]$} &\leq& \scalebox{0.94}{$\Delta+\varepsilon, $}\label{eq:Distortion_Q_1_fixed_codebook_joint_one_sided} \IEEEeqnarraynumspace\\
    \scalebox{0.99}{$\|Q^{(1)}_{Y^n,Z^n|\mathcal{C}^{(n)}{=}c^{(n)}_{*}} - p_{X,Z}^{\otimes n}\|_{TV}$} &\leq& \scalebox{0.94}{$\varepsilon. $}\label{eq:TV_Q_1_Y_fixed_codebook_joint_one_sided} \IEEEeqnarraynumspace\\
    \scalebox{0.99}{$\|Q^{(1)}_{J, X^n, Z^n|\mathcal{C}^{(n)}{=}c^{(n)}_{*}} - p^{\mathcal{U}}_{[2^{nR_c}]}p_{X, Z}^{\otimes n}\|_{TV}$} &\leq& \scalebox{0.94}{$\varepsilon. $}\label{eq:TV_Q_1_J_X_Z_fixed_codebook_joint_one_sided} \IEEEeqnarraynumspace\\
    \scalebox{0.94}{$\substack{\scalebox{1.0}{$Q^{(1)} \; (\hat{M}'\neq M') $} \\ M', \hat{M}' |\mathcal{C}^{(n)}{=}c_*^{(n)}}$} &\leq& \scalebox{0.94}{$\varepsilon. $}\label{eq:Q_1_decoding_error_fixed_codebook_joint} \IEEEeqnarraynumspace
\end{IEEEeqnarray}
We construct distribution $Q^{(2)}$ and D-code $P^{(1)}$ exactly as in Section \ref{sec:achievability}, and by the same arguments that brought inequalities \eqref{eq:TV_Q_1_Q_2} and \eqref{eq:TV_P_1_Q_2}, we obtain
\begin{IEEEeqnarray}{rCl}
\IEEEeqnarraymulticol{3}{l}{
\big\|P^{(1)}_{\scalebox{0.6}{$M, M', J, X^n, Z^n, \hat{M}', Y^n$}} - Q^{(1)}_{\scalebox{0.6}{$M, M', J, X^n, Z^n, \hat{M}', Y^n$}}\big\|_{TV}\leq 2\varepsilon.}\label{eq:TV_P_1_Q_1_joint_one_sided}
\end{IEEEeqnarray}Hence, $P^{(1)}$ achieves near-perfect joint realism. We conclude as in Section \ref{sec:conclusion_of_achievability_two_sided}, but using Proposition \ref{proposition:to_perfect_joint_realism} instead of Proposition \ref{proposition:to_perfect_realism}, with \eqref{eq:Distortion_Q_1_fixed_codebook_joint_one_sided}, \eqref{eq:TV_Q_1_Y_fixed_codebook_joint_one_sided} and \eqref{eq:TV_P_1_Q_1_joint_one_sided}.

\section{Proofs of Theorems \ref{thm:common_part} and \ref{thm:D_infty_rates_region_marginal_realism} and Corollaries \ref{cor:independent_side_info_two_sided} and \ref{cor:E_D_infty_region_finite_alphabets}}\label{sec:special_cases}

\subsection{Proof of Theorem \ref{thm:common_part}}\label{subsec:proof_thm_common_part}

Achievability follows from Theorem \ref{theorem:D_rates_region_marginal_realism}. Moving to the converse,
we first show that D-achievability with near-perfect realism marginal for joint source and side information distribution $p_{X,Z}$ implies D-achievability with near-perfect marginal realism for joint source and side information distribution $p_{X,\phi(X)}.$ This is the object of Subsection \ref{subsec:Z_is_useless_beyond_common_part}. For the latter type of achievability we have the outer bound of Theorem \ref{theorem:E_D_rates_region_marginal_realism} (for E-D-codes with marginal realism). We conclude, in Subsection \ref{subsec:using_E_D_on_common_part}, by rewriting the latter bound into the form appearing in \eqref{eq:def_S_D}. Throughout this section, we manipulate Markov chains in a way that is correct for general alphabets (see Appendix \ref{app:subsec:measure_theory_general}), but do not justify this fact explicitly for the sake of conciseness.

\vspace{10pt}
\subsubsection{Using only the common part is sufficient}\label{subsec:Z_is_useless_beyond_common_part}
\hfill\\
\indent Let $(R,R_c,\Delta)$ be achievable for $p_{X,Z}$ and $P
$ be induced by a corresponding code. Then, $P$ satisfies
\begin{IEEEeqnarray}{rCl}
    Z^n-(X^n,J)-M, &\text{ hence }& ((\psi(Z_t))_{t \in [n]},Z^n)-(X^n,J)-M\label{eq:short_Markov_in_proof_common_part}\\
    X^n-(M,J,Z^n)-Y^n, &\text{ hence }& X^n-(M,J,Z^n,(\psi(Z_t))_{t \in [n]})-Y^n.\label{eq:long_Markov_in_proof_common_part}
\end{IEEEeqnarray}
We add to the decoder a variable $\Tilde{Z}^n$ defined by
\begin{IEEEeqnarray}{c}
    P_{\Tilde{Z}^n|X^n,Z^n,J,M,Y^n} := \prod_{t{=}1}^n p_{Z|\psi(Z)=\psi(z_t)}.\label{eq:def_Z_Tilde_in_proof_common_part}
\end{IEEEeqnarray}
We use the initial decoder $P_{Y^n|M,J,Z^n},$ applied to $\Tilde{Z}^n$ instead of $Z^n,$ outputting variable $\Tilde{Y}^n$ as follows:
\begin{IEEEeqnarray}{c}
    P_{\Tilde{Y}^n|X^n,Z^n,J,M,Y^n,\Tilde{Z}^n} :=P_{Y^n|M{=}m,J{=}j,Z^n{=}\Tilde{z}^n}.\label{eq:def_Y_Tilde_in_proof_common_part}
\end{IEEEeqnarray}By construction 
$P$ satisfies
\begin{IEEEeqnarray}{c}
    X^n - (M,J,(\psi(Z_t))_{t \in [n]})-(\Tilde{Z}^n,\Tilde{Y}^n), \text{ hence } X^n - (M,J,(\psi(Z_t))_{t \in [n]},\Tilde{Z}^n) - \Tilde{Y}^n.\label{eq:obvious_Markov_chain_Tilde_Y_in_proof_common_part}
\end{IEEEeqnarray}
We claim that replacing $(Z^n,Y^n)$ by $(\Tilde{Z}^n,\Tilde{Y}^n)$ in the initial code does not change the joint distribution:
\begin{claim}\label{claim:distrib_unchanged_by_Tilde_in_proof_common_part}
\begin{IEEEeqnarray}{c}
    P_{X^n,\Tilde{Z}^n,J,M,\Tilde{Y}^n,(\psi(Z_t))_{t \in [n]}} \equiv P_{X^n,Z^n,J,M,Y^n,(\psi(Z_t))_{t \in [n]}}.
\end{IEEEeqnarray}
\end{claim}
\begin{IEEEproof}
From \eqref{eq:long_Markov_in_proof_common_part}, \eqref{eq:def_Y_Tilde_in_proof_common_part} and \eqref{eq:obvious_Markov_chain_Tilde_Y_in_proof_common_part},
it is sufficient to prove that
\begin{IEEEeqnarray}{c}
    P_{X^n,\Tilde{Z}^n,J,M,(\psi(Z_t))_{t \in [n]}} \equiv P_{X^n,Z^n,J,M,(\psi(Z_t))_{t \in [n]}}.
\end{IEEEeqnarray}By definition \eqref{eq:def_Z_Tilde_in_proof_common_part} of $\Tilde{Z}^n$ and the fact that $P_{Z^n,(\psi(Z_t))_{t \in [n]}} \equiv p_{Z,\psi(Z)}^{\otimes n},$ it is sufficient to show that $P$ satisfies $Z^n - (\psi(Z_t))_{t \in [n]} - (X^n,M,J).$ We denote $(\psi(Z_t))_{t \in [n]}$ by $S^n.$ By the chain rule for general alphabets (Proposition \ref{prop:chain_rule}), we have
\begin{IEEEeqnarray}{rCl}
    I(Z^n; X^n,M,J,S^n) &=&  I(Z^n; X^n,J,S^n) + I(Z^n; M|X^n,J,S^n)\nonumber\\*
    &=& I(Z^n; X^n,S^n) + I(Z^n;J|X^n,S^n) + I(Z^n; M|X^n,J,S^n)\nonumber\\*
    &=& I(Z^n; S^n) + I(Z^n; X^n|S^n) + I(Z^n;J|X^n,S^n) + I(Z^n; M|X^n,J,S^n).\nonumber
\end{IEEEeqnarray}By assumption, $S^n$ is finite-valued, hence the first term is finite (by basic Lemma \ref{lemma:entropy_geq_mutual_info}). It remains to prove that the last three terms are null. 
Since $P_{X^n,Z^n}\sim p_{X,Z}^{\otimes},$ then by assumption $P$ satisfies \begin{IEEEeqnarray}{c}
\forall t \in [n], \ Z_t-\psi(Z_t)-X_t, \text{ thus } Z^n-S^n-X^n
\end{IEEEeqnarray}by the chain rule (Proposition \ref{prop:chain_rule}). Since $J$ is independent from $(X^n,Z^n)$ and $\psi$ is deterministic, then $J$ is independent from $(X^n,Z^n,S^n).$ Hence, $I(Z^n;J|X^n,S)=0$ by the chain rule. From \eqref{eq:short_Markov_in_proof_common_part} and the chain rule we have $I(Z^n;M|X^n,J,S^n)=0.$
\end{IEEEproof}
From \eqref{eq:short_Markov_in_proof_common_part}, \eqref{eq:obvious_Markov_chain_Tilde_Y_in_proof_common_part}, Claim \ref{claim:distrib_unchanged_by_Tilde_in_proof_common_part} and Lemma \ref{lemma:conditions_for_P_to_define_a_code},
distribution $P_{X^n,J,M,\Tilde{Y}^n,(\psi(Z_t))_{t \in [n]}}$
is the distribution induced by a D-code for joint source and side information distribution $p_{X,\phi(X)}.$ Moreover, by Claim \ref{claim:distrib_unchanged_by_Tilde_in_proof_common_part}, the message and common randomness rates are the same as for $P.$ Thus,
$(R,R_c,\Delta)$ is D-achievable with near-perfect marginal realism for joint source and side information distribution $p_{X,\phi(X)}.$
\vspace{10pt}
\subsubsection{Conclusion}\label{subsec:using_E_D_on_common_part}
\hfill\\
\indent Hence, since $\phi(X)$ is finite-valued, triplet $(R,R_c,\Delta)$ satisfies the converse of Theorem \ref{theorem:E_D_rates_region_marginal_realism} (for E-D-codes with marginal realism). Therefore, there exists $\varepsilon>0$ and a distribution $p_{X,\phi(X),V,Y}$ satisfying
\begin{IEEEeqnarray}{rCl}
    R+\varepsilon &\geq& I_p(X;V|\phi(X))\label{eq:ineq_on_R_with_common_part}\\
    R+R_c+\varepsilon &\geq& I_p(Y;V|\phi(X)) - H_p(\phi(X)|Y)\label{eq:ineq_on_R_and_R_c_with_common_part}\\
    \Delta + \varepsilon &\geq& \mathbb{E}_p[d(X,Y)]
    \nonumber\\
    \IEEEeqnarraymulticol{3}{l}{
    \quad
    X-(\phi(X),V)-Y
    }
    \label{eq:single_letter_Markov_with_common_part}\\
    p_Y &\equiv& p_X.\label{eq:single_letter_realism_with_common_part}
\end{IEEEeqnarray}It remains to show that these properties can be rewritten into the form appearing in definition \eqref{eq:def_S_D} of region $\mathcal{S}^{(m)}_D$ --- for joint source and side information distribution $p_{X,Z}.$ Define variable $\Tilde{Z}$ by
\begin{IEEEeqnarray}{c}
    p_{\Tilde{Z}|X,V,\phi(X),Y} := p_{Z|\psi(Z)=\phi(x)}.\label{eq:def_Tilde_Z_single_letter}
\end{IEEEeqnarray}Hence,
\begin{IEEEeqnarray}{c}
    p_{\Tilde{Z},X,\phi(X)} \equiv p_{X,\phi(X)} \cdot p_{Z|\psi(Z)=\phi(x)} \equiv p_{Z,X,\phi(X)},\label{eq:Tilde_Z_and_X_have_correct_joint}
\end{IEEEeqnarray}
where the last equality follows from
$X-\phi(X)-Z.$ Define $\Tilde{V}=(V,\phi(X)).$
By definition \eqref{eq:def_Tilde_Z_single_letter} of $\Tilde{Z},$ we have $X-(V,Y,\phi(X))-\Tilde{Z}.$ Therefore, from \eqref{eq:single_letter_Markov_with_common_part} we have $X-(V,\phi(X))-(Y,\Tilde{Z}),$ hence
\begin{IEEEeqnarray}{c}
    X-(\Tilde{V},\Tilde{Z})-Y.\label{eq:intermediate_single_letter_long_Markov}
\end{IEEEeqnarray}
By definition \eqref{eq:def_Tilde_Z_single_letter} of $\Tilde{Z},$ we have $\Tilde{Z}-X-V.$ We also trivially have $\Tilde{Z}-(X,V)-\phi(X),$ hence
\begin{IEEEeqnarray}{c}
\Tilde{Z}-X-\Tilde{V}.\label{eq:single_letter_short_Markov}
\end{IEEEeqnarray}
Moreover,
\begin{IEEEeqnarray}{rCl}
I(\Tilde{Z};\Tilde{V})=I(\Tilde{Z};V,\phi(X))
=
I(\Tilde{Z};\phi(X))+I(\Tilde{Z};V|\phi(X))
&=&I(\Tilde{Z};\psi(\Tilde{Z}))\nonumber\\
&=&H(\psi(\Tilde{Z}))\nonumber\\
&=&I(X;\phi(X))\label{eq:form1}\\
&=&I(\phi(X);Y) + H(\phi(X)|Y),\label{eq:form2}
\end{IEEEeqnarray}
where \eqref{eq:form2} follows from the fact that $\phi(X)$ is finite-valued. This implies that $I(\Tilde{Z};\Tilde{V})<\infty.$
Combining this with \eqref{eq:single_letter_realism_with_common_part}, \eqref{eq:Tilde_Z_and_X_have_correct_joint}, 
\eqref{eq:intermediate_single_letter_long_Markov} and \eqref{eq:single_letter_short_Markov}
gives $p_{X,\Tilde{Z},\Tilde{V},Y} \in \mathcal{D}^{(m)}_{D}$ (Eq. \ref{eq:def_D_D}).
To conclude, from \eqref{eq:single_letter_short_Markov},
\begin{IEEEeqnarray}{rCl}
I(X;\Tilde{V})=I(X,\Tilde{Z};\Tilde{V})
=
I(X;V,\phi(X))+I(\Tilde{Z};V,\phi(X)|X)
&=& I(X;V,\phi(X))\nonumber\\
&=& I(X;\phi(X))+I(X;V|\phi(X))\nonumber
\end{IEEEeqnarray}
and
\begin{IEEEeqnarray}{rCl}
    I(Y;\Tilde{V})= I(Y;\phi(X))+I(Y;V|\phi(X)),\nonumber
\end{IEEEeqnarray}
hence from \eqref{eq:ineq_on_R_with_common_part}, \eqref{eq:ineq_on_R_and_R_c_with_common_part}, \eqref{eq:form1}, and \eqref{eq:form2}, we obtain
\begin{IEEEeqnarray}{rCl}
R+\varepsilon
\geq
I(X;\Tilde{V}) - I(\Tilde{Z};\Tilde{V})
\ \ \text{and} \ \
R+R_c+\varepsilon &\geq& I(Y;\Tilde{V}) - I(\Tilde{Z};\Tilde{V}).\nonumber
\end{IEEEeqnarray}
This concludes the proof.

\subsection{Proof of Theorem \ref{thm:D_infty_rates_region_marginal_realism}}\label{subsec:proof_D_infty_marginal}

\subsubsection{Achievability}
\hfill\\
\indent Consider a point $(R,\Delta)$ in $\mathcal{S}^{(m)}_{D,\infty}.$ By definition there exists $p_{X,Z,V,Y}\in \mathcal{D}^{(m)}_{D}$ (Eq. \ref{eq:def_D_D}) such that
\begin{IEEEeqnarray}{rCl}
R
\geq
I_p(X;V|Z)
\ \text{and} \
\infty
>
I_p(Y;V)
\ \text{and} \
\Delta &\geq& \mathbb{E}_p[d(X,Y)]\nonumber.
\end{IEEEeqnarray}Since $p_{X,Z,V,Y}\in \mathcal{D}^{(m)}_{D}$ we also have $I_p(Z;V)<\infty$
and $Z-X-V,$ 
hence
\begin{IEEEeqnarray}{rCl}
I_p(X;V|Z)
=
I_p(X;V) - I_p(Z;V)
\ \ \text{and} \ \
\infty &>& I_p(Y;V)-I_p(Z;V).\nonumber
\end{IEEEeqnarray}Fix some $R_c \geq \max(0,I_p(Y;V)-I_p(Z;V)).$ Then, $(R,R_c,\Delta) \in \mathcal{S}^{(m)}_{D}.$ By Theorem \ref{theorem:D_rates_region_marginal_realism}, we have $\mathcal{S}^{(m)}_{D} \subseteq \overline{\mathcal{A}}^{(m)}_{D}.$ Hence, $(R,\Delta) \in \overline{\mathcal{A}}^{(m)}_{D, \infty}.$

\vspace{10pt}
\subsubsection{Converse}
\hfill\\
\indent We are to prove that $\overline{\mathcal{A}}^{(m)}_{D,\infty} \subseteq \overline{\mathcal{S}}^{(m)}_{D,\infty}.$
To do so, we can incorporate quantization on the side information in the proof of the converse of Theorem \ref{theorem:E_D_rates_region_marginal_realism} (E-D-codes with marginal realism), where bounds on rates and distortion do not rely on any Markov chain properties. We only add the following arguments, related to the difference between the constraints \eqref{eq:def_D_E_D} and \eqref{eq:def_D_D} regarding the single-letter distributions. First, the Markov chain $Z_T-X_T-V$ is satisfied by Lemma \ref{lemma:Markov_chain_X_T_Z_T} (Appendix \ref{app:subsec:Markov_chains_precise_reference_for_converse}). Second, we have $I(Z_T;V)<\infty$ by the exact same argument as in the converse proof of Theorem \ref{theorem:D_rates_region_joint_realism} (D-codes with joint realism), which does not rely on joint realism. Third, we have $I(Y_T;V)<\infty$ simply if either of $\mathcal{X}$ or $\mathcal{Z}$ is finite. Indeed, in that case, either $Y_T$ or $V$ is finite-valued. Finally, the quantization argument for the lower bound on the rate is done with
three quantizers, on $\mathcal{X},\mathcal{Z},$ and $\mathcal{Z}^{n-1},$
rather than just on $\mathcal{X}.$

\subsection{Proof of Corollary \ref{cor:E_D_infty_region_finite_alphabets}}\label{subsec:proof_corollary_large_CR_two_sided}
\subsubsection{Optimality of region \eqref{eq:def_S_E_D_infty}}
\hfill\\
\indent The converse is a direct consequence of the converse direction of Theorem \ref{theorem:E_D_rates_region_marginal_realism},
and the fact that the alphabet of $V$ is finite.
The achievability follows from that of Theorem \ref{theorem:E_D_rates_region_marginal_realism} by taking
$R_c=H(Z)+H(V)$ --- which is finite because the alphabets of $Z$ and $V$ are finite.

\vspace{10pt}
\subsubsection{Optimality of region \eqref{eq:def_S_E_D_infty_finite_alphabet}}
\hfill\\
\indent For the achievability, it is sufficient to provide a proof in the case of finite-rate common randomness. Achievability follows from that of region \eqref{eq:def_S_E_D_infty}, and by defining $V$ as equal to $Y.$ For the converse, it is sufficient to prove it in the case of unconstrained common randomness. Consider a couple $(R,\Delta)$ E-D achievable with near-perfect marginal realism with unconstrained common randomness. In the converse proof of Theorem \ref{theorem:E_D_rates_region_marginal_realism}, except for the finiteness of $\mathcal{V}$ which is not needed here, the assumption of finite-rate common randomness is not necessary as long as $R_c$ is replaced by $\infty.$ In particular, note that Theorem \ref{theorem:equivalence_perfect_realism} (on the equivalence between perfect and near-perfect realism) is stated with both unconstrained and finite-rate common randomness. The only change that is needed is to replace the common randomness variable $J$ by a quantized version in the derivation --- which preserves independence relations ---, then take a limit. Hence,
for any $\varepsilon>0$ there exists a point in $\mathcal{D}^{(m)}_{E\text{-}D}$ (Eq. \ref{eq:def_D_E_D}) such that $(R+\varepsilon, \Delta+\varepsilon)$ satisfies the first and third inequality in definition \eqref{eq:def_S_E_D} of $\mathcal{S}^{(m)}_{E\text{-}D}.$ The conclusion follows from the data processing inequality, using the Markov chain property appearing in definition \eqref{eq:def_D_E_D} of $\mathcal{D}^{(m)}_{E\text{-}D}.$

\subsection{Proof of Corollary \ref{cor:independent_side_info_two_sided}}\label{subsec:proof_cor_independent_side_info_two_sided}

Consider an element $(R,R_c,\Delta)$ of $\mathcal{S}^{(m)}_{E\text{-}D}.$ Let $p_{X,Z,V,Y}$ be a corresponding element of $\mathcal{D}^{(m)}_{E\text{-}D}$ (Eq. \ref{eq:def_D_E_D}). Then, all the constraints appearing in \eqref{eq:def_S} and \eqref{eq:def_D} are satisfied with auxiliary random variable $\Tilde{V}=(V,Z),$ and common randomness rate $R_c+H(Z).$ Thus, $(R,R_c+H(Z),\Delta)\in\mathcal{S}^{(m)}.$ Hence,
\begin{IEEEeqnarray}{c}
\mathcal{S}^{(m)}_{E\text{-}D} + (0,H(Z),0) \subseteq \mathcal{S}^{(m)} \cap \mathbb{R}_{\geq 0} \times [H(Z),\infty] \times \mathbb{R}_{\geq 0}.\nonumber
\end{IEEEeqnarray}
Moving to the reverse direction, consider an element $(R,R_c,\Delta)$ of $\mathcal{S}^{(m)} \cap \mathbb{R}_{\geq 0} \times [H(Z),\infty] \times \mathbb{R}_{\geq 0},$ and a corresponding $p_{X,V,Y}$ in $\mathcal{D}^{(m)}$ (Eq. \ref{eq:def_D}). Define
$p_{Z,X,V,Y} := p_Z \cdot p_{X,V,Y}.$ Then, we have $I_p(X;V)=I_p(X;V|Z),$ $I_p(Y;V) = I_p(Y;V|Z) - H_p(Z|Y) +H_p(Z),$ and
\begin{IEEEeqnarray}{c}
Z-X-V-Y, \text{ thus } (X,Z)-V-Y, \text{ hence } X-(Z,V)-Y.\nonumber
\end{IEEEeqnarray}
Therefore, all constraints in \eqref{eq:def_S_E_D} and \eqref{eq:def_D_E_D} are satisfied with common randomness rate $\Tilde{R}_c = R_c-H(Z).$
Thus, $(R,R_c-H(Z),\Delta) \in \mathcal{S}^{(m)}_{E\text{-}D}.$ Hence,
\begin{IEEEeqnarray}{c}
\mathcal{S}^{(m)} \cap \mathbb{R}_{\geq 0} \times [H(Z),\infty] \times \mathbb{R}_{\geq 0} \subseteq \mathcal{S}^{(m)}_{E\text{-}D} + (0,H(Z),0).\nonumber
\end{IEEEeqnarray}

\section{The Gaussian case}\label{sec:gaussian}

\subsection{The role of common randomness for a Normal source in the absence of side information}\label{subsec:normal_source_no_side_info}
We now prove Proposition \ref{prop:standard_normal_source}.
Without loss of generality, we consider $\rho$ to be non-negative. Simple algebra shows that $\rho$ is the unique solution in $[0,1)$ to
\begin{equation}
\label{eq:Normal_rho}
    1 - \frac{\Delta}{2} = \rho \sqrt{1 - 2^{-2 R_c} (1-\rho^2)},
\end{equation}which is a quadratic equation in $\rho^2.$\\
Consider $\rho$ satisfying (\ref{eq:Normal_rho}), and $R$ satisfying \eqref{eq:Normal_R_min}.
We choose a distribution $p_{X,V,Y}$ with $V$ standard Normal such that $(X,V,Y)$ are jointly 
Gaussian, satisfy the Markov chain $X - V - Y,$ and
\begin{align}
    \mathbb{E}[XV] & = \rho, \\
    \mathbb{E}[VY] & = \sqrt{1 - 2^{-2R_c}(1-\rho^2)} =: \tilde{\rho},
\end{align}
with $Y$ also being standard Normal. Then,
$p_{X,V,Y
}$
satisfies all the conditions of set
$\mathcal{D}^{(m)}$ (Eq. \ref{eq:def_D}),
except for the finiteness of the alphabet of $V.$
Nevertheless, from Corollary \ref{cor:region_with_no_side_info} and Theorem \ref{theorem:D_rates_region_marginal_realism} in the case of constant side information, the closure of optimal region $\mathcal{S}^{(m)}$ (Eq. \ref{eq:def_S}) would not change if the alphabet of
the auxiliary random variable
$V$ were not constrained to be finite, but simply Polish.
\begin{claim}\label{claim:Normal_achievability}
Through simple computations, one can obtain
\begin{IEEEeqnarray}{rCl}
\mathbb{E}[(X-Y)^2]
=
\Delta
\ \ \text{and} \ \
I(X;V)
=
\frac{1}{2} \log \frac{1}{1 - \rho^2}
\ \ \text{and} \ \
I(Y;V)
&=&
R_c + \frac{1}{2} \log \frac{1}{1 - \rho^2}.\nonumber
\end{IEEEeqnarray}
\end{claim}

The proof of this claim is deferred to Appendix \ref{app:subsec:Gaussian}.
Therefore, $(R,R_c,\Delta)\in\overline{\mathcal{S}}^{(m)}.$ Thus, $(R,R_c,\Delta)\in\overline{\mathcal{S}}^{(m)}_{E\text{-}D},$ assuming constant side information.
The latter property is the starting point, in Subsection \ref{subsec:introducing_R_R_c_Delta_achievability_marginal_two_sided}, of the achievability proof of Theorem \ref{theorem:E_D_rates_region_marginal_realism}. From that proof, and in particular from its conclusion in Subsection \ref{sec:conclusion_of_achievability_two_sided}, we have that for any $\varepsilon>0,$ the triplet $(R+\varepsilon,R_c,\Delta+\varepsilon)$ is achievable with perfect marginal realism. By combining codes with different $\varepsilon,$ we obtain that for any $\delta>0,$ the triplet $(R+\delta,R_c,\Delta)$ is achievable with perfect marginal realism.
Since $R$ satisfies
\eqref{eq:Normal_R_min},
the direct part of the proposition is proved.\\

Moving to the reverse direction, suppose $(R,R_c,\Delta) \in \mathcal{A}^{(m)}.$
Since $(d,p_X)$ is uniformly integrable (Claim \ref{claim:uniform_integrability_if_MSE}),
then by 
Corollary~\ref{cor:region_with_no_side_info}
we have
$(R,R_c,\Delta) \in \overline{\mathcal{S}}^{(m)}.$
Therefore, for arbitrarily small $\varepsilon > 0,$ we have 
$(R+ \varepsilon, R_c + \varepsilon, \Delta + \varepsilon) \in \mathcal{S}^{(m)}.$ Fix such a $\varepsilon>0$ satisfying $\Delta + \varepsilon < 2,$ which exists since $\Delta \in (0,2).$ Then, by definition of $\mathcal{S}^{(m)}$
there exists $p_{X,V,Y} \in \mathcal{D}^{(m)}$ (Eq. \ref{eq:def_D}) such that
\begin{align}
  R + \varepsilon & \geq I(X;V) \\
  R_c + R + 2\varepsilon & \geq I(Y;V) \\
  \Delta + \varepsilon & \geq \mathbb{E}[(X-Y)^2].
\end{align}
Now define
\begin{align}
    \rho & = \sqrt{\mathbb{E}[\mathbb{E}[X|V]^2]} \\
    \tilde{\rho} & = \sqrt{\mathbb{E}[\mathbb{E}[Y|V]^2]}.
\end{align}
Since $R$ and $R_c$ are finite but $I(X;X)$ and $I(Y;Y)$ are infinite, then each of $X$ and $Y$ is not a deterministic function of $V.$ Since $\mathbb{E}[X^2] = \mathbb{E}[Y^2] =1,$ this implies
\begin{IEEEeqnarray}{c}
    \rho,\Tilde{\rho} \in [0,1).\label{eq:in_proof_Normal_rho_less_than_1}
\end{IEEEeqnarray}
\begin{claim}\label{claim:Normal_converse_entropy_maximizing}
Using the entropy-maximizing property
of the Gaussian distribution, one can obtain
\begin{IEEEeqnarray}{c}
R + \varepsilon \geq \frac{1}{2} \log \frac{1}{1 - \rho^2}\label{eq:in_proof_Normal_converse_R_first_ineq}\\*
R + R_c + 2\varepsilon \ge \frac{1}{2} \log \frac{1}{1 - \tilde{\rho}^2}.\nonumber
\end{IEEEeqnarray}
\end{claim}

The proof of this claim is deferred to Appendix \ref{app:subsec:Gaussian}. The second inequality can be written as
\begin{equation}
    R + 2 \varepsilon \ge \frac{1}{2} \log \frac{2^{-2 R_c}}{1 - \tilde{\rho}^2}.\label{eq:in_proof_Normal_converse_R_second_ineq}
\end{equation}
Turning to the distortion constraint, we have --- see Appendix \ref{app:subsec:Gaussian}:
\begin{claim}\label{claim:Normal_converse_distortion}
By Cauchy-Schwarz inequality, one can obtain
\begin{IEEEeqnarray}{c}
    \Delta + \varepsilon \geq 2 - 2 \rho \tilde{\rho}.\label{eq:in_proof_Normal_converse_Delta_ineq}
\end{IEEEeqnarray}
\end{claim}

Thus, from \eqref{eq:in_proof_Normal_rho_less_than_1}, \eqref{eq:in_proof_Normal_converse_R_first_ineq}, \eqref{eq:in_proof_Normal_converse_R_second_ineq} and \eqref{eq:in_proof_Normal_converse_Delta_ineq} we have
\begin{align}
    R + 2\varepsilon & \ge \inf_{\rho,\tilde{\rho} \in [0,1) }
       \max \left(\frac{1}{2} \log \frac{1}{1 - \rho^2},
       \frac{1}{2} \log \frac{2^{-2 R_c}}{1 - \tilde{\rho}^2} \right) \nonumber\\
      \text{s.t.} \ \ & \Delta + \varepsilon \ge 2 - 2 \rho \tilde{\rho}.\label{eq:in_proof_Normal_optim_problem}
\end{align}
Since $s \mapsto -\log(1-s^2)$ is continuous and diverges when $s \to 1,$ and the constraint $\Delta+\varepsilon \geq 2-2\rho\Tilde{\rho}$ does not prevent any of $\rho,\Tilde{\rho}$ from increasing with the other being fixed, then the optimization space can be reduced to the set of $(\rho,\Tilde{\rho}) \in [0,1)^2$ such that
\begin{equation}
    \frac{1}{2} \log \frac{1}{1 - \rho^2}
       = \frac{1}{2} \log \frac{2^{-2 R_c}}{1 - \tilde{\rho}^2},
       \nonumber
\ \ \text{i.e.,} \ \
       \tilde{\rho} = \sqrt{1 - 2^{-2 R_c}(1-\rho^2)}.\label{eq:in_proof_Normal_form_of_optimal_rho_tilde}
\end{equation}
Then, there exists such a $(\rho_{\varepsilon},\Tilde{\rho}_\varepsilon)$ achieving near-optimality:
\begin{equation}
    R+3\varepsilon \geq \frac{1}{2} \log \frac{1}{1 - \rho_{\varepsilon}^2}.\label{eq:in_proof_Nornal_final_ineq_R}
\end{equation}
\begin{claim}\label{claim:Normal_converse_continuity_monotonicity_argument}
By taking $\varepsilon \to 0$ and using a continuity and monotonicity argument, we obtain the desired:
\begin{IEEEeqnarray}{c}
R \geq \frac{1}{2} \log \frac{1}{1 - (\rho^*)^2},\nonumber
\end{IEEEeqnarray}where $\rho^*$ is the unique solution in $[0,1)$ to
\begin{equation}
    1 - \frac{\Delta}{2} = \rho \sqrt{1 - 2^{-2 R_c} (1-\rho^2)}.
\end{equation}
\end{claim}

The proof is provided in Appendix \ref{app:subsubsec_proof_claim:Normal_converse_continuity_monotonicity_argument}.

\subsection{Bi-dimensional Gaussian source and side information}\label{subsec:bi_dimensional_gaussian}

We now prove Proposition \ref{prop:gaussian_infinite_common_randomness}.
Fix $\Delta \in (0,2-2|\eta|].$
Since we know that $R_{D}(\Delta) \geq R_{E\text{-}D}(\Delta),$
it is sufficient to prove that $R_{E\text{-}D}(\Delta)$ is lower bounded by the
right hand side of \eqref{eq:statement_R_E_D_equals_R_D_bi_dimensional_Gaussian}
and that $R_{D}(\Delta)$ is upper bounded by
the latter.
We start with the lower bound.
\begin{claim}\label{claim:remark_in_converse_bi_dimensional_gaussian}
$R_{E\text{-}D}(\Delta) \geq \tfrac{1}{2}\log((1-\eta^2)/(1-\rho^2)).$
\end{claim}
\begin{IEEEproof}
Fix $\varepsilon>0.$
Consider a triplet $(R,R_c,\Delta)$ E-D-achievable with near-perfect marginal realism.
From Theorem \ref{theorem:equivalence_perfect_realism}, $(R,R_c,\Delta)$ is E-D-achievable with perfect marginal realism.
Then, there exists $n\in\mathbb{N}$ and a $(n,R,R_c)$
E-D-code inducing a joint distribution $P$ such that $P_{Y^n} \, {\equiv} \, p_X^{\otimes n}$ and $\mathbb{E}_P[d(X^n,Y^n)] \, {\leq} \, \Delta \, + \, \varepsilon,$ i.e., $(1/n)\sum_{t{=}1}^n\mathbb{E}[X_t Y_t] \, \geq \rho_\varepsilon,$ where $\rho_\varepsilon = 1 - (\Delta+\varepsilon)/2.$ 
Then, similarly to Claim \ref{claim:Normal_converse_entropy_maximizing} --- see its proof in Appendix \ref{app:subsec:Gaussian} ---,
we can lower bound $I(X_t;Y_t)$ as follows.
\begin{IEEEeqnarray}{rCl}
nR
\geq I(M;Y^n|Z^n)
\geq I(X^n;Y^n|Z^n)
&\geq& I(X^n;Y^n) - I(X^n;Z^n)
\nonumber\\
&\geq& \sum_{t{=}1}^n I(X_t;Y_t) + \frac{n}{2} \log(1 - \eta^2)
\label{eq:using_that_Xn_iid}\\
&\geq& \sum_{t{=}1}^n\frac{1}{2} \log \frac{1-\eta^2}{1 - \mathbb{E}[\mathbb{E}[X_t|Y_t]^2]}
\nonumber\\
&\geq& \sum_{t{=}1}^n\frac{1}{2} \log \frac{1-\eta^2}{1 - \mathbb{E}[Y_t\mathbb{E}[X_t|Y_t]]^2}
\label{eq:Cauchy_and_monotonicity}
\\
&\geq& \frac{n}{2} \log \frac{1-\eta^2}{1 - ( \tfrac{1}{n}\sum_{t{=}1}^n \mathbb{E}[X_t Y_t])^2}
\label{eq:using_Jensen_on_log_1_over_1_minus_squared_u}\\
&\geq& \frac{n}{2} \log \frac{1-\eta^2}{1 - \rho_\varepsilon^2}
,
\nonumber
\end{IEEEeqnarray}
where \eqref{eq:using_that_Xn_iid} follows from the chain rule for mutual information and from the fact that $P_{X^n} \equiv p_X^{\otimes n}$; and \eqref{eq:Cauchy_and_monotonicity} follows from Cauchy-Schwarz inequality; and \eqref{eq:using_Jensen_on_log_1_over_1_minus_squared_u} follows from Jensen inequality.
This being true for every $\varepsilon>0,$ and any triplet $(R,R_c,\Delta)$ E-D-achievable with near-perfect marginal realism, the claim is proved.
\end{IEEEproof}
It remains to prove that $R_{D}(\Delta)$ is upper bounded by the right hand side
of
\eqref{eq:statement_R_E_D_equals_R_D_bi_dimensional_Gaussian}.

\subsubsection{Case 1: $\eta=0$}
\hfill\\
\indent We have
\begin{IEEEeqnarray}{c}
    R(\Delta,R_c) \underset{R_c \to \infty}{\longrightarrow} \tfrac{1}{2}\log(1/(1-\rho^2)),
\end{IEEEeqnarray}with $R(\Delta,R_c)$ as defined in Proposition \ref{prop:standard_normal_source} and $\rho=1-\Delta/2.$
Therefore, by Proposition \ref{prop:standard_normal_source}, for any $\varepsilon>0$ there exists a finite rate $R_c^{(\varepsilon)}$ such that $(R^{(\varepsilon)},R_c^{(\varepsilon)},\Delta)$
is D-achievable with perfect marginal realism, where
\begin{IEEEeqnarray}{c}
    R^{(\varepsilon)} = \tfrac{1}{2}\log(1/(1-\rho^2)) + \varepsilon.
\end{IEEEeqnarray}
Hence, we have
\begin{equation*}
    \scalebox{0.95}{$R_{D}(\Delta) \leq \tfrac{1}{2}\log(1/(1-\rho^2)).$}
\end{equation*}
\subsubsection{Case 2: $\eta \neq 0$}
\hfill\\
\indent Fix a $\Delta$ in $(0, 2-2|\eta|].$ Let 
\begin{equation}\label{eq:def_Z_X_V}
    \scalebox{0.9}{$(Z,X,V) = \big(\eta \Tilde{X} + \sqrt{1-\eta^2} \Tilde{Z}, \Tilde{X}, b \Tilde{X} + \sqrt{1-b^2} \Tilde{V}\big)$}
\end{equation}
\begin{equation}
    Y=\rho^{-1}\mathbb{E}[X|Z,V],\label{eq:def_Y_Gaussian_achievability}
\end{equation}where $(\Tilde{Z}, \Tilde{X}, \Tilde{V})$ is standard Gaussian, with the real $b$ chosen as follows. 
\begin{claim}\label{claim:choice_of_b}
For a suitable choice of $b$ in $(0,1),$
one can find
\begin{IEEEeqnarray}{c}
    \scalebox{0.93}{$\mathbb{E}\big[ \mathbb{E}[X|Z,V]^2 \big] = \rho^2$}\label{eq:rho_carre}\\*
    \scalebox{0.93}{$p_{X,Z,V,Z} \in \mathcal{D}^{(m)}_{D} \ \text{(Eq. \ref{eq:def_D_D})}$}.
\end{IEEEeqnarray}
\end{claim}
\begin{claim}\label{claim:Y_MMSE_is_a_good_choice}
\begin{IEEEeqnarray}{c}
    \scalebox{1.0}{$I_p(X;V|Z) = \tfrac{1}{2}\log((1 - \eta^2)/(1 - \rho^2)),$}\label{eq:gaussian_conditional_mutual_info}\IEEEeqnarraynumspace\\*
    \mathbb{E}[d(X,Y)]=\Delta\\* 
    I(Y;V) < \infty.
\end{IEEEeqnarray}
\end{claim}

The proofs are deferred to Appendices
\ref{app:subsubsec_proof_claim:choice_of_b}
and
\ref{app:subsubsec_proof_claim:Y_MMSE_is_a_good_choice}.
Moreover, by Claim \ref{claim:uniform_integrability_if_MSE} $(d, p_{X})$ is uniformly integrable. 
Therefore, by Theorem \ref{theorem:D_rates_region_marginal_realism}
there exists a finite $R_c$ such that $(I_p(X;V|Z),R_c,\Delta) \in \overline{\mathcal{A}^{(m)}}_D.$
Then, for any $\varepsilon>0,$ there exists a sequence of $(n,I_p(X;V|Z)+\varepsilon,R_c+\varepsilon)$ D-codes achieving average distortion $\Delta+\varepsilon$ asymptotically, and perfect marginal realism. By considering a vanishing sequence $(\varepsilon_n)_{n \geq 0}$ and combining the latter codes, we obtain that for any $\delta>0,$ the triplet $(I_p(X;V|Z)+\delta,R_c+\delta,\Delta)$ is D-achievable with perfect marginal realism.
Hence, the proof is concluded:
\begin{equation*}
    \scalebox{0.95}{$R_{D}(\Delta) \leq \tfrac{1}{2}\log((1-\eta^2)/(1-\rho^2)).$}
\end{equation*}

\section{Conclusion}
\label{sec:conclusion}

We have studied the role of side information and common randomness in the rate-distortion-perception trade-off with strong perfect and near-perfect realism constraints. Our work complements previous results on the key role of randomization and common randomness in this setting. We have considered the traditional problem of compressing a memoryless source $X^n$ in the presence of
side information $Z^n,$ studied by Wyner and Ziv, with the additional requirement of strong perfect or near-perfect realism on the reconstruction. We also introduced and studied a constraint of joint realism of the reconstruction and the side information. We have characterized the rate-distortion trade-off under perfect or near-perfect realism constraints
when common randomness is available between the encoder and the decoder at a limited rate or is absent.
Our results show how the similarity or dissimilarity between our problem and that of Wyner and Ziv depends on the common randomness rate.
Our work also extends previous results in the absence of side information to general alphabets, under a new kind of assumption which holds for most cases of interest.
The entropy term present in our single-letter characterization when $Z^n$ is available at both terminals somewhat quantifies the role of the side information as a source of common randomness. This provides insight into the problem of video compression in the absence of (another) common source of randomness. We have also showed that
when $X$ and $Z$ are jointly Gaussian and large enough common randomness rate is available, the rate-distortion trade-off is the same whether $Z^n$ is available only at the encoder or at both terminals, but that this breaks down in the absence of common randomness, thereby further highlighting the role of common randomness. We have studied the latter role in more detail for Normal $X$ in the absence of side information.

\appendices

\section{Measure-theoretic justifications}\label{app:sec:measure_theoretic_justifications}

\subsection{General results}\label{app:subsec:measure_theory_general}

\subsubsection{Conditional probabilities on general alphabets and sequential definitions}
\hfill\\
\indent The family of Polish spaces includes all discrete spaces and all finite-dimensional real vector spaces, and any finite product of Polish spaces is a Polish space. Every measurable space we consider that is of the form $(\mathcal{W},\mathcal{B})$ with $\mathcal{W}$ a Polish space satisfies that $\mathcal{B}$ is its Borel $\sigma$-algebra.
A discrete alphabet is always endowed with its power set as topology, and hence its Borel $\sigma$-algebra is its power set. Finite-dimensional real vector spaces are endowed with the standard topology (induced by any norm). A product of the two measurable spaces will be implicitly endowed with the product $\sigma$-algebra. If the two spaces are Polish, then the product of (Borel) $\sigma$-algebras is equal to the Borel $\sigma$-algebra of the product Polish space.
As is standard practice (see, e.g., \cite{1995ProbabilityAndMeasure}), all conditional probability kernels we consider are defined for all elements of the input alphabet (not just almost everywhere):
\begin{definition}\label{def:regular_kernel}
    Given any two measurable spaces $(\mathcal{W}, \mathcal{B})$ and $(\mathcal{W}', \mathcal{B}'),$ a \textit{regular channel} $\rho,$ also called a \textit{regular conditional probability kernel} or \textit{regular transition kernel} from the former to the latter is a function from $\mathcal{W}\times \mathcal{B}'$ to $[0,1]$ such that for every $w \in \mathcal{W},$ function $\rho(w,\cdot)$ is a probability measure on $(\mathcal{W}', \mathcal{B}')$ and for every $B' \in \mathcal{B}',$ the function $\rho(\cdot, B')$ is $\mathcal{B}$-measurable.
\end{definition}
\vspace{5pt}

The well-known Fubini theorem generalizes beyond product measures as follows.
\begin{theorem}\label{thm:generalized_fubini}(Generalized Fubini theorem) \cite[Proposition~III-2-1]{1970JacquesNeveuBasesProbabilites}\\
Consider any two measurable spaces $(\mathcal{W}, \mathcal{B})$ and $(\mathcal{W}', \mathcal{B}'),$ a probability measure $\mu$ on the former and a regular transition kernel $\rho(\cdot, \cdot)$ from $(\mathcal{W}, \mathcal{B})$ to $(\mathcal{W}', \mathcal{B}').$ Then, there is a unique measure $Q$ on the product space $(\mathcal{W}\times\mathcal{W}', \mathcal{B}\otimes\mathcal{B}')$ with\begin{equation}
    \forall B\in \mathcal{B}, \forall B' \in \mathcal{B}', \ Q(B\times B') = \int_B \rho(w, B') d\mu(w).\label{eq:def_of_composition}
\end{equation} We denote this distribution by $\mu \cdot \rho.$ Its marginal on $\mathcal{W}$ is $\mu.$ Moreover, for any non-negative measurable function $\varphi,$ the map $w \mapsto \int \varphi \ d\rho(w, \cdot)$ is well-defined on all $\mathcal{W}$ and is $\mathcal{B}$-measurable, and we have\begin{equation*}
    \int \varphi \ dQ = \int \bigg(\int \varphi \ d\rho(w, \cdot) \bigg) d\mu(w).
\end{equation*}When $\rho$ is constant in its first argument and equal to a distribution $\nu$ on $\mathcal{W}',$ the distribution $\mu \cdot \rho$ --- or $\mu \cdot \nu$ --- is the product distribution.
Given a distribution $Q$ on $(\mathcal{W}\times\mathcal{W}', \mathcal{B}\otimes\mathcal{B}')$ with marginal $\mu$ on $\mathcal{W},$ then any regular transition kernel $\rho(\cdot,\cdot)$ satisfying \eqref{eq:def_of_composition} is called \textit{a regular transition kernel of $Q$ from $\mathcal{W}$ to $\mathcal{W}'.$}
If $(W,W')$ is a random tuple with distribution $Q,$ then we denote $\mu$ by $Q_W$ and we denote $\rho$ by $Q_{W'|W}$ --- although such a transition kernel might not be unique. For any $w\in\mathcal{W},$ we denote the distribution $\rho(w,\cdot)$ by $Q_{W'|W{=}w}.$ We also use, with abuse of notation: $Q \equiv Q_W \cdot Q_{W'|W{=}w},$ where $w$ is a dummy variable.
\end{theorem}
\begin{remark}\label{rem:a_notion_of_uniqueness_of_the_transition_kernel}
Moreover, in the setting of Theorem \ref{thm:generalized_fubini}, consider a distribution $Q$ on $(\mathcal{W}\times\mathcal{W}', \mathcal{B}\otimes\mathcal{B}')$ and two regular transition kernels $\rho(\cdot,\cdot)$ and $\Tilde{\rho}(\cdot,\cdot)$ of $Q$ from $\mathcal{W}$ to $\mathcal{W}'.$ Then, for any measurable set $B',$ the maps $w \mapsto \rho(w, B')$ and $w \mapsto \Tilde{\rho}(w, B')$ are equal outside of a $\mu$-null subset of $\mathcal{W},$ where $\mu$ is the marginal of $Q$ on $\mathcal{W}.$
Although there is no functional uniqueness of a regular transition kernel of $Q$ from $\mathcal{W}$ to $\mathcal{W}',$ in many cases an integral involving such a kernel has a unique value.
\end{remark}

The claim in the above remark is proved in Section \ref{app:subsec:proofs_of_measure_theory_lemmas}. Moreover, Theorem \ref{thm:generalized_fubini} implies the following corollary.
We do not provide the proof, which is rather straightforward, for the sake of conciseness.
\begin{corollary}\label{cor:transition_kernel_with_few_active_variables}
Consider four Polish measurable spaces $(\mathcal{W}_1, \mathcal{B}_1), (\mathcal{W}_2, \mathcal{B}_2), (\mathcal{W}_3, \mathcal{B}_3), (\mathcal{W}_4, \mathcal{B}_4),$ a random triplet $(W_1,W_2,W_3)$ taking values in $\mathcal{W}_1\times\mathcal{W}_3\times\mathcal{W}_3$ and a transition kernel $\rho(\cdot,\cdot)$ from $(\mathcal{W}_3, \mathcal{B}_3)$ to $(\mathcal{W}_4, \mathcal{B}_4).$ Denote the distribution of $(W_1,W_2,W_3)$ by $\mu_{W_1,W_2,W_3}.$
The following map
\begin{IEEEeqnarray}{c}
\Tilde{\rho}: \mathcal{W}_1 \times \mathcal{W}_2 \times \mathcal{W}_3 \times \mathcal{B}_4 \to [0,1], \ (w_1,w_2,w_3,B_4) \mapsto \rho(w_3,B_4)
\end{IEEEeqnarray}
is
a regular transition kernel
from $(\mathcal{W}_1\times\mathcal{W}_2\times\mathcal{W}_3, \mathcal{B}_1\otimes\mathcal{B}_2\otimes\mathcal{B}_3)$ to $(\mathcal{W}_4, \mathcal{B}_4).$
Define
\begin{IEEEeqnarray}{c}
Q_{W_1,W_2,W_3,W_4} := \mu_{W_1,W_2,W_3} \cdot \Tilde{\rho}.\nonumber
\end{IEEEeqnarray}
Then,
$\rho$ is
a regular transition kernel of
$Q_{W_3,W_4}$ from $(\mathcal{W}_3, \mathcal{B}_3)$ to $(\mathcal{W}_4, \mathcal{B}_4),$
and
$(w_2,w_3,B_4)\mapsto \rho(w_3,B_4)$
is
a regular transition kernel of
of $Q_{W_2,W_3,W_4}$ from $(\mathcal{W}_2\times\mathcal{W}_3, \mathcal{B}_2\otimes\mathcal{B}_3)$ to $(\mathcal{W}_4, \mathcal{B}_4).$
With some abuse of notation, we shall denote all the aforementioned maps by $\rho,$ or $Q_{W_4|W_3}.$
\end{corollary}

When given a joint distribution, such as from \eqref{eq:def_D_D}, we extensively use corresponding conditional distributions, which is justified if the alphabets are Polish spaces according to the following theorem, sometimes referred to as Jirina's theorem.
\begin{theorem}\cite[Theorem~9.2.2]{BookWithJirinaTheorem2010}\label{theorem:jirina}
Let $\mathcal{W}$ and $\mathcal{W}'$ be Polish spaces and let $\Omega=\mathcal{W}\times \mathcal{W}'.$
Let $Q$ be a probability measure on $\Omega.$ 
Then, there exists a regular probability transition kernel $\rho(\cdot, \cdot)$
of $Q$
from $\mathcal{W}$ to $\mathcal{W}'.$ 
\end{theorem}

We will also use the following conditional version of Theorem \ref{thm:generalized_fubini}, valid for Polish spaces, which allows us to build transition kernels sequentially. The proof is deferred to Section \ref{app:subsec:proofs_of_measure_theory_lemmas}.
\begin{proposition}\label{prop:associativity_of_kernel_composition}
    Consider three Polish measurable spaces $(\mathcal{W}_1, \mathcal{B}_1), (\mathcal{W}_2, \mathcal{B}_2), (\mathcal{W}_3, \mathcal{B}_3),$ a transition kernel $\rho(\cdot,\cdot)$ from $(\mathcal{W}_1, \mathcal{B}_1)$ to $(\mathcal{W}_2, \mathcal{B}_2)$ and a transition kernel $\gamma(\cdot, \cdot, \cdot)$ from $(\mathcal{W}_1 \times \mathcal{W}_2, \mathcal{B}_1 \otimes \mathcal{B}_2)$ to $(\mathcal{W}_3, \mathcal{B}_3).$ 
    Define $\lambda: \mathcal{W}_1 \times (\mathcal{B}_2 \otimes \mathcal{B}_3) \to [0,1],$ 
    \begin{equation}\label{eq:conditional_transition_kernel} (w_1, B_{2,3}) \mapsto \Big( \rho(w_1, \cdot) \cdot \gamma(w_1, \cdot, \cdot) \Big)(B_{2,3}),\end{equation}
    with the notation of Theorem \ref{thm:generalized_fubini}.
    Then, $\lambda$ is a regular transition kernel from $(\mathcal{W}_1 ,\mathcal{B}_1)$ to $(\mathcal{W}_2 \times \mathcal{W}_3, \mathcal{B}_2 \otimes \mathcal{B}_3).$ We denote it $\rho \cdot \gamma.$ Moreover, for any probability measure $\mu$ on $(\mathcal{W}_1, \mathcal{B}_1)$ we have associativity:$$(\mu \cdot \rho) \cdot \gamma \equiv \mu \cdot (\rho \cdot \gamma).$$ Consequently, for the resulting distribution, say $Q,$ kernel $\rho$ is a transition kernel of $Q$ from $\mathcal{W}_1$ to $\mathcal{W}_2,$ kernel $\gamma$ is a transition kernel of $Q$ from $\mathcal{W}_1 \times \mathcal{W}_2$ to $\mathcal{W}_3$ and $\lambda$ is a transition kernel of $Q$ from $\mathcal{W}_1$ to $\mathcal{W}_2 \times \mathcal{W}_3.$
    We use the product notation $\prod$ to denote sequential composition.
\end{proposition}
\vspace{5pt}

We provide a proof in Section \ref{app:subsec:proofs_of_measure_theory_lemmas}.
\vspace{10pt}
\subsubsection{The notion of Markov chain for general alphabets}
\hfill\\
\indent We follow the definition of the notion of Markov chain for 
general alphabets given in \cite[page~196]{2011GrayGeneralAlphabetsRelativeEntropy}, which we state in a slightly different but equivalent form:
\begin{definition}\label{def:Markov_chain}
    Consider Polish spaces $\mathcal{W}_1,$ $\mathcal{W}_2$ and $\mathcal{W}_3.$
    Let $(W_1,W_2,W_3)$ be a random tuple taking values in $\mathcal{W}_1 \times \mathcal{W}_2 \times \mathcal{W}_3,$ with distribution denoted $Q.$ 
    Then, we say that $Q$ \textit{satisfies the Markov chain} $W_1-W_2-W_3$ iff for some choice of transition kernel of $Q,$ we have
    \begin{IEEEeqnarray}{c}
        Q_{W_1,W_2,W_3} \equiv Q_{W_1,W_2} \cdot Q_{W_3|W_2{=}w_2}.
    \end{IEEEeqnarray}
    Moreover, this property does not depend on the choice of transition kernel.
    The notion extends to tuples of more than three variables. For any $n \geq 4,$ we say that a family $(W_i)_{1\leq i \leq n}$ satisfies the Markov chain $W_1 - W_2 -...-W_n$ iff for any $k \in [n]\backslash \{1,n\},$ we have the Markov chain $W_{1:k-1}-W_k-W_{k+1:n}.$
\end{definition}
\vspace{5pt}

The independence on the choice of kernels is a direct application of Remark \ref{rem:a_notion_of_uniqueness_of_the_transition_kernel}.
The notion of Markov chain is related to conditional mutual information, as stated in Lemma \ref{lemma:Markov_chain_is_null_conditional_mutual_info}.

\subsubsection{Density functions}
\hfill\\
\indent When a joint distribution over Polish spaces is dominated by another, conditional densities exist and form a regular transition kernel, as stated in the following corollary of Theorems \ref{thm:generalized_fubini} and \ref{theorem:jirina}.

\begin{corollary}\label{cor:conditional_density_exists_and_gives_regular_kernel}
Consider Polish spaces $\mathcal{W}$ and $\mathcal{W}'.$
Let $P$ and $Q$ be probability measures on $\mathcal{W} \times \mathcal{W}',$ with marginals on $\mathcal{W}$ denoted by $\mu$ and $\nu$ respectively. Assume that $Q $ dominates $P$ (then $\nu$ dominates $\mu$). Fix a density function $\tfrac{dP}{dQ}(\cdot, \cdot)$ and a regular transition kernel
$Q_{W'|W}$
of $Q$ from $\mathcal{W}$ to $\mathcal{W}'.$
Then,
\begin{equation*}
 f: \mathcal{W} \to [0,\infty], \ w \mapsto \int \dfrac{dP}{dQ}(w,w') \ 
 dQ_{W'|W{=}w}(w')
\end{equation*}
defines a density of $\mu$ with respect to $\nu.$
Define
\begin{equation*}
    g: \mathcal{W} \times \mathcal{W}' \to [0,\infty], \ (w,w') \mapsto \begin{cases}\dfrac{dP}{dQ}(w,w') \Big/ f(w) \text{ if } f(w) >0,\\ \ 1 \qquad \qquad \qquad \ \; \text{ if } f(w) =0.\end{cases}
\end{equation*}
Then, $g$ is measurable and $Q$-integrable and the following map defines a regular probability transition kernel of $P$ from $\mathcal{W}$ to $\mathcal{W}':$
\begin{equation*}
    (w,B) \mapsto \int g(w,w') \cdot \mathbf{1}_B(w') \ 
    dQ_{W'|W{=}w}(w')
    .
\end{equation*}The corresponding distribution for a given $w \in \mathcal{W}$ admits $g(w,\cdot)$ as a density function with respect to $Q_{W'|W{=}w}.$
\end{corollary}

The proof is provided in Section \ref{app:subsec:proofs_of_measure_theory_lemmas}.
The total variation distance between distributions $P,Q$ on general alphabets is defined by \begin{equation}
   \sup_{A}\{P(A)-Q(A)\} = \dfrac{1}{2} \int |p-q|d\mu = \int (p-q)^+d\mu,
   \label{eq:def_TVD}
\end{equation}for any common dominating $\mu$ and corresponding densities $p,q.$

\subsection{Information quantities on general alphabets}

\subsubsection{Fundamental definitions and results}\label{app:subsubsec:info_theory_for_general_spaces}

\begin{definition}\cite[Section~7.1]{2011GrayGeneralAlphabetsRelativeEntropy}
Given a probability space $(\Omega,\mathcal{B},P)$ and another probability measure $Q$ on the same space, define the relative entropy or divergence of $P$ with respect to $Q$ as
\begin{IEEEeqnarray}{c}
D(P\|Q) = \sup_{X} \sum_{x \in \mathcal{X}} P_X(x)\log\Big(\dfrac{P_X(x)}{Q_X(x)}\Big),
\end{IEEEeqnarray}where the supremum is over all random variables $X$ defined from $(\Omega,\mathcal{B})$ into any finite-valued alphabet $\mathcal{X},$ and where we assume the convention $a/0 = + \infty,$ for $a>0.$
\end{definition}
\begin{lemma}\label{lemma:mutual_info_geq_0}\cite[Lemma~7.1]{2011GrayGeneralAlphabetsRelativeEntropy}
For any distributions $P$ and $Q$ we have $D(P\|Q) \geq 0,$ with equality iff $P \equiv Q.$
\end{lemma}
\begin{definition}\label{def:mutual_information}\cite[Equations~7.28,~7.29,~7.30]{2011GrayGeneralAlphabetsRelativeEntropy}
Let $X,$ $Y$ and $Z$ denote random variables defined on the same probability space and taking values in some Polish spaces. Define the mutual information, entropy and conditional mutual information by
\begin{IEEEeqnarray}{rCl}
I(X;Y) &=& D(P_{X,Y} \| P_X \cdot P_Y)
\nonumber\\
H(X) &=& I(X;X)
\nonumber\\
I(X;Y|Z) &=& D(P_{X,Y,Z} \| P_Z \cdot P_{X|Z} \cdot P_{Y|Z}),
\nonumber
\end{IEEEeqnarray}where $P_{X,Y,Z}$ denotes the distribution of $(X,Y,Z).$ The distribution $P_Z \cdot P_{X|Z} \cdot P_{Y|Z}$ does not depend on the choice of transition kernels $P_{X|Z}$ and $P_{Y|Z}.$ Since these definitions depend only on distributions, we also denote $H(X)$ by $H(P_X).$
\end{definition}These quantities satisfy the following properties.
\begin{lemma}\label{lemma:entropy_geq_mutual_info}\cite[Corollary~7.11,~Lemma~7.20]{2011GrayGeneralAlphabetsRelativeEntropy}
Let $X$ and $Y$ denote random variables defined on the same probability space and taking values in some Polish spaces. Then,
\begin{IEEEeqnarray}{c}
    H(X) \geq I(X;Y).
\end{IEEEeqnarray}If $I(X;Y)<\infty,$ define $H(X|Y):=H(X)-I(X;Y).$ Let $P_{X,Y}$ denote the distribution of $(X,Y).$ If $X$ is finite-valued, then for any regular transition kernel $P_{X|Y}$ of $P,$ we have
\begin{IEEEeqnarray}{c}
    H(X|Y) = \int H(P_{X|Y{=}y})dP_Y(y).
\end{IEEEeqnarray}
\end{lemma}
\begin{lemma}\cite[Corollary~7.16]{2011GrayGeneralAlphabetsRelativeEntropy}\label{lemma:deterministic_map_reduces_I}
    Let $X$ and $Y$ denote random variables defined on the same probability space and taking values in some Polish spaces. For any measurable functions $f$ and $g$ we have
    \begin{IEEEeqnarray}{c}
        I(f(X);g(Y)) \leq I(X;Y).\nonumber
    \end{IEEEeqnarray}
\end{lemma}
\begin{proposition}\label{prop:chain_rule}\cite[Corollaries~7.14~\&~7.15]{2011GrayGeneralAlphabetsRelativeEntropy}
    Let $X,$ $Y,$ $Z$ and $W$ be finite-valued random variables. Then, we have
    \begin{IEEEeqnarray}{c}
        I(X;Y,Z|W) = I(X;Y|W) + I(X;Z|Y,W).\label{eq:conditional_chain_rule}
    \end{IEEEeqnarray}Moreover, for random variables $X',$ $Y'$ and $Z'$ taking values in Polish spaces, we have
    \begin{IEEEeqnarray}{c}
        I(X';Y',Z') = I(X';Y') + I(X';Z'|Y'),\nonumber
    \end{IEEEeqnarray}with the conventions $\infty+u=\infty$ for any (finite) real $u$ and $\infty+\infty=\infty.$
\end{proposition}
\begin{lemma}\label{lemma:Markov_chain_is_null_conditional_mutual_info}\cite[Lemma~7.1~\&~Corollaries~7.14~\&~7.15]{2011GrayGeneralAlphabetsRelativeEntropy}
Let $X,Y,Z$ be random variable taking values in some Polish spaces. Then, $X$ and $Y$ are independent iff $I(X;Y)=0.$ Moreover, the Markov chain $X-Z-Y$ holds iff $I(X;Y|Z)=0.$ In that case, we have the \textit{data processing inequality}:
\begin{IEEEeqnarray}{c}
    I(X;Z) \geq I(X;Y).\nonumber
\end{IEEEeqnarray}
\end{lemma}

The conditional chain rule \eqref{eq:conditional_chain_rule} generalizes to infinite alphabets under certain conditions, but in this paper we prefer to perform mutual information computations with quantized variables and take the limit.
\textit{A quantizer} is any finite-valued measurable map. Under the following conditions, information quantities of variables on general alphabets are limits of the corresponding quantities for quantized variables, and the sequence of quantizers can be chosen independently of the original variables' distribution.
\begin{proposition}\label{prop:limit_on_sequence_of_quantizers}\cite[Corollaries~7.14~\&~7.15~\&~Lemma~7.18]
{2011GrayGeneralAlphabetsRelativeEntropy}
Let $X,$ $Y$ and $Z$ denote random variables defined on the same probability space and taking values in some Polish spaces $(\mathcal{W}_1, \mathcal{B}_1),$
$(\mathcal{W}_2, \mathcal{B}_2),$
and $(\mathcal{W}_3,\mathcal{B}_3),$
respectively. For each
$i \in \{1,2
,3
\},$
let $(\kappa_i^{(n)})_{n \geq 1}$ be a sequence of quantizers such that the corresponding partitions asymptotically generate $\mathcal{B}_i.$ If for some
$i \in \{1,2
,3
\}$
the set $\mathcal{W}_i$ is finite, then $(\kappa_i^{(n)})_{n \geq 1}$ can be a constant sequence with $\kappa_i^{(1)}$ being the identity map. Then, whatever the distribution of $(X,Y,Z)$ we have
\begin{IEEEeqnarray}{rCl}
    I(X;Y,Z) &=& \underset{n \to \infty}{\lim} I\big(\kappa_1^{(n)}(X);\kappa_2^{(n)}(Y),\kappa_3^{(n)}(Z)\big) \nonumber\\
    H(X) &=& \underset{n \to \infty}{\lim} H\big(\kappa_1^{(n)}(X)\big).\nonumber
\end{IEEEeqnarray}Moreover, if $I(X;Z)<\infty$ or $I(Y;Z)<\infty,$ then
\begin{IEEEeqnarray}{rCl}
    I(X;Y|Z) &=& \underset{n \to \infty}{\lim} I\big(\kappa_1^{(n)}(X);\kappa_2^{(n)}(Y)
    |
    \kappa_3^{(n)}(
    Z
    )
    \big)
    \label{eq:limit_of_conditional_mutual_info_over_sequence_of_quantizers}
\end{IEEEeqnarray}
\end{proposition}
\begin{IEEEproof}
The set of partitions of quantizer $(y,z)\mapsto(\kappa_2^{(n)}(Y),
\kappa_3^{(n)}(Z))$ is the product of the respective sets of partitions of $\kappa_2^{(n)}$ and $\kappa_3^{(n)}.$ Then, that set asymptotically generates the product $\sigma$-algebra $\mathcal{B}_2 \otimes \mathcal{B}_3,$ which is the Borel $\sigma$-algebra of the Polish product space $\mathcal{Y}\times\mathcal{Z}.$
Hence,
the first two properties of Proposition \ref{prop:limit_on_sequence_of_quantizers} follow from \cite[Lemma~7.18]{2011GrayGeneralAlphabetsRelativeEntropy}.
Assume that $I(X;Z)<\infty.$
Then, from Proposition \ref{prop:chain_rule},
\begin{IEEEeqnarray}{rCl}
\forall n\in\mathbb{N}, \ 
I\big(\kappa_1^{(n)}(X);\kappa_2^{(n)}(Y)|
\kappa_3^{(n)}(Z)
\big)
&=&
I\big(\kappa_1^{(n)}(X);\kappa_2^{(n)}(Y),
\kappa_3^{(n)}(Z)
\big)
-
I\big(\kappa_1^{(n)}(X);
\kappa_3^{(n)}(Z)
\big)
,\nonumber\\
I(X;Y|Z) &=& I(X;Y,Z) - I(X;Z)
.\nonumber
\end{IEEEeqnarray}
Then, 
\eqref{eq:limit_of_conditional_mutual_info_over_sequence_of_quantizers} follows
from the first property of Proposition \ref{prop:limit_on_sequence_of_quantizers}.
By symmetry, \eqref{eq:limit_of_conditional_mutual_info_over_sequence_of_quantizers} also holds if $I(Y;Z)<\infty.$
\end{IEEEproof}
\vspace{5pt}

Moreover, we have the following fundamental link to the \textit{information density}.
\begin{proposition}\label{prop:information_density_exists}\cite[Equation~7.28,~Lemma~7.4]{2011GrayGeneralAlphabetsRelativeEntropy}
Let $(X,Y)$ be a couple of joint distribution $P_{X,Y}.$ If the product distribution $P_X\cdot P_Y$ dominates $P_{X,Y},$ we define the \textit{information density} $$i_{P_{X,Y}} = \log\Bigg(\dfrac{dP_{X,Y}}{d(P_X\cdot P_Y)}\Bigg),$$ with values in $[-\infty, +\infty].$ Then, $\min(0,i_{P_{X,Y}})$ is $P_{X,Y}$-integrable and we have \begin{equation}
    I(X;Y) = \int i_{P_{X,Y}}(x,y)dP_{X,Y}(x,y).
\end{equation} Moreover, $P_X\cdot P_Y$ dominates $P_{X,Y}$ whenever $I(X;Y)<\infty,$ and in that case the information density is well-defined and $P_{X,Y}$-integrable.
\end{proposition}
\vspace{5pt}

We omit known results on entropy and mutual information for finite-valued variables.
\vspace{10pt}
\subsubsection{Decoder for bin index $M'$ in the case of infinite alphabets}\label{app:decoder_M_prime}
\hfill\\
\indent We prove Claim \ref{claim:decoding_virtual_message_general_alphabets}.
By definition of $R',$ if $I_p(Z;V)=0$ then $R'=0$ and the alphabet of $M'$ is a singleton, and therefore $M'$ can be decoded perfectly for any choice of decoder. Suppose the contrary. Then, by definition of $R'$ we have $R'<I_p(Z;V).$
Then, from Proposition \ref{prop:limit_on_sequence_of_quantizers} there exists a (measurably) quantized tuple $([Z], [V])$ such that $I_p([Z]; [V]) > R'.$ 
As depicted in Figure \ref{fig:quantization_setup}, we extend the definition of $Q^{(1)}$ from \eqref{eq:def_Q_1_two_sided} by:
\begin{figure}[t!]
    \centering\includegraphics[width=0.7\textwidth]{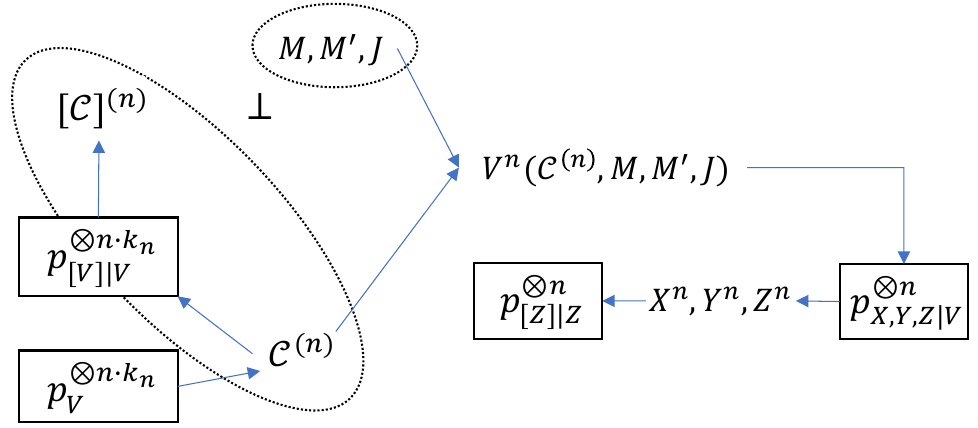}
    \caption{Graphical model for $Q^{(1)}$ with quantized variables, where $k_n = \lfloor 2^{n(R+\varepsilon)}\rfloor \times \lfloor 2^{nR'}\rfloor \times \lfloor 2^{nR_c}\rfloor.$ The variables inside a dashed oval are independent from variables inside another oval.}
    \label{fig:quantization_setup}
\end{figure}
\begin{equation}\label{eq:def_single_letter_quantization}
    p_{V,X,Y,Z,[V],[Z]} = p_{V,X,Y,Z}\cdot p_{[V]|V} \cdot p_{[Z]|Z} \text{ and}
\end{equation}\begin{IEEEeqnarray}{rCl}
\IEEEeqnarraymulticol{3}{l}{
\substack{\scalebox{1.0}{$Q^{(1)} \; ([z]^n, [c]^{(n)}, \hat{m}') \qquad \qquad \qquad \qquad \qquad \qquad \qquad \quad \;$} \\ [Z]^n, [\mathcal{C}]^{(n)}, \hat{M}' |\mathcal{C}^{(n)}\text{=}c^{(n)}, M\text{=}m, M'\text{=}m', J\text{=}j, X^n\text{=}x^n, Y^n\text{=}y^n, Z^n\text{=}z^n}
}\nonumber\\*
\quad & = & \prod_{k=1}^n \raisebox{2pt}{\scalebox{1.1}{$p$}} \substack{\scalebox{1.0}{$ \: ([z]_k)$} \\ \scalebox{0.7}{$[Z]|Z\text{=}z_k$} } \prod_{\Tilde{j},\Tilde{m},\Tilde{m}'} \prod_{k=1}^n \raisebox{2pt}{\scalebox{1.1}{$p$}} \substack{\scalebox{1.0}{$ ([v]_k) \qquad \qquad $} \\ \scalebox{0.7}{$[V]|V\text{=}v_k(c^{(n)}, \Tilde{m}, \Tilde{m}', \Tilde{j})$} }\nonumber\\*
&& \substack{ \raisebox{-2pt}{$P^{D}$} \scalebox{1.0}{$ \; (\hat{m}') \qquad \qquad \qquad \qquad \qquad \; \ $} \\ \hat{M}' | [\mathcal{C}]^{(n)}=[c]^{(n)}, M{=}m, J{=}j, [Z]^n=[z]^n}, \label{eq:def_quantized_codebook_and_P_D}
\end{IEEEeqnarray}where $P^D$ represents, for each $(m,j),$ the joint typicality decoder of \cite[page~201]{Cover&Thomas2006} for the joint distribution $p_{[V],[Z]}$ and codebook $([V]^n([\mathcal{C}]^{(n)}, m, a, j))_a.$ Note that all transition kernels in the right-hand-side are regular, because each represents a measurable deterministic mapping.
It can be checked that this setting is that of the channel random coding proof of \cite[Section~7.7]{Cover&Thomas2006}.
Moreover, for any $(m,j)$ the distribution of the corresponding random sub-codebook of $[\mathcal{C}]^{(n)}$ does not depend on $(m,j).$
Therefore,
we have
\begin{IEEEeqnarray}{c}
\sup_{(m,j)}
\substack{\scalebox{1.0}{$Q^{(1)} \; (\hat{M}'\neq \ M') $} \\ \hat{M}', M' |M{=}m, J{=}j } \underset{n \to \infty}{\longrightarrow} 0. \label{eq:Q_1_decoding_error_fixed_sub_codebook_general_sources} \IEEEeqnarraynumspace
\end{IEEEeqnarray}

\subsection{Proofs of lemmas stated in Appendix \ref{app:sec:measure_theoretic_justifications} and in achievability proofs}\label{app:subsec:proofs_of_measure_theory_lemmas}

\subsubsection{Proof of Remark \ref{rem:a_notion_of_uniqueness_of_the_transition_kernel}}
\hfill\\
\indent Fix a measurable set $B'\subset \mathcal{W}'.$ For any measurable set $B\subset \mathcal{W},$ we have $$Q(B\times B') = \int_B \rho(w, B') d\mu$$ by taking $\varphi=\mathbf{1}_{B \times B'}.$
If $Q(\mathcal{W}\times B')=0,$ then $\rho(\cdot, B')=0$ $\mu$-almost everywhere. If $Q(\mathcal{W}\times B')>0,$ then $w \mapsto \rho(w, B')$ is a density of $Q(\cdot|B')$ w.r.t. $\mu$ up to a multiplicative factor. Hence, it is unique up to a $\mu$-null set.

\subsubsection{Proof of Proposition \ref{prop:associativity_of_kernel_composition}}
\hfill\\
\indent First we justify that for any $w_1 \in \mathcal{W}_1,$ the object $\rho(w_1, \cdot) \cdot \gamma(w_1, \cdot, \cdot)$ is well defined. We just need to prove that $\gamma(w_1, \cdot, \cdot)$ is a regular transition kernel. Only the measurability is (highly) non-trivial. Let $B_3$ be in $\mathcal{B}_3$ and $A$ be a real Borel set. 
Then, $\gamma(\cdot, \cdot, B_3)^{-1}(A) \in \mathcal{B}_1 \otimes \mathcal{B}_2$ by definition of $\gamma.$ Since $\mathcal{W}_1$ is Polish, it is metrizable, therefore T4 (normal), hence T1: its singletons are closed. Since $\mathcal{B}_1$ is the Borel $\sigma$-algebra of $\mathcal{W}_1,$ we have $\{w_1\} \in \mathcal{B}_1,$ hence $ \{w_1\}\times \mathcal{W}_2 \in \mathcal{B}_1 \otimes \mathcal{B}_2.$ Therefore, $\gamma(\cdot, \cdot, B_3)^{-1}(A) \cap (\{w_1\} \times \mathcal{W}_2) \in \mathcal{B}_1 \otimes \mathcal{B}_2.$ 
The pre-image of the latter by the map $\pi: w_2 \mapsto (w_1,w_2)$ gives the desired set $\gamma(w_1, \cdot, B_3)^{-1}(A).$ The canonical bijection $\pi$ is continuous. 
Since $\mathcal{B}_1$ and $\mathcal{B}_2$ are the Borel $\sigma$-algebras of Polish spaces $\mathcal{W}_1$ and $\mathcal{W}_2,$ then
$\mathcal{B}_1 \otimes \mathcal{B}_2$ is the Borel $\sigma$-algebra of $\mathcal{W}_1 \times \mathcal{W}_2,$ that is, the $\sigma$-algebra generated by its open sets. From the continuity of $\pi,$ the pre-image of an open set by $\pi$ is an open set of $\mathcal{W}_2$. Therefore, the pre-image of $\mathcal{B}_1 \otimes \mathcal{B}_2$ by $\pi$ is the $\sigma$-algebra generated by the open sets of $\mathcal{W}_2$ (e.g. \cite[Section~II.1]{1970JacquesNeveuBasesProbabilites}), i.e. $\mathcal{B}_2.$ 
This proves that $\gamma(w_1, \cdot, B_3)^{-1}(A) \in \mathcal{B}_2.$
This being true for any real Borel set $A,$ then $w_2\mapsto \gamma(w_1,w_2,B_3)$ is $\mathcal{B}_2$-measurable. This being true for any $B_3 \in \mathcal{B}_3,$ then $\gamma(w_1, \cdot, \cdot)$ is a regular transition kernel from $\mathcal{W}_2$ to $\mathcal{W}_3.$
Since $\rho(w_1, \cdot)$ is a distribution on $\mathcal{W}_2,$ then the object $\rho(w_1, \cdot) \cdot \gamma(w_1, \cdot, \cdot)$ is well defined, as per Theorem \ref{thm:generalized_fubini}. This is true for any $w_1 \in \mathcal{W}_1.$

Second, consider a probability measure $\mu$ on $(\mathcal{W}_1, \mathcal{B}_1)$ and define the distribution $Q = (\mu \cdot \rho) \cdot \gamma.$ Then, by Theorem \ref{thm:generalized_fubini}, for any non-negative and $\mathcal{B}_1 \otimes \mathcal{B}_2 \otimes \mathcal{B}_2$-measurable $\varphi,$ the map $(w_1,w_2) \mapsto \int \varphi d\gamma(w_1,w_2, \cdot)$ is $\mathcal{B}_1 \otimes \mathcal{B}_2$-measurable and 
$$ \int \varphi dQ = \int \Bigg( \int \varphi(w_1,w_2,\cdot) d\gamma(w_1,w_2, \cdot) \Bigg) d(\mu \cdot \rho).$$
Hence, we can apply Theorem \ref{thm:generalized_fubini} on $Q' = \mu \cdot \rho$ and obtain that the map 
\begin{equation}\label{eq:conditional_Tonelli} w_1 \mapsto \int \Bigg( \int \varphi(w_1,w_2,\cdot) d\gamma(w_1,w_2, \cdot)\Bigg) d\rho(w_1, \cdot)\end{equation}
is $\mathcal{B}_1$-measurable and that $\int \varphi dQ$ rewrites as 
\begin{IEEEeqnarray}{c} \label{eq:triple_generalized_Fubini} \int \Bigg[ \int \Bigg( \int \varphi(w_1,w_2,\cdot) d\gamma(w_1,w_2, \cdot) \Bigg) d\rho(w_1, \cdot) \Bigg] d\mu.\IEEEeqnarraynumspace\end{IEEEeqnarray}
Fix some $w_1$ in $\mathcal{W}_1.$ By Theorem \ref{thm:generalized_fubini} on the double integral in \eqref{eq:conditional_Tonelli} with joint distribution $\lambda(w_1, \cdot) =\rho(w_1, \cdot) \cdot \gamma(w_1, \cdot, \cdot),$ the measurable map in \eqref{eq:conditional_Tonelli} rewrites as $$ w_1 \mapsto \int \varphi(w_1, \cdot, \cdot) d\lambda(w_1, \cdot).$$ This shows that $\lambda$ is a regular transition kernel. 
Furthermore, equation \eqref{eq:triple_generalized_Fubini} becomes 
$$\int \varphi dQ = \int \Bigg(\int \varphi(w_1, \cdot, \cdot) d\lambda(w_1, \cdot) \Bigg) d\mu.$$
Since $\lambda$ is a regular transition kernel, then by Theorem \ref{thm:generalized_fubini} this becomes:
$$\int \varphi dQ =\int \varphi d(\mu \cdot \lambda).$$
This being true for any non-negative $\mathcal{B}_1 \otimes \mathcal{B}_2 \otimes \mathcal{B}_2$-measurable $\varphi,$ we obtain $$ (\mu \cdot \rho) \cdot \gamma \equiv Q \equiv \mu \cdot \lambda.$$

\subsubsection{Proof of Corollary \ref{cor:conditional_density_exists_and_gives_regular_kernel}}
\hfill\\
\indent The fact that $f$ is a density of $\mu$ with respect to $\nu$ follows from a simple manipulations of integrals using Theorem \ref{thm:generalized_fubini}.
For any $a\in\mathbb{R},$ we have
\begin{IEEEeqnarray}{c}
\Big\{(w,w')\in\mathcal{W}\times\mathcal{W}' \ | \ f(w)>0 \text{ and } a \geq \dfrac{dP}{dQ}(w,w') \Big/ f(w)\Big\}
\nonumber\\*
=
\Big\{(w,w')\in\mathcal{W}\times\mathcal{W}' \ | \ f(w)>0 \text{ and } 0 \geq \dfrac{dP}{dQ}(w,w') - af(w)\Big\}
\nonumber
\end{IEEEeqnarray}
Therefore, $\{g\in (-\infty,a]\}$ is measurable because $dP/dQ$ and $f$ are. Hence, $g$ is measurable.
Fix $w\in\mathcal{W}.$ If $f(w)=0,$ then $\int g(w,\cdot)dQ_{W'|W{=w}}=1.$ This is also the case if $f(w)>0$ by definition of $f$ and $g.$ Hence, $g$ is $Q$-integrable, and for any $w\in\mathcal{W},$ the measure with density $g(w,\cdot)$ with respect to $Q_{W'|W{=}w}$ is a probability measure. Fix $B\in\mathcal{B}(\mathcal{W}').$ Since $(w,w') \mapsto g(w,w') \cdot \mathbf{1}_{B}(w')$ is measurable, then from Theorem \ref{thm:generalized_fubini}, the map
\begin{equation*}
    w \mapsto \int g(w,w')\cdot \mathbf{1}_B(w') \
    dQ_{W'|W{=}w}(w')
\end{equation*}
is measurable. Hence, the map
\begin{equation*}
    (w,B) \mapsto \int g(w,w')\cdot \mathbf{1}_B(w') \
    dQ_{W'|W{=}w}(w')
\end{equation*}
defines a regular transition kernel.

\subsubsection{Proof of Lemma \ref{lemma:get_expectation_out_of_TV}}
\hfill\\
\indent Define $\mu :=\Pi_W( \Pi_{L|W}+ \Gamma_{L|W})/2,$ which is
a probability measure dominating both $\Pi_W \Pi_{L|W}$ and $\Pi_W \Gamma_{L|W}.$
Then, $(\Pi_{L|W}+\Gamma_{L|W})/2$ is a regular transition kernel of $\mu$ from $\mathcal{W}$ to $\mathcal{L},$ which we denote by $\mu_{L|W}.$
Note that for any $w\in\mathcal{W},$ the distribution
$\mu_{L|W{=}w}$
dominates $\Pi_{L|W{=}w}$ (resp. $\Gamma_{L|W{=}w}$).
Moreover, we have $\mu_W \equiv \Pi_W.$
Then, it can be checked through Theorem \ref{thm:generalized_fubini} that
\begin{IEEEeqnarray}{c}
(w,l) \mapsto
\dfrac{d \Pi_{L|W{=}w}}{d \mu_{L|W{=}w}}(l)
\quad
\bigg(
\text{resp. }
\dfrac{d \Gamma_{L|W{=}w}}{d \mu_{L|W{=}w}}(l)
\bigg)
\end{IEEEeqnarray}
is a density of $\Pi_W \Pi_{L|W}$ (resp. $\Pi_W \Gamma_{L|W}$) with respect to $\mu.$
Moreover, from
Theorem \ref{thm:generalized_fubini} and the definition of the TVD (Eq. \ref{eq:def_TVD})
we obtain
\begin{IEEEeqnarray}{rCl}
\|\Pi_W \Pi_{L|W} - \Pi_W \Gamma_{L|W}\|_{TV}
&=&
\int
\bigg|
\dfrac{d \Pi_{L|W{=}w}}{d \mu_{L|W{=}w}}(l)
-
\dfrac{d \Gamma_{L|W{=}w}}{d \mu_{L|W{=}w}}(l)
\bigg|
\ d\mu_{W,L}(w,l)
\nonumber\\*
&=&
\int \bigg(\int \bigg|
\dfrac{d \Pi_{L|W{=}w}}{d \mu_{L|W{=}w}}(l)
-
\dfrac{d \Gamma_{L|W{=}w}}{d \mu_{L|W{=}w}}(l)
\bigg| \ d\mu_{L|W{=}w}(l) \bigg) d\Pi_W(w)
\nonumber\\*
&=&
\mathbb{E}_{\Pi_W}[\|
\Pi_{L|W} - \Gamma_{L|W}
\|_{TV}],
\label{eq:using_domination_and_def_TVD}
\end{IEEEeqnarray}
where
\eqref{eq:using_domination_and_def_TVD} follows from the definition of the TVD (Eq. \ref{eq:def_TVD}).

\subsubsection{Proof of Lemma \ref{lemma:TV_joint_to_TV_marginal}}
\hfill\\
\indent Taking the supremum of the same quantity over a smaller set reduces the supremum. Therefore,
\begin{IEEEeqnarray}{rCl}
\|\Pi_{W,L}-\Gamma_{W,L}\|_{TV} =
\sup_{A \subset \mathcal{W}\times\mathcal{L}}\{\Pi_{W,L}(A)-\Gamma_{W,L}(A)\}
&\geq&
\sup_{B \subset \mathcal{W}}\{\Pi_{W,L}(B\times\mathcal{L})-\Gamma_{W,L}(B\times\mathcal{L})\}
\nonumber\\*
&=&
\sup_{B \subset \mathcal{W}}\{\Pi_{W}(B)-\Gamma_{W}(B)\}
\nonumber\\*
&=&
\|\Pi_{W}-\Gamma_{W}\|_{TV}.
\nonumber
\end{IEEEeqnarray}

\subsubsection{Proof of Lemma \ref{lemma:TV_same_channel}}
\hfill\\
\indent Define the probability measure $\mu$ with marginal $\mu_W:=(\Pi_W + \Gamma_W)/2$ and regular transition kernel $\Pi_{L|W}.$ Then, $\Pi_W \Pi_{L|W}$ and $\Gamma_W\Pi_{L|W}$ are dominated by $\mu,$ and $\mu_W$ dominates $\Pi_W$ (resp. $\Gamma_W$).
Moreover, $\mu,\Pi,\Gamma$ share the same transition kernel from $\mathcal{W}$ to $\mathcal{L}.$
Then, it can be checked through Theorem \ref{thm:generalized_fubini} that
\begin{IEEEeqnarray}{c}
(w,l) \mapsto
\dfrac{d \Pi_W}{d \mu_W}(w)
\quad
\bigg(
\text{resp. }
\dfrac{d \Gamma_W}{d \mu_W}(w)
\bigg)
\end{IEEEeqnarray}is a density of $\Pi_W \Pi_{L|W}$ (resp. $\Gamma_W \Pi_{L|W}$) with respect to $\mu.$
Then,
from
Theorem \ref{thm:generalized_fubini} and the definition of the TVD (Eq. \ref{eq:def_TVD})
we obtain
\begin{IEEEeqnarray}{rCl}
\|\Pi_W \Pi_{L|W} - \Gamma_W \Pi_{L|W}\|_{TV} &=&
\int
\bigg|
\dfrac{d \Pi_W}{d \mu_W}(w)
-
\dfrac{d \Gamma_W}{d \mu_W}(w)
\bigg|
\ d\mu_{W,L}(w,l)
\nonumber\\*
&=&
\int
\bigg|
\dfrac{d \Pi_W}{d \mu_W}(w)
-
\dfrac{d \Gamma_W}{d \mu_W}(w)
\bigg|
\ d\mu_{W}(w)
\nonumber\\*
&=&\|
\Pi_{W} - \Gamma_{W}
\|_{TV},
\label{eq:in_proof_same_channel_using_domination_and_def_TVD}
\end{IEEEeqnarray}
where
\eqref{eq:in_proof_same_channel_using_domination_and_def_TVD} follows from the definition of the TVD (Eq. \ref{eq:def_TVD}).

\section{Equivalence between achievability with near-perfect and perfect realism}\label{app:equivalence_perfect_realism}

\subsection{Marginal realism}\label{app:subsec:equivalence_perfect_marginal_realism}
In order to prove the part of Theorem \ref{theorem:equivalence_perfect_realism} pertaining to marginal realism, we prove the following result,
which we also use in Sections \ref{sec:achievability_two_sided} and \ref{sec:achievability}.
\begin{proposition}\label{proposition:to_perfect_realism}
Let $n$ be a positive integer and $\varepsilon_1, \varepsilon_2, \varepsilon_3, \varepsilon_4$ be positive reals.
Let $\mathcal{X}$
and $\mathcal{W}$ be two 
Polish alphabets and $p_{X}$ be a distribution on $\mathcal{X}.$  
Let $d$ be a distortion measure on $\mathcal{X}^2.$
Let $Q_{X^n, Y^n}$ be a distribution on $\mathcal{X}^n \times \mathcal{X}^n$ satisfying
\begin{IEEEeqnarray}{rCl}
\mathbb{E}_Q[d(X^n, Y^n)] \leq \Delta + \varepsilon_1
\text{ and } \| Q_{Y^n} - p_X^{\otimes n} \|_{TV} \leq \varepsilon_2,
\nonumber
\end{IEEEeqnarray}for some real $\Delta.$ 
Let $P_{X^n,W,Y^n}$ be a distribution on $\mathcal{X}^n\times\mathcal{W}\times\mathcal{X}^n$
satisfying $\hat{P}_{X^n}\equiv p_X$
and the Markov chain property $X^n-W-Y^n.$
Moreover,
assume that
\begin{IEEEeqnarray}{rCl}
\| P_{Y^n} - Q_{Y^n} \|_{TV}
\leq
\varepsilon_3
\text{ and }
\| \hat{P}_{X^n,Y^n} - \hat{Q}_{X^n,Y^n} \|_{TV} &\leq& \varepsilon_4.
\nonumber
\end{IEEEeqnarray}
Then, there
exists a conditional distribution $P'_{Y^n|W}$
such that the distribution $P'$ defined by 
\begin{IEEEeqnarray}{c}
P'_{X^n,W,Y^n} := P_{X^n,W} \cdot P'_{Y^n|W}
\nonumber
\end{IEEEeqnarray}
satisfies
\begin{IEEEeqnarray}{rCl}
\mathbb{E}_{P'}[d(X^n, Y^n)] \leq \Delta + \varepsilon_1 + \sup_{\mathbb{P}_{X,Y,\xi}} \mathbb{E}[d(X,Y) \cdot \xi]
\ \text{ and } \
P'_{Y^n} \equiv p_X^{\otimes n},
\nonumber
\end{IEEEeqnarray}
where the supremum is taken over all distributions $\mathbb{P}_{X,Y,\xi}$ on $\mathcal{X}^2 \times \{0,1\}$ satisfying $\mathbb{P}_X \equiv \mathbb{P}_Y \equiv p_X$ and $\mathbb{P}(\{\xi=1\}) \leq \varepsilon_2 + \varepsilon_3 + \varepsilon_4.$
\end{proposition}

Proposition \ref{proposition:to_perfect_realism} implies the part of Theorem \ref{theorem:equivalence_perfect_realism} pertaining to marginal realism, by taking $Q\equiv P$ at fixed blocklength, and $\mathcal{W}=[2^{nR}]\times[2^{nR_c}]\times\mathcal{Z}^n,$ $W=(M,J,Z^n).$
Proposition \ref{proposition:to_perfect_realism} is proved in the next subsection.

\subsection{Joint realism}\label{app:subsec:equivalence_perfect_joint_realism}

In order to prove the part of Theorem \ref{theorem:equivalence_perfect_realism} pertaining to joint realism, we prove the following result,
which we also use in Sections \ref{sec:achievability_joint_two_sided} and \ref{sec:achievability_joint_one_sided}.
\begin{proposition}\label{proposition:to_perfect_joint_realism}
Let $n$ be a positive integer and $\varepsilon_1, \varepsilon_2, \varepsilon_3, \varepsilon_4$ be positive reals. Let $\mathcal{X},\mathcal{Z},\mathcal{W}$ be
Polish alphabets and $p_{X,Z}$ be a joint distribution on $\mathcal{X} \times \mathcal{Z}.$
Let $d$ be a distortion measure on $\mathcal{X}^2.$
Let $Q_{X^n, Y^n,Z^n}$ be a distribution on $\mathcal{X}^n \times \mathcal{X}^n \times \mathcal{Z}^n$ satisfying
\begin{IEEEeqnarray}{rCl}
\IEEEeqnarraymulticol{3}{l}{
\mathbb{E}_Q[d(X^n, Y^n)] \leq \Delta + \varepsilon_1 
} \label{eq:distortion_assumption_theorem_to_perfect_joint_realism}\\*
&\text{ and } \| Q_{Y^n,Z^n} - p_{X,Z}^{\otimes n} \|_{TV} \leq \varepsilon_2,& \label{eq:perception_assumption_theorem_to_perfect_joint_realism}
\end{IEEEeqnarray}for some real $\Delta.$
Let $P_{X^n,Z^n,W,Y^n}$ be a distribution on $\mathcal{X}^n\times\mathcal{Z}^n\times\mathcal{W}\times\mathcal{X}^n$
satisfying $\hat{P}_{X^n}\equiv p_X,$
$P_{Z^n}\equiv p_{Z}^{\otimes n},$ and the Markov chain property $X^n-(Z^n,W)-Y^n.$
Moreover,
assume that
\begin{IEEEeqnarray}{rCl}
\| P_{Y^n,Z^n} - Q_{Y^n,Z^n} \|_{TV} &\leq& \varepsilon_3\label{eq:total_variation_Q_P_on_Y_assumption_theorem_to_perfect_joint_realism}\\
\| \hat{P}_{X^n,Y^n} - \hat{Q}_{X^n,Y^n} \|_{TV} &\leq& \varepsilon_4.\label{eq:total_variation_empirical_Q_P_assumption_theorem_to_perfect_joint_realism}
\end{IEEEeqnarray}
Then, there
exists a conditional distribution $P'_{Y^n|Z^n,W}$
such that the distribution $P'$ defined by 
\begin{IEEEeqnarray}{c}
P'_{X^n,Z^n,W,Y^n} := P_{X^n,Z^n,W} \cdot P'_{Y^n|Z^n,W}
\end{IEEEeqnarray}
satisfies
\begin{IEEEeqnarray}{rCl}
\mathbb{E}_{P'}[d(X^n, Y^n)] \leq \Delta + \varepsilon_1 + \sup_{\mathbb{P}_{X,Y,\xi}} \mathbb{E}[d(X,Y) \cdot \xi]
\ \text{ and } \
P'_{Y^n,Z^n} \equiv p_{X,Z}^{\otimes n},
\nonumber
\end{IEEEeqnarray}
where the supremum is taken over all distributions $\mathbb{P}_{X,Y,\xi}$ on $\mathcal{X}^2 \times \{0,1\}$ satisfying $\mathbb{P}_X \equiv \mathbb{P}_Y \equiv p_X$ and $\mathbb{P}(\{\xi=1\}) \leq \varepsilon_2 + \varepsilon_3 + \varepsilon_4.$
\end{proposition}

Proposition \ref{proposition:to_perfect_joint_realism} implies
the part of Theorem \ref{theorem:equivalence_perfect_realism} pertaining to joint realism,
by taking $Q\equiv P$ at fixed blocklength, and $\mathcal{W}=[2^{nR}]\times[2^{nR_c}],$ $W=(M,J).$ Moreover,
Proposition \ref{proposition:to_perfect_realism} is a direct consequence of Proposition \ref{proposition:to_perfect_joint_realism} applied with
a singleton $\mathcal{Z}.$
The remainder of Appendix \ref{app:equivalence_perfect_realism} is dedicated to the proof of Proposition \ref{proposition:to_perfect_joint_realism}.

\subsection{Modified distribution}

Consider the setting of Proposition \ref{proposition:to_perfect_joint_realism}. If $P_{Y^n,Z^n} \equiv p_{X,Z}^{\otimes n},$ we just set $P':= P.$ Assume this is not the case.
From
the Markov property $X^n-(W,Z^n)-Y^n$
the conditional distribution of $Y^n$ knowing all other variables reduces to $P_{Y^n|Z^n,W}.$
In order to ensure measurability properties regarding conditional densities,
we consider a
transition kernel $
P_{Y^n|Z^n
,W
}
$
constructed as follows.
Define the $\sigma$-finite measure $\mu_{Y^n,Z^n
,W
} = P_{Y^n,Z^n
,W
}+p_{X,Z}^{\otimes n}
\cdot P_W
.$ Then, $\mu_{Y^n,Z^n
,W
}$ dominates $P_{Y^n,Z^n
,W
}
,
$ and
$\mu_{Y^n,Z^n}$ dominates
$p_{Y,Z}^{\otimes n}.$
Fix
densities
$d P_{Y^n,Z^n
,W
}/d\mu_{Y^n,Z^n
,W
}$
and $d p_{Y,Z}^{\otimes n}/d\mu_{Y^n,Z^n}$ and
transition kernels
$\mu_{W|Z^n}$ and
$\mu_{Y^n|Z^n
,W
}.$
This determines a unique
transition kernel
$
\mu_{Y^n
,W
|Z^n
}
,$
and thus a unique $\mu_{Y^n|Z^n}.$
From Corollary \ref{cor:conditional_density_exists_and_gives_regular_kernel} (Appendix \ref{app:subsec:measure_theory_general}), we can construct
a density function
$dP_{Z^n,W}/d\mu_{Z^n,W}$
and
a
regular transition
kernel
$
P_{Y^n|Z^n
,W
}
$
of $P_{Y^n,Z^n
,W
}$
such that
the following holds:
for all $(z^n,w)\in\mathcal{Z}^n\times\mathcal{W},$ the measure $\mu_{Y^n|Z^n{=}z^n,W{=}w}$ dominates $P_{Y^n|Z^n{=}z^n,W{=}w},$ and there exists a choice of conditional density functions such that the map
\begin{IEEEeqnarray}{c}
(y^n,z^n,w) \mapsto \dfrac{dP_{Y^n|Z^n{=}z^n,W{=}w}}{d\mu_{Y^n|Z^n{=}z^n,W{=}w}} (y^n)
\label{eq:measurability_density_P_Yn_knowing_Zn_W}
\end{IEEEeqnarray}
is measurable.
Similarly, one can construct
a
transition
kernel
$
P_{
W
|Z^n
}
$
such that
the following holds:
for all $z^n\in\mathcal{Z}^n,$ the measure $\mu_{W|Z^n{=}z^n}$ dominates $P_{W|Z^n{=}z^n},$
and there exists a choice of conditional density functions such that the map
\begin{IEEEeqnarray}{c}
(z^n,w) \mapsto \dfrac{dP_{W|Z^n{=}z^n}}{d\mu_{W|Z^n{=}z^n}} (w)
\label{eq:measurability_density_P_W_knowing_Zn}
\end{IEEEeqnarray}
is measurable.
The transition kernels $P_{Y^n|Z^n,W}$ and $P_{W|Z^n}$ determine a unique transition kernel $P_{Y^n,W|Z^n},$ and thus a unique $P_{Y^n|Z^n}.$
Then, from
the measurability of the maps in \eqref{eq:measurability_density_P_Yn_knowing_Zn_W} and \eqref{eq:measurability_density_P_W_knowing_Zn}, and  from the generalized Fubini theorem (Theorem \ref{thm:generalized_fubini} in Appendix \ref{app:subsec:measure_theory_general}), one can readily check that for any $z^n\in\mathcal{Z}^n,$
the measure $\mu_{Y^n|Z^n{=}z^n}$ dominates $P_{Y^n|Z^n{=}z^n},$ with the following map
\begin{IEEEeqnarray}{c}
y^n \mapsto \int_{\mathcal{W}}
\dfrac{dP_{W|Z^n{=}z^n}}{d\mu_{W|Z^n{=}z^n}}(w)
\dfrac{dP_{Y^n|Z^n{=}z^n,W{=}w}}{d\mu_{Y^n|Z^n{=}z^n,W{=}w}}(y^n)
d\mu_{W|Y^n{=}y^n,Z^n{=}z^n}
\nonumber
\end{IEEEeqnarray}
as density, denoted $dP_{Y^n|Z^n{=}z^n}/d\mu_{Y^n|Z^n{=}z^n}.$ Moreover, from
the measurability of the maps in \eqref{eq:measurability_density_P_Yn_knowing_Zn_W} and \eqref{eq:measurability_density_P_W_knowing_Zn}, and from
Theorem \ref{thm:generalized_fubini}, the map
\begin{IEEEeqnarray}{c}
(y^n,z^n) \mapsto \dfrac{dP_{Y^n|Z^n{=}z^n}}{d\mu_{Y^n|Z^n{=}z^n}}(y^n)
\label{eq:measurability_density_P_Yn_knowing_Z^n}
\end{IEEEeqnarray}
is measurable.
Similarly, one can construct
a transition kernel $p_{Y^n|Z^n}$
such that the following holds:
for all $z^n\in\mathcal{Z}^n,$ the measure $\mu_{Y^n|Z^n{=}z^n}$ dominates $p_{Y^n|Z^n{=}z^n},$
and there exists a choice of conditional density functions such that the map
\begin{IEEEeqnarray}{c}
(y^n,z^n) \mapsto \dfrac{dp_{Y^n|Z^n{=}z^n}}{d\mu_{Y^n|Z^n{=}z^n}} (y^n)
\label{eq:measurability_density_p_Yn_knowing_Z^n}
\end{IEEEeqnarray}
is measurable.
For any $z^n\in \mathcal{Z}^n,$
let
\begin{IEEEeqnarray}{c}
\mathcal{X}^+_{z^n} = \bigg\{ y^n \in \mathcal{X}^n \text{ s.t. } \dfrac{dP_{Y^n|Z^n{=}z^n}}{d\mu_{Y^n|Z^n{=}z^n}}(y^n) > \dfrac{dp_{Y^n|Z^n{=}z^n}}{d\mu_{Y^n|Z^n{=}z^n}}(y^n) \bigg\} \nonumber\\*
\text{and }\theta_{z^n}: \mathcal{X}^+_{z^n} \to [0,1), \ y^n \mapsto \dfrac{dp_{Y^n|Z^n{=}z^n}/d\mu_{Y^n|Z^n{=}z^n}(y^n)}{dP_{Y^n|Z^n{=}z^n}/d\mu_{Y^n|Z^n{=}z^n}(y^n)}.\nonumber
\end{IEEEeqnarray}
Then,
\begin{IEEEeqnarray}{c}
(y^n,z^n)\mapsto \mathbf{1}_{\mathcal{X}^+_{z^n}}(y^n) \text{ is measurable, and }
(y^n,z^n)\mapsto \mathbf{1}_{\mathcal{X}^+_{z^n}}(y^n)\theta_{z^n}(y^n) \text{ is measurable.}
\label{eq:measurability_X_+_and_theta}
\end{IEEEeqnarray}
In the remainder of Appendix \ref{app:equivalence_perfect_realism},
we implicitly use the following lemma, which follows from the measurability of singletons in Polish spaces (which are metrizable, therefore T4, thus T1).
\begin{lemma}\label{lemma:projection_is_measurable}
Let $\mathcal{W}_1$ and $\mathcal{W}_2$ be Polish spaces. Let $\phi:\mathcal{W}_1\times\mathcal{W}_2\to\mathbb{R}$ be a measurable map. Then, for any $w_1\in\mathcal{W}_1,$ the map $w_2\mapsto\phi(w_1,w_2)$ is measurable.
\end{lemma}

Let $B_1$ denote the set of all $z^n\in\mathcal{Z}^n$ such that
$P_{Y^n|Z^n{=}z^n} \neq p_{Y^n|Z^n{=}z^n}.$
From the definition of the total variation distance (Eq. \ref{eq:def_TVD}), we have, for any $z^n\in\mathcal{Z}^n,$
\begin{IEEEeqnarray}{c}
\|P_{Y^n|Z^n{=}z^n} - p_{Y^n|Z^n{=}z^n}\|_{TV} = \int_{\mathcal{X}^n \backslash \mathcal{X}^+_{z^n} } \bigg(
\dfrac{dp_{Y^n|Z^n{=}z^n}}{d\mu_{Y^n|Z^n{=}z^n}}(y^n)
-
\dfrac{dP_{Y^n|Z^n{=}z^n}}{d\mu_{Y^n|Z^n{=}z^n}}(y^n)
\bigg)
d\mu_{Y^n|Z^n{=}z^n}.
\label{eq:denominator_of_Gamma_is_TVD}
\IEEEeqnarraynumspace
\end{IEEEeqnarray}
Hence, from the measurability of the maps in \eqref{eq:measurability_density_P_Yn_knowing_Z^n} and \eqref{eq:measurability_density_p_Yn_knowing_Z^n}, and from Theorem \ref{thm:generalized_fubini},
the map
$z^n \mapsto \|P_{Y^n|Z^n{=}z^n} - p_{Y^n|Z^n{=}z^n}\|_{TV}$ is measurable.
Therefore, the set $B_1$ is measurable.
For any $w,$ and any $z^n\notin B_1,$
let
$P'_{Y^n|Z^n{=}z^n,W{=}w}(A)=P_{Y^n|Z^n{=}z^n,W{=}w}(A)$ for any Borel set $A \subseteq \mathcal{X}^n.$ Otherwise, for any Borel set $A$ we define
$P'_{Y^n|Z^n{=}z^n,W{=}w}(A)$ as
\begin{IEEEeqnarray}{rCl}
\IEEEeqnarraymulticol{3}{l}{
P_{Y^n|Z^n{=}z^n,W{=}w}(A \backslash \mathcal{X}^+_{z^n}) \ +
\int_{A \cap \mathcal{X}^+_{z^n} } \theta_{z^n}(y^n) dP_{Y^n|Z^n{=}z^n,W{=}w}(y^n) + \phi_{z^n, w} \Gamma_{z^n}(A), \text{ where}
}
\nonumber\\*
&& \Gamma_{z^n}(A) = \dfrac{\int_{A \backslash \mathcal{X}^+_{z^n} } (dp_{Y^n|Z^n{=}z^n}/d\mu_{Y^n|Z^n{=}z^n}(y^n) - dP_{Y^n|Z^n{=}z^n}/d\mu_{Y^n|Z^n{=}z^n}(y^n))
d\mu_{Y^n|Z^n{=}z^n}}
{\int_{\mathcal{X}^n \backslash \mathcal{X}^+_{z^n} } (dp_{Y^n|Z^n{=}z^n}/d\mu_{Y^n|Z^n{=}z^n}(y^n) - dP_{Y^n|Z^n{=}z^n}/d\mu_{Y^n|Z^n{=}z^n}(y^n))
d\mu_{Y^n|Z^n{=}z^n}}\nonumber\\*
&& \text{and } \phi_{z^n, w} = \int_{\mathcal{X}^+_{z^n}} 1 - \theta_{z^n}(y^n) dP_{Y^n|Z^n{=}z^n,W{=}w}(y^n). \label{eq:def_phi_in_proof_equivalence_perfect_joint_realism}
\end{IEEEeqnarray}
\noindent This is well-defined by the measurability of the maps in \eqref{eq:measurability_density_P_Yn_knowing_Z^n} and \eqref{eq:measurability_density_p_Yn_knowing_Z^n}, and by \eqref{eq:measurability_X_+_and_theta} (and Lemma \ref{lemma:projection_is_measurable}).
From \eqref{eq:denominator_of_Gamma_is_TVD}, the denominator in the definition of $\Gamma_{z^n}(A)$ is non-zero for $z^n\in B_1.$
We check that this defines a probability measure for every $(z^n, w)$ such that $z\in B_1.$ Fix such a pair.
By definition of $\mathcal{X}^+_{z^n}$ and $\theta_{z^n},$ we have $P'_{Y^n|Z^n{=}z^n,W{=}w}(A)\geq 0$ for any Borel set $A.$
In addition, $P'_{Y^n|Z^n{=}z^n,W{=}w}(\mathcal{X}^n)$ equals
\begin{IEEEeqnarray}{rCl}
\IEEEeqnarraymulticol{3}{l}{
P_{Y^n|Z^n{=}z^n,W{=}w}(\mathcal{X}^n\backslash \mathcal{X}^+_{z^n}) \ +
\int_{
\mathcal{X}^+_{z^n} } \theta_{z^n}
dP_{Y^n|Z^n{=}z^n,W{=}w}
+
\int_{
\mathcal{X}^+_{z^n} } 1-\theta_{z^n}
dP_{Y^n|Z^n{=}z^n,W{=}w}
}
\nonumber\\*
&=&  P_{Y^n|Z^n{=}z^n,W{=}w}(\mathcal{X}^n\backslash \mathcal{X}^+_{z^n}) \ +
\int_{
\mathcal{X}^+_{z^n} } 1
dP_{Y^n|Z^n{=}z^n,W{=}w}
\nonumber\\*
&=&1.\nonumber
\end{IEEEeqnarray}
Thus, $P'_{Y^n|Z^n{=}z^n,W{=}w}$ is a probability measure.
Moreover, for any Borel set $A,$ the map $(z^n, w) \mapsto P'_{Y^n|Z^n{=}z^n,W{=}w}(A)$ is measurable because
\begin{itemize}
    \item $B_1$ is measurable
    \item $P_{Y^n|Z^n,W}$ is a regular transition kernel
    \item All other integrals are measurbale functions of $(z^n,w)$ from Theorem \ref{thm:generalized_fubini},
    \eqref{eq:measurability_X_+_and_theta}, and the measurability of the maps in \eqref{eq:measurability_density_P_Yn_knowing_Z^n} and \eqref{eq:measurability_density_p_Yn_knowing_Z^n}.
\end{itemize}
Hence, $P'_{Y^n|Z^n,W}$ is a regular transition kernel.
We define the joint distribution for the new code as
\begin{equation}\label{eq:def_P_2_in_proof_equivalence_perfect_joint_realism}
    P'_{\scalebox{0.7}{$X^n, Z^n, W, Y^n$}} = P_{\scalebox{0.7}{$X^n, Z^n, W$}} \cdot P'_{\scalebox{0.7}{$Y^n| Z^n, W$}}.
\end{equation}

\subsection{Proving perfect joint realism}
We show that $P'$ satisfies $P'_{Y^n,Z^n}\equiv p_{X,Z}^{\otimes n}.$
By construction, this is true if $P_{Y^n,Z^n}\equiv p_{X,Z}^{\otimes n}.$
Assume that this is not the case.
Fix measurable measurable sets $A\subseteq \mathcal{X}^n$ and $B\subseteq \mathcal{Z}^n.$
From the
definition \eqref{eq:def_phi_in_proof_equivalence_perfect_joint_realism} of $P'_{Y^n|Z^n{=}z^n,W{=}w},$
we have,
for any $(w,z^n)\in\mathcal{W}\times\mathcal{Z}^n,$
\begin{IEEEeqnarray}{rCl}
\IEEEeqnarraymulticol{3}{l}{
\mathbf{1}_{
B
\cap B_1
}(
z^n)
P'_{Y^n|Z^n{=}z^n,W{=}w}(
A
)
}\nonumber\\*
&=& \mathbf{1}_{B
\cap B_1
}(z^n) \cdot \Big[ P_{Y^n|Z^n{=}z^n,W{=}w}(A \backslash \mathcal{X}^+_{z^n}) \ + \int_{A \cap \mathcal{X}^+_{z^n} } \theta_{z^n}(y^n) dP_{Y^n|Z^n{=}z^n,W{=}w}(y^n) + \phi_{z^n, w} \Gamma_{z^n}(A) \Big]\nonumber\\*
&=& \mathbf{1}_{B
\cap B_1
}(z^n) \cdot \int \Big[
\mathbf{1}_{
A \backslash \mathcal{X}^+_{z^n}
}
(y^n)
+ \mathbf{1}_{
A \cap \mathcal{X}^+_{z^n}
}(y^n) \ \theta_{z^n}(y^n) + \Gamma_{z^n}(A) \mathbf{1}_{ \mathcal{X}^+_{z^n}
} (1 - \theta_{z^n}(y^n)) \Big] dP_{Y^n|Z^n{=}z^n,W{=}w}
.\nonumber
\end{IEEEeqnarray}
From \eqref{eq:def_P_2_in_proof_equivalence_perfect_joint_realism},
similarly to $P$
we have $P'_{Z^n}\equiv p_Z^{\otimes n}.$ Then, by the generalized Fubini theorem (Theorem \ref{thm:generalized_fubini}), integrating both sides according to $P_{Z^n,W}$ gives
\begin{IEEEeqnarray}{rCl}
\IEEEeqnarraymulticol{3}{l}{
P'_{Y^n,Z^n}(A \times (B\cap B_1)) = P'_{Y^n,Z^n,W}(A \times (B\cap B_1))
}\nonumber\\*
&=& \int \Big[ \mathbf{1}_{(A \backslash \mathcal{X}^+_{z^n}) \times (B\cap B_1)} + \mathbf{1}_{(A \cap \mathcal{X}^+_{z^n}) \times (B\cap B_1)} \ \theta_{z^n}(y^n) + \Gamma_{z^n}(A) \mathbf{1}_{ \mathcal{X}^+_{z^n} \times (B\cap B_1)} (1 - \theta_{z^n}(y^n)) \Big] dP_{Y^n,Z^n,W}\nonumber\\
&=& \int \Big[ \mathbf{1}_{(A \backslash \mathcal{X}^+_{z^n}) \times (B\cap B_1)} + \mathbf{1}_{(A \cap \mathcal{X}^+_{z^n}) \times (B\cap B_1)} \ \theta_{z^n}(y^n) + \Gamma_{z^n}(A) \mathbf{1}_{ \mathcal{X}^+_{z^n} \times (B\cap B_1)} (1 - \theta_{z^n}(y^n)) \Big] dP_{Y^n,Z^n}\nonumber\\
&=& \int \mathbf{1}_{B\cap B_1}(z^n) \int \Big[ \mathbf{1}_{A \backslash \mathcal{X}^+_{z^n}} + \mathbf{1}_{A \cap \mathcal{X}^+_{z^n}} \ \theta_{z^n}(y^n) \Big] dP_{Y^n|Z^n{=}z^n} dP_{Z^n}(z^n)\nonumber\\* 
&&+ \int \mathbf{1}_{B\cap B_1}(z^n) \Gamma_{z^n}(A)  \int \mathbf{1}_{ \mathcal{X}^+_{z^n}} (1 - \theta_{z^n}(y^n)) dP_{Y^n|Z^n{=}z^n} dP_{Z^n}(z^n)\nonumber\\
&=& \int \mathbf{1}_{B\cap B_1}(z^n) \Big[ \int \mathbf{1}_{A \backslash \mathcal{X}^+_{z^n}} dP_{Y^n|Z^n{=}z^n} + \int \mathbf{1}_{A \cap \mathcal{X}^+_{z^n}} dp_{Y^n|Z^n{=}z^n} \Big] dP_{Z^n}(z^n)\nonumber\\*
&&+ \int \mathbf{1}_{B\cap B_1}(z^n) \Gamma_{z^n}(A)  \int_{\mathcal{X}^+_{z^n}}
\Big[dp_{Y^n|Z^n{=}z^n}/d\mu_{Y^n|Z^n{=}z^n}(y^n)
\nonumber\\*
&&
\qquad\qquad\qquad\qquad\qquad\qquad\quad
- dP_{Y^n|Z^n{=}z^n}/d\mu_{Y^n|Z^n{=}z^n}(y^n)\Big]d\mu_{Y^n|Z^n{=}z^n} dP_{Z^n}(z^n)\nonumber\\
&=& \int \mathbf{1}_{B\cap B_1}(z^n) \Big[ \int \mathbf{1}_{A \backslash \mathcal{X}^+_{z^n}} dP_{Y^n|Z^n{=}z^n} + \int \mathbf{1}_{A \cap \mathcal{X}^+_{z^n}} dp_{Y^n|Z^n{=}z^n}\nonumber\\*
&&+ \int \mathbf{1}_{A \backslash \mathcal{X}^+_{z^n}} (dp_{Y^n|Z^n{=}z^n} - dP_{Y^n|Z^n{=}z^n}) \Big] dP_{Z^n}(z^n)\nonumber\\
&=& p_{X,Z}^{\otimes n}(A\times (B\cap B_1)),\nonumber
\end{IEEEeqnarray}where throughout we omit null terms corresponding to values of $z^n$
which are not in $B_1.$
Moreover,
\begin{IEEEeqnarray}{rCl}
P'_{Y^n,Z^n}(A \times (B\backslash B_1)) &=& P'_{Y^n,Z^n,W}(A \times (B\backslash B_1))
\nonumber\\
&=& \int \mathbf{1}_{B\backslash B_1}(z^n) \int \mathbf{1}_{A}(y^n) dP'_{Y^n|Z^n{=}z^n,W{=}w}(y^n) dP'_{Z^n,W}(z^n,w)\nonumber\\
&=& \int \mathbf{1}_{B\backslash B_1}(z^n) \int \mathbf{1}_{A}(y^n) dP_{Y^n|Z^n{=}z^n,W{=}w}(y^n) dP_{Z^n,W}(z^n,w)\nonumber\\
&=& \int \mathbf{1}_{B\backslash B_1}(z^n) \int \mathbf{1}_{A}(y^n) dP_{Y^n|Z^n{=}z^n}(y^n) dP_{Z^n}(z^n)\nonumber
\\
&=& \int \mathbf{1}_{B\backslash B_1}(z^n) \int \mathbf{1}_{A}(y^n) dp_{Y^n|Z^n{=}z^n}(y^n) dp_Z^{\otimes n}(z^n)\nonumber\\
&=& p^{\otimes n}_{X,Z}(A \times (B\backslash B_1)).\nonumber
\end{IEEEeqnarray}
Hence, $P'_{Y^n,Z^n}(A \times B) = p^{\otimes n}_{X,Z}(A \times B).$
This being true for any measurable (Borel) sets $A\subseteq \mathcal{X}^n$ and $B\subseteq \mathcal{Z}^n,$ then, from the uniqueness property in Theorem \ref{thm:generalized_fubini}, $P'_{Y^n,Z^n}$ and $p_{X,Z}^{\otimes n}$ are equal on the product $\sigma$-algebra of the (Borel) $\sigma$-algebras of $\mathcal{X}^n$ and $\mathcal{Z}^n.$
As mentioned in Appendix \ref{app:subsec:measure_theory_general}, the latter is equal to the Borel $\sigma$-algebra of the product Polish space $\mathcal{X}^n\times\mathcal{Z}^n.$
Hence, $P'_{Y^n,Z^n}\equiv p_{X,Z}^{\otimes n}.$
The last step consists of obtaining the distortion bound by coupling $P'$ with $Q.$

\subsection{Coupling with $Q$}

We first compare $P'$ to $P.$ For every $z^n, w$ and every set $A \in \mathcal{X}^n$ we have:
\begin{IEEEeqnarray}{rCl}
P_{Y^n|Z^n\text{=}z^n, W\text{=}w}(A) - P'_{Y^n|Z^n\text{=}z^n, W\text{=}w}(A)
&\leq& \int_{A \cap \mathcal{X}^+_{z^n}} 1 - \theta_{z^n}(y^n) dP_{Y^n|Z^n\text{=}z^n, W\text{=}w}(y^n) - \phi_{z^n, w} \Gamma_{z^n}(A). \nonumber
\end{IEEEeqnarray}
Both terms in the right hand side being non-negative and smaller than or equal to $\phi_{z^n, w},$ we have
\begin{IEEEeqnarray}{c}
    \big\|P'_{Y^n|Z^n\text{=}z^n, W\text{=}w} - P_{Y^n|Z^n\text{=}z^n, W\text{=}w}\big\|_{TV} \leq \phi_{z^n, w}.\nonumber
\end{IEEEeqnarray}
When integrating this over all variables except $Y^n,$ we obtain by definition \eqref{eq:def_P_2_in_proof_equivalence_perfect_joint_realism} of $P'$ and Lemma \ref{lemma:get_expectation_out_of_TV} on the one hand and by definition \eqref{eq:def_phi_in_proof_equivalence_perfect_joint_realism} of $\phi$ and Fubini's theorem on the other hand:
\begin{IEEEeqnarray}{rCl}
\big\|P'_{\scalebox{0.6}{$X^n, Z^n, W, Y^n$}} - P_{\scalebox{0.6}{$X^n, Z^n, W, Y^n$}}\big\|_{TV}
&\leq& \int \mathbf{1}_{\mathcal{X}^+_{z^n}} [1 - \theta^n(y^n)] dP_{Y^n,Z^n}(y^n,z^n)\nonumber\\
&=&\int \int \mathbf{1}_{\mathcal{X}^+_{z^n}} [1 - \theta^n(y^n)] dP_{Y^n|Z^n{=}z^n} dP_{Z^n}(z^n)\nonumber\\
&=&\int \big\|P_{Y^n|Z^n{=}z^n} - p_{Y^n|Z^n{=}z^n} \big\|_{TV} dP_{Z^n}(z^n)\nonumber\\
&=& \big\|P_{Y^n,Z^n} - p_{X,Z}^{\otimes n} \big\|_{TV}.\nonumber
\end{IEEEeqnarray}
Therefore, 
by the triangle inequality and assumptions \eqref{eq:perception_assumption_theorem_to_perfect_joint_realism} and \eqref{eq:total_variation_Q_P_on_Y_assumption_theorem_to_perfect_joint_realism} we obtain
\begin{IEEEeqnarray}{c}
    \big\|P'_{\scalebox{0.6}{$X^n, Z^n, W, Y^n$}} - P_{\scalebox{0.6}{$X^n, Z^n, W, Y^n$}}\big\|_{TV} \leq \varepsilon_2 + \varepsilon_3.\nonumber
\end{IEEEeqnarray}
Therefore, by Lemma \ref{lemma:TV_joint_to_TV_marginal} 
and the triangle inequality we obtain
\begin{IEEEeqnarray}{c}
    \big\|\hat{P}'_{X^n,Y^n} - \hat{P}_{X^n,Y^n}\big\|_{TV} \leq \varepsilon_2 + \varepsilon_3.
    \nonumber
\end{IEEEeqnarray}
By applying again the triangle inequality together with assumption \eqref{eq:total_variation_empirical_Q_P_assumption_theorem_to_perfect_joint_realism} we obtain
\begin{IEEEeqnarray}{c}
    \big\|\hat{P}'_{X^n,Y^n} - \hat{Q}_{X^n,Y^n}\big\|_{TV} \leq \varepsilon_2 + \varepsilon_3 + \varepsilon_4. \IEEEeqnarraynumspace \label{eq:TV_empirical_on_X_and_Y_P_2_Q_in_proof_equivalence_perfect_joint_realism}
\end{IEEEeqnarray}
We then use the following standard coupling lemma (see, e.g., \cite[Chapter~I, Theorem~5.2]{1992Coupling_Book}):
from \eqref{eq:TV_empirical_on_X_and_Y_P_2_Q_in_proof_equivalence_perfect_joint_realism} there exists a quadruple $(X, Y, \Tilde{X}, \Tilde{Y})$ with distribution denoted by $\rho$ having marginals $\rho_{X,Y}
\equiv
\hat{P}'_{X^n,Y^n}$ and $\rho_{\Tilde{X},\Tilde{Y}}
\equiv
\hat{Q}_{X^n,Y^n}$ and such that $\rho((X, Y)\neq (\Tilde{X}, \Tilde{Y})) \leq \varepsilon_2 + \varepsilon_3 + \varepsilon_4.$ We then have
\begin{IEEEeqnarray}{rCl}
\mathbb{E}_{P'}[d(X^n, Y^n)] 
=
\mathbb{E}_{\hat{P}'_{X^n,Y^n}}[d(X, Y)]
&=& \mathbb{E}_{\rho}[d(X, Y)]\nonumber\\*
&=& \mathbb{E}_{\rho}\Big[d(\Tilde{X}, \Tilde{Y})\substack{\scalebox{1.0}{$\mathbf{1} \qquad \qquad \quad$}  \\ (X, Y) = (\Tilde{X}, \Tilde{Y})}\Big] \nonumber \\*
&&+ \mathbb{E}_{\rho}\Big[d(X, Y)\substack{\scalebox{1.0}{$\mathbf{1} \qquad \qquad \quad$}  \\ (X, Y) \neq (\Tilde{X}, \Tilde{Y})}\Big] \nonumber\\*
&\leq& \mathbb{E}_{\hat{Q}_{X^n,Y^n}}[d(X, Y)] + 
\mathbb{E}_{\rho}\Big[d(X, Y)\substack{\scalebox{1.0}{$\mathbf{1} \qquad \qquad \quad$}  \\ (X, Y) \neq (\Tilde{X}, \Tilde{Y})}\Big]\nonumber\\*
&\leq& \mathbb{E}_{Q}[d(X^n, Y^n)] + 
\sup_{\mathbb{P}_{X,Y,\xi}}\mathbb{E}\Big[d(X, Y)
\cdot \xi\Big], \nonumber
\end{IEEEeqnarray} where the supremum is over
all distributions $\mathbb{P}_{X,Y,\xi}$ on $\mathcal{X}^2 \times \{0,1\}$ satisfying $\mathbb{P}_X \equiv \mathbb{P}_Y \equiv p_X$ and $\mathbb{P}(\{\xi=1\}) \leq \varepsilon_2 + \varepsilon_3 + \varepsilon_4,$
and the last inequality holds because $P'$ satisfies perfect realism $P'_{Y^n} = p_X^{\otimes n}$ and from \eqref{eq:def_P_2_in_proof_equivalence_perfect_joint_realism} we have $\hat{P}'_{X^n} \equiv \hat{P}_{X^n} \equiv p_X.$ We conclude by assumption \eqref{eq:distortion_assumption_theorem_to_perfect_joint_realism}.

\section{Further justifications}

\subsection{Use of the soft covering lemma when the side information is available only at the decoder}\label{app:subsec:use_of_soft_covering_marginal_one_sided}

\subsubsection{Statement}
\hfill\\
\indent We state the soft covering lemma for general alphabets as it appears in \cite{2013PaulCuffDistributedChannelSynthesis}, but specialize it to memoryless distributions and to the case of finite mutual information: 
\begin{lemma}\label{lemma:soft_covering_basic}\cite[Corollary~VII.4]{2013PaulCuffDistributedChannelSynthesis} Let $\mathcal{V}$ be an alphabet and $\rho_V$ a distribution on $\mathcal{V}.$ Let $R$ be a non-negative real number and let $(k_n)_{n {\geq} 1}$ be a sequence of positive integers satisfying $k_n \; \substack{\raisebox{-4pt}{$\sim$} \\ \scalebox{0.5}{$n$$\to$$\infty$}} \; 2^{nR}.$ For every positive integer $n$ let $\mathcal{E}^{(n)}$ be a randomly generated collection of $k_n$ mutually independent sequences in $\mathcal{V}^n$ each drawn according to $\rho_V^{\otimes n}.$ Its distribution is denoted $\Gamma_{\mathcal{E}^{(n)}}.$ The sequences are indexed by some set $\mathcal{I}$ of size $k_n$ and in a realization $e^{(n)}$ of $\mathcal{E}^{(n)}$ the sequence with index $i$ is denoted by $v^n(e^{(n)}, i).$ A memoryless channel $(\rho_{W|V\text{=}v})_{v \in \mathcal{V}}$ induces an output distribution defined as
\begin{IEEEeqnarray}{c}
\text{$\Gamma$}
_{
\scalebox{0.6}{$
\mathcal{E}^{(n)},
I, V^n, W^n 
$}
}  
:=
\Gamma_{\mathcal{E}^{(n)}}
\cdot
p^{\mathcal{U}}_{[k_n]}
\cdot
\substack{\scalebox{1.2}{$\mathbf{1} \qquad \quad \quad $} \\ V^n = v^n(e^{(n)}, i)} 
\cdot
\prod_{t=1}^n \rho \substack{\scalebox{0.8}{$ (w_t) $} \\ \scalebox{0.55}{$W|V{=}v_t$} } \ .
\end{IEEEeqnarray}
Assume that $I_{\rho}(V;W)<\infty.$ If $R > \overline{I}_{\rho}(V;W)$ then
\begin{IEEEeqnarray}{c}
\mathbb{E}_{\mathcal{E}^{(n)}}\big[\|\Gamma_{W^n|\mathcal{E}^{(n)}} - \rho_{W}^{\otimes n}\|_{TV}\big] \underset{n \to \infty}{\longrightarrow} 0,
\label{eq:in_basic_soft_covering_lemma_vanishing_TV}
\end{IEEEeqnarray}
where $\rho_W$ is the marginal of $\rho_{V,W} = \rho_V \cdot \rho_{W|V}.$
\end{lemma}
\vspace{5pt}

Note that the result
is stated with the spectral sup-mutual information:\begin{equation*}
    \overline{I}(V;W) = \inf\Big\{\tau: \ \rho_{V,W}^{\otimes n} \Big( (1/n) \  i_{\rho_{V,W}^{\otimes n}}(V^n,W^n) > \tau\Big) \to 0\Big\},
\end{equation*}where $i_{p_{V,W}^{\otimes n}}$ is as in Proposition \ref{prop:information_density_exists}.
Since it is assumed that
$I_\rho(V;W)<\infty,$ and hence $I_{\rho^{\otimes n}}(V;W)<\infty$ by the chain rule (Proposition \ref{prop:chain_rule}), then by Proposition \ref{prop:information_density_exists} $i_{p_{Z,V}}$ and $i_{p_{Z,V}^{\otimes n}}$ are well-defined and $i_{p_{Z,V}}$ is $p_{Z,V}$-integrable. Hence, the law of large numbers applies: 
by the multiplicativity of the Radon-Nikodym derivative for product distributions we have 
$$\forall (z^n,v^n), \quad i_{p_{Z,V}^{\otimes n}}(z^n,v^n) = \sum_{t=1}^n i_{p_{Z,V}}(z_t,v_t),$$ 
thus $\overline{I}_\rho(V;W)=I_\rho(V,W).$
\vspace{10pt}
\subsubsection{Use of the lemma}
\hfill\\
\indent We proceed as in \cite{2013PaulCuffDistributedChannelSynthesis} for the proof of our Claim \ref{claim:use_of_soft_covering_marginal_one_sided}. We first use Lemma \ref{lemma:soft_covering_basic} with $\rho = p,$ $\mathcal{E}^{(n)}=\mathcal{C}^{(n)},$ $k_n = \lfloor 2^{n(R+\varepsilon)}\rfloor \times \lfloor 2^{nR'}\rfloor \times \lfloor 2^{nR_c}\rfloor,$ $I=(M,M',J),$ $W^n=Y^n$ and $R$ replaced by $R+\varepsilon+R'+R_c.$ 
The corresponding output distribution is exactly $Q^{(1)}_{\mathcal{C}^{(n)},M,M',J, V^n, Y^n},$ by definition \eqref{eq:def_Q_1_one_sided} of the latter. Indeed, from \eqref{eq:def_Q_1_one_sided} distribution $Q^{(1)}$ satisfies $(\mathcal{C}^{(n)},M,M',J)-V^n-Y^n$ and from \eqref{eq:Q_1_averaged_over_codebook_gives_p} we have $Q^{(1)}_{V^n, Y^n} \equiv p_{V,Y}^{\otimes n}.$ 
Note that,
from Lemma \ref{lemma:get_expectation_out_of_TV}, the expectation in \eqref{eq:in_basic_soft_covering_lemma_vanishing_TV} does not depend on the choice of transition kernel $\Gamma_{W^n|\mathcal{E}^{(n)}}.$
Thus, from \eqref{eq:sum_rate_large_enough_for_Y} and Lemma \ref{lemma:soft_covering_basic} together with the realism property $p_Y \equiv p_X$ from \eqref{eq:def_D_D} we obtain \eqref{eq:TV_Q_1_Y_one_sided}.

Second, we use Lemma \ref{lemma:soft_covering_basic} as follows. We fix a positive integer $j.$ For every $n$ such that $2^{nR_c}\geq j,$ we set $\mathcal{E}^{(n)}$ to be the sub-codebook of $\mathcal{C}^{(n)}$ corresponding to $j,$ denoted $\mathcal{C}^{(n)}_j,$ and set $\rho = p,$ $k_n = \lfloor 2^{n(R+\varepsilon)}\rfloor \times \lfloor 2^{nR'}\rfloor.$ We further set $I=(M,M',j),$ $W=(X,Z)$ and replace $R$ in the lemma by $R+\varepsilon+R'.$ The corresponding output distribution is exactly $Q^{(1)}_{\mathcal{C}^{(n)}_{j},M,M',J, V^n, X^n, Z^n |J{=}j},$ by definition \eqref{eq:def_Q_1_one_sided} of the latter. Indeed, from \eqref{eq:def_Q_1_one_sided} distribution $Q^{(1)}$ satisfies $(\mathcal{C}^{(n)},M,M',J)-V^n-(X^n,Z^n)$ and from \eqref{eq:Q_1_averaged_over_codebook_gives_p} we have $Q^{(1)}_{V^n,X^n, Z^n} \equiv p_{V,X,Z}^{\otimes n}.$
Thus, from \eqref{eq:sum_rate_large_enough_for_X_and_Z} and Lemma \ref{lemma:soft_covering_basic} we have
\begin{equation}\label{eq:use_of_soft_covering_TV_Q_1_X_Z}
    \forall j, \quad \mathbb{E}_{\mathcal{C}^{(n)}_j}\big[\|Q^{(1)}_{X^n, 
Z^n|\mathcal{C}^{(n)}_{j}, J{=}j} - p_{X, Z}^{\otimes n}\|_{TV}\big] \underset{n \to \infty}{\longrightarrow} 0.
\end{equation}

Moreover, since $Q^{(1)}_{\mathcal{C}^{(n)}, J} \equiv \mathbb{Q}_{\mathcal{C}^{(n)}}p^{\mathcal{U}}_{[2^{nR_c}]},$ then by Lemma \ref{lemma:get_expectation_out_of_TV} the total variation distance between\\
$Q^{(1)}_{\mathcal{C}^{(n)}, J, X^n, Z^n}$ and $\mathbb{Q}_{\mathcal{C}^{(n)}}p^{\mathcal{U}}_{[2^{nR_c}]}p_{X, Z}^{\otimes n}$ rewrites as
\begin{IEEEeqnarray}{c}
    \sum_{j=1}^{\lfloor 2^{nR_c} \rfloor} \tfrac{\text{\normalsize 1}}{\lfloor 2^{nR_c} \rfloor} \mathbb{E}_{\mathcal{C}^{(n)}_j}\big[\|Q^{(1)}_{X^n, Z^n|\mathcal{C}^{(n)}_{j}, J{=}j} - p_{X, Z}^{\otimes n}\|_{TV}\big]. \nonumber
\end{IEEEeqnarray}
Finally, by construction of $\mathcal{C}^{(n)}$ the distribution of sub-codebook $\mathcal{C}^{(n)}_j=(V^n(\mathcal{C}^{(n)}, m,m',j))_{m,m'}$ is independent of $j.$ Hence, all expectations in the above sum are identical. Using \eqref{eq:use_of_soft_covering_TV_Q_1_X_Z} we obtain \eqref{eq:TV_Q_1_J_X_Z_one_sided}.

\subsection{Use of the soft covering lemma when the side information is available at both encoder and decoder}\label{app:subsec:use_of_soft_covering_marginal_two_sided}

\subsubsection{Statement}
\begin{lemma}\label{lemma:soft_covering_with_side_info}\cite[Corollary~VII.5]{2013PaulCuffDistributedChannelSynthesis} Let $\mathcal{U}, \mathcal{V}$ be alphabets, with $\mathcal{U}$ discrete, and $\rho_{U,V}$ a distribution on $\mathcal{U} \times \mathcal{V}.$ Let $R$ be a positive real number and let $(k_n)_{n \geq 1}$ be a sequence of positive integers satisfying $k_n$ $\substack{\raisebox{-4pt}{$\sim$} \\ \scalebox{0.5}{$n$$\to$$\infty$}}$ $2^{nR}.$ For every positive integer $n$ and every sequence $u^n$ in $\mathcal{U}^n,$ let $\mathcal{E}_{u^n}$ be a randomly generated collection of $k_n$ mutually independent sequences in $\mathcal{V}^n,$ each drawn according to $\prod_{t=1}^n \rho_{V|U{=}u_t}.$ The sequences are indexed by some set $\mathcal{I}$ of size $k_n$ and in a realization $e_{u^n}$ of $\mathcal{E}_{u^n},$ the sequence with index $i$ is denoted by $v^n(e_{u^n}, i).$ We also assume that for every $n,$ the collections $\mathcal{E}_{u^n}$ are mutually independent. We denote this random family by $\mathcal{E}_{\mathcal{U}^n}$ and a realization by $e_{\mathcal{U}^n}.$
Its distribution is denoted $\Gamma_{\mathcal{E}_{\mathcal{U}^n}}.$
A memoryless channel $(\rho_{W|U{=}u,V{=}v})_{u \in \mathcal{U}, v \in \mathcal{V}}$ induces an output distribution defined as
    $$\text{$\Gamma$}_{
    \scalebox{0.6}{$\mathcal{E}_{\mathcal{U}^n}, I,U^n, V^n, W^n
    $}
    }
    := \Gamma_{\mathcal{E}_{\mathcal{U}^n}} \cdot
    p^{\mathcal{U}}_{[k_n]}
    \cdot \rho_U^{\otimes n}
    \cdot \substack{\scalebox{1.2}{$\mathbf{1} \qquad \qquad $} \\ V^n {=} v^n(e_{u^n}, i)} 
    \cdot
    \prod_{t=1}^n 
    \rho_{
    \scalebox{0.55}{$W|U{=}u_t,V{=}v_t$} } \ .$$
    If $R + H_{\rho}(U) > I_{\rho}(U,V;W)$ then
    $$\mathbb{E}_{\mathcal{E}_{\mathcal{U}^n}}\big[\|\Gamma_{W^n|\mathcal{E}_{\mathcal{U}^n}} - \rho_{W}^{\otimes n}\|_{TV}\big] \underset{n \to \infty}{\longrightarrow} 0,$$
    where $\rho_W$ is the marginal of $\rho_{U,V,W} = \rho_{U,V} \cdot \rho_{W|U,V}.$
\end{lemma}

\vspace{10pt}
\subsubsection{Use of the lemma}
\hfill

We proceed as in the case of marginal realism (Appendix \ref{app:subsec:use_of_soft_covering_marginal_one_sided}) for the proof of our Claim \ref{claim:use_of_soft_covering_marginal_two_sided}.
First, we use Lemma \ref{lemma:soft_covering_with_side_info} with $\mathcal{U}=\mathcal{Z},$ $\mathcal{E}^{(n)}_{\mathcal{U}^n}=\mathcal{C}^{(n)}_{\mathcal{Z}^n},$ $k_n = \lfloor 2^{n(R+\varepsilon)}\rfloor \times \lfloor 2^{nR_c}\rfloor,$ $I=(M,J),$ $\mathcal{W}=\mathcal{X},$ $\rho_{U,V,W} = p_{Z,V,Y},$ and $R$ replaced by $R+\varepsilon+R_c.$ 
The corresponding output distribution $\Gamma_{\mathcal{E}_{\mathcal{U}^n}, I,U^n, V^n, W^n}$ is exactly $Q^{(1)}_{\mathcal{C}^{(n)}_{\mathcal{Z}^n},M,J, Z^n, V^n, Y^n},$ by definition \eqref{eq:def_Q_1_two_sided} of the latter. Indeed, from \eqref{eq:def_Q_1_two_sided} distribution $Q^{(1)}$ satisfies $(\mathcal{C}^{(n)}_{\mathcal{Z}^n},M,J)-V^n-Y^n$ and from \eqref{eq:Q_1_two_sided_averaged_over_codebook_gives_p} we have $Q^{(1)}_{V^n, Y^n} \equiv p_{V,Y}^{\otimes n}.$
Thus, from \eqref{eq:introducing_R_c_two_sided}, \eqref{eq:replace_conditional_entropy_by_entropy_in_proof} and Lemma \ref{lemma:soft_covering_with_side_info} together with the realism property $p_Y \equiv p_X$ from \eqref{eq:def_D_D} we obtain \eqref{eq:TV_Q_1_Y_two_sided}.

Second, we use Lemma \ref{lemma:soft_covering_with_side_info} as follows. We fix a positive integer $j.$ For every $n$ such that $2^{nR_c}\geq j$ we set $\mathcal{E}^{(n)}_{\mathcal{U}^n}$ to be the sub-codebook of $\mathcal{C}^{(n)}_{\mathcal{Z}^n}$ corresponding to $j,$ denoted $\mathcal{C}^{(n)}_{\mathcal{Z}^n,j},$ and set $\rho = p,$ $k_n = \lfloor 2^{n(R+\varepsilon)}\rfloor.$ We further set $I=(M,j),$ $W=(X,Z).$ The corresponding output distribution $\Gamma_{\mathcal{E}_{\mathcal{U}^n}, I,U^n, V^n, W^n}$ is exactly $Q^{(1)}_{\mathcal{C}^{(n)}_{\mathcal{Z}^n,j},M,J, V^n, X^n, Z^n | J{=}j},$ thus from \eqref{eq:introducing_R_two_sided}, \eqref{eq:rewrite_conditional_info_with_entropy_in_proof} and Lemma \ref{lemma:soft_covering_with_side_info} we have
\begin{equation}\label{eq:use_of_soft_covering_with_side_info_TV_Q_1_X_Z}
    \forall j, \quad \mathbb{E}_{\mathcal{C}^{(n)}_{\mathcal{Z}^n,j}}\big[\|Q^{(1)}_{X^n, 
Z^n|\mathcal{C}^{(n)}_{\mathcal{Z}^n,j}, J{=}j} - p_{X, Z}^{\otimes n}\|_{TV}\big] \underset{n \to \infty}{\longrightarrow} 0.
\end{equation}
Moreover, since $Q^{(1)}_{\mathcal{C}^{(n)}_{\mathcal{Z}^n}, J} \equiv \mathbb{Q}_{\mathcal{C}^{(n)}_{\mathcal{Z}^n}}p^{\mathcal{U}}_{[2^{nR_c}]},$ then by Lemma \ref{lemma:get_expectation_out_of_TV} the total variation distance between\\
$Q^{(1)}_{\mathcal{C}^{(n)}_{\mathcal{Z}^n}, J, X^n, Z^n}$ and $\mathbb{Q}_{\mathcal{C}^{(n)}_{\mathcal{Z}^n}}p^{\mathcal{U}}_{[2^{nR_c}]}p_{X, Z}^{\otimes n}$ rewrites as
\begin{IEEEeqnarray}{c}
    \sum_{j=1}^{\lfloor 2^{nR_c} \rfloor} \tfrac{\text{\normalsize 1}}{\lfloor 2^{nR_c} \rfloor} \mathbb{E}_{\mathcal{C}^{(n)}_{\mathcal{Z}^n,j}}\big[\|Q^{(1)}_{X^n, Z^n|\mathcal{C}^{(n)}_{\mathcal{Z}^n,j}, J{=}j} - p_{X, Z}^{\otimes n}\|_{TV}\big]. \nonumber
\end{IEEEeqnarray}
Finally, by construction of $\mathcal{C}^{(n)}_{\mathcal{Z}^n}$ the distribution of sub-codebook $\mathcal{C}^{(n)}_{\mathcal{Z}^n,j}=(V^n(\mathcal{C}^{(n)}_{\mathcal{Z}^n}, m,j))_{m}$ is independent of $j.$ Hence, all expectations in the above sum are identical. Using \eqref{eq:use_of_soft_covering_with_side_info_TV_Q_1_X_Z} we obtain \eqref{eq:TV_Q_1_J_X_Z_two_sided}.

\subsection{Computations in the Gaussian case}\label{app:subsec:Gaussian}

\subsubsection{Proof of Claim \ref{claim:uniform_integrability_if_MSE}}
\label{app:subsubsec_proof_claim:uniform_integrability_if_MSE}
\hfill\\
\indent Let $p$ be a distribution on $\mathbb{R}$ with finite fourth moment. Let $\mathbb{P}_{X,Y,\xi}$ be a distribution on $\mathbb{R}^2 \times \{0,1\}$ such that $\mathbb{P}_X \equiv \mathbb{P}_Y \equiv p$ and $\mathbb{P}(\{\xi{=}1\}) \leq \tau.$ Then, by the Cauchy-Schwarz inequality (applied several times), we have:
\begin{IEEEeqnarray}{rCl}
    \mathbb{E}_{\mathbb{P}}[(X-Y)^2\xi] &=& \mathbb{E}_{\mathbb{P}}[X^2\xi]+\mathbb{E}_{\mathbb{P}}[Y^2\xi] - 2\mathbb{E}_{\mathbb{P}}[(X\xi)(Y\xi)] \nonumber \\*
    &\leq& \mathbb{E}_{\mathbb{P}}[X^2\xi]+ \mathbb{E}_{\mathbb{P}}[Y^2\xi]+ 2\sqrt{\mathbb{E}_{\mathbb{P}}[X^2\xi] \mathbb{E}_{\mathbb{P}}[Y^2\xi]}\nonumber \\*
    &\leq& \big(\mathbb{E}_p[X^4]\mathbb{P}\{\xi{=}1\}\big)^{1/2} +  \big(\mathbb{E}_p[Y^4]\mathbb{P}\{\xi{=}1\}\big)^{1/2} + 2\big(\mathbb{E}_p[X^4]\mathbb{P}\{\xi{=}1\} \mathbb{E}_p[Y^4]\mathbb{P}\{\xi{=}1\}\big)^{1/4}\nonumber\\*
    &\leq&
    4\sqrt{\tau\mathbb{E}_p[X^4]}.\nonumber
\end{IEEEeqnarray}
This concludes the proof since $p$ has a finite fourth moment.

\vspace{10pt}
\subsubsection{Proof of Claim \ref{claim:Normal_achievability}}
\label{app:subsubsec_proof_claim:Normal_achievability}
\begin{align}
    \mathbb{E}[(X-Y)^2] & = \mathbb{E}[(X - \rho V + \rho V - \tilde{\rho} V
    + \tilde{\rho} V - Y)^2] \nonumber\\
    & = \mathbb{E}[(X - \rho V)^2] + \mathbb{E}[(\rho V - \tilde{\rho} V)^2]
    + \mathbb{E}[(\tilde{\rho} V - Y)^2] \nonumber\\
    & = 1 - \rho^2 + (\rho - \tilde{\rho})^2 + 1 - \tilde{\rho}^2 \nonumber\\
    & = 2 - 2 \rho \tilde{\rho} \nonumber\\
    & = 2 - 2 \rho \sqrt{1 - 2^{-2 R_c} (1-\rho^2)} \nonumber\\
    & = \Delta \nonumber
\end{align}
\begin{align}
    I(X;V) & = \frac{1}{2} \log \frac{1}{1 - \rho^2} \nonumber\\
    I(Y;V) & = \frac{1}{2} \log \frac{1}{1 - \tilde{\rho}^2}
    = \frac{1}{2} \log \frac{1}{2^{-2 R_c} (1 -\rho^2)}
    = R_c + \frac{1}{2} \log \frac{1}{1 - \rho^2}.\nonumber
\end{align}

\subsubsection{Proof of Claim \ref{claim:Normal_converse_entropy_maximizing}}
\label{app:subsubsec_proof_claim:Normal_converse_entropy_maximizing}
\hfill\\
\indent
Although the alphabet $\mathcal{V}$ of $V$ is finite --- from the definition of $\mathcal{D}^{(m)}$ (Eq. \ref{eq:def_D}) ---,
we provide a general proof for a Polish alphabet $\mathcal{V},$ to which we refer in other parts of the paper.
A sufficient condition is that $I(X;V) \, {<} \, \infty.$
% and $p_X$ is a Normal distribution.
Since $I(X;V) \, {<} \, \infty,$ then
from Proposition \ref{prop:information_density_exists}, $p_X \times p_V$ dominates $p_{X,V}.$
Therefore, from Corollary \ref{cor:conditional_density_exists_and_gives_regular_kernel},
there exists a regular transition kernel $p_{X|V}$ such that
for every $v\in\mathcal{V},$ $p_X$ dominates $p_{X|V{=}v},$
and
$(x,v)\mapsto dp_{X|V{=}v}/dp_X (x)$ is measurable.
Thus, for any $v,$ the Lebesgue measure $\mu$ dominates $p_{X|V{=}v},$
and
$(x,v)\mapsto dp_{X|V{=}v}/d\mu (x)$ is measurable.
Hence,
from Theorem \ref{thm:generalized_fubini} for the joint distribution of $(V,X-\mathbb{E}[X|V]),$
the distribution of $X-\mathbb{E}[X|V]$
has a density with respect to the Lebesgue measure.
Therefore, from Proposition \ref{prop:information_density_exists}, and since the density of $p_X$ is non-zero $p_X$-almost everywhere, we have
\begin{align}
    R + \varepsilon
    \ge I(X;V)
    &
    = h(X) -
    \mathbb{E}_{v \sim p_V}[ h(p_{X|V{=}v})]
    \nonumber\\
    &
    =
    h(X) -
    \mathbb{E}_{v \sim p_V}[h(p_{X-E[X|V]|V{=}v})]
    \nonumber\\
    & \ge \frac{1}{2} \log (2 \pi e)
    -
    \mathbb{E}
    \Big[\frac{1}{2} \log \Big(2 \pi e \mathbb{E}\big[(X - \mathbb{E}[X|V])^2 \big| V
    \big]\Big)\Big]
    \nonumber\\
    &
    \ge
    \frac{1}{2} \log \frac{1}{\mathbb{E}[(X - \mathbb{E}[X|V])^2]} \label{eq:using_Jensen_on_log}\\
    & = \frac{1}{2} \log \frac{1}{1 - \mathbb{E}[\mathbb{E}[X|V]^2]} \nonumber\\
    & = \frac{1}{2} \log \frac{1}{1 - \rho^2},\nonumber
\end{align}
where \eqref{eq:using_Jensen_on_log} follows from Jensen's inequality
for $-\log.$
The second inequality of Claim \ref{claim:Normal_converse_entropy_maximizing} is obtained similarly.
\vspace{10pt}
\subsubsection{Proof of Claim \ref{claim:Normal_converse_distortion}}
\label{app:subsubsec_proof_claim:Normal_converse_distortion}
\begin{align}
    \Delta + \varepsilon
    \ge \mathbb{E}[(X - Y)^2]
    &
    = \mathbb{E}[(X - \mathbb{E}[X|V])^2] + \mathbb{E}[(\mathbb{E}[X|V] - \mathbb{E}[Y|V])^2] + \mathbb{E}[(\mathbb{E}[Y|V] - Y)^2] \nonumber\\
    & \ge (1 - \rho^2) + (1- \tilde{\rho}^2) + \left(\sqrt{\mathbb{E}[\mathbb{E}[X|V]^2]} -
    \sqrt{\mathbb{E}[\mathbb{E}[Y|V]^2]}\right)^2 \nonumber\\
    & = (1 - \rho^2) + (1- \tilde{\rho}^2)
             + (\rho - \tilde{\rho})^2 \nonumber\\
    & = 2 - 2 \rho \tilde{\rho},\nonumber
\end{align}
where the second inequality follows from Cauchy-Schwarz.
\vspace{10pt}
\subsubsection{Proof of Claim \ref{claim:Normal_converse_continuity_monotonicity_argument}}
\label{app:subsubsec_proof_claim:Normal_converse_continuity_monotonicity_argument}
\hfill\\
\indent Define the function
\begin{IEEEeqnarray}{c}
\phi: [0,1) \to [0,1), \ s \mapsto s \sqrt{1 - 2^{-2 R_c}(1-s^2)},
\end{IEEEeqnarray}which is strictly increasing and bijective. Thus, its inverse $\phi^{-1}$ is continuous. Since $\Delta \in (0,2)$ and we have chosen $\varepsilon$ such that $\Delta+\varepsilon \in (0,2),$ we can define $\rho^*=\phi^{-1}(1 - \Delta/2)$ and $\rho^*_\varepsilon=\phi^{-1}(1 - (\Delta+\varepsilon)/2).$ Since $(\rho_{\varepsilon},\Tilde{\rho}_\varepsilon)$ satisfies the constraint of \eqref{eq:in_proof_Normal_optim_problem} and the equality \eqref{eq:in_proof_Normal_form_of_optimal_rho_tilde}, we have $\phi(\rho^*_\varepsilon) \leq \phi(\rho_\varepsilon),$ which implies $\rho_\varepsilon \geq \rho^*_\varepsilon.$ Therefore, from \eqref{eq:in_proof_Nornal_final_ineq_R}, we have
\begin{IEEEeqnarray}{c}
R +3\varepsilon \geq \frac{1}{2} \log \frac{1}{1 - (\rho^*_\varepsilon)^2}.\nonumber
\end{IEEEeqnarray}
This being true for arbitrarily small $\varepsilon,$ then by continuity of $\phi^{-1},$ we obtain the desired inequality:
\begin{IEEEeqnarray}{c}
R \geq \frac{1}{2} \log \frac{1}{1 - (\rho^*)^2}.\nonumber
\end{IEEEeqnarray}

\vspace{10pt}
\subsubsection{Proof of Claim \ref{claim:choice_of_b}}
\label{app:subsubsec_proof_claim:choice_of_b}
\hfill\\
\indent Since $\rho \geq |\eta|,$ we can choose
\begin{equation}\label{eq:def_b}
    b= \sqrt{(\rho^2 - \eta^2)/(1+\eta^2\rho^2-2\eta^2)}.
\end{equation}
From \eqref{eq:def_Z_X_V}, vector $(Z,X,V)$ is Gaussian and we have $(X,Z) \sim p_{X,Z},$ the chain $Z-X-V$ and $I(Z;V)<\infty$ (since $|\eta b|<1$). Then, $Y$ has a Normal distribution with zero mean and we have $X-(Z,V)-Y.$
Hence,
$p_{X,Z,V,Z} \in \mathcal{D}^{(m)}_{D}.$
To finish the proof of Claim \ref{claim:choice_of_b}, it only remains to compute the variance of $Y,$ i.e. to prove \eqref{eq:rho_carre}.
\begin{lemma}
Let $(\Vec{x}_1, \Vec{x}_2)$ be a Gaussian vector with mean $(\Vec{\mu}_1, \Vec{\mu}_2)$ and covariance matrix \begin{equation*}
    \scalebox{0.9}{$\begin{pmatrix}
\Sigma_{11} & \Sigma_{12} \\
\Sigma_{21} & \Sigma_{12}
\end{pmatrix}$}
\end{equation*} Then, conditioned on $\Vec{x}_2=\Vec{a},$ variable $\Vec{x}_1$ is Gaussian with:
\begin{IEEEeqnarray}{c}
    \overline{\mu}_{|\Vec{a}} = \Vec{\mu}_1 + \Sigma_{12}\Sigma_{22}^{-1}(\Vec{a}-\Vec{\mu}_2)\label{eq:formula_conditiona_mean}\\
    \overline{\Sigma} = \Sigma_{11} - \Sigma_{12}\Sigma_{22}^{-1}\Sigma_{21},\label{eq:formula_conditiona_variance}
\end{IEEEeqnarray} with the covariance matrix $\overline{\Sigma}$ not depending on $\Vec{a}.$
\end{lemma}

For the centered Gaussian vector $(Z,X,V),$ we obtain
\begin{IEEEeqnarray}{c}
\text{Var}(X|Z) = 1 - \eta \cdot 1 \cdot \eta = 1-\eta^2 \text{ and}\label{eq:proved_Gaussian_Var_X_knowing_Z}\\
\mathbb{E}[X|Z,V]= 0 + \scalebox{0.9}{$\begin{pmatrix}
\eta & b
\end{pmatrix}$} \scalebox{0.9}{$\begin{pmatrix}
\tfrac{1}{1-\eta^2b^2} & \tfrac{-\eta b}{1-\eta^2b^2} \\
\tfrac{-\eta b}{1-\eta^2b^2} & \tfrac{1}{1-\eta^2b^2}
\end{pmatrix}$} \scalebox{0.9}{$\begin{pmatrix}
Z \\
V
\end{pmatrix}$}.\label{eq:gaussian_proof_conditional_mean}
\end{IEEEeqnarray}
From \eqref{eq:def_Z_X_V}, $\mathbb{E}[Z^2]=\mathbb{E}[V^2]=1$ and $\mathbb{E}[ZV]=\eta b.$ Therefore, from \eqref{eq:gaussian_proof_conditional_mean} we have
\begin{IEEEeqnarray}{rCl}
\mathbb{E}[ \mathbb{E}[X|Z,V]^2]
&=&
\dfrac{1}{(1-\eta^2 b^2)^2} \mathbb{E}\big[ [\eta(1-b^2)Z + b(1-\eta^2)V]^2 \big] \nonumber \\*
&=&
\dfrac{1}{(1-\eta^2 b^2)^2} \mathbb{E}\big[ \eta^2(1-b^2)^2 + b^2(1-\eta^2)^2 + 2\eta^2 b^2 (1-\eta^2)(1-b^2) \big]\IEEEeqnarraynumspace\label{eq:gaussian_developper_carre_rho_2}
\end{IEEEeqnarray}
Define $\lambda=1-\eta^2$ and $\tilde{\rho}=1-\rho^2.$ Then, from \eqref{eq:def_b} we have
\begin{IEEEeqnarray}{c}
    (1+\eta^2\rho^2-2\eta^2)b^2=\lambda-\tilde{\rho}, (1+\eta^2\rho^2-2\eta^2)(1-b^2)=\tilde{\rho}\lambda, \nonumber\\*
    (1+\eta^2\rho^2-2\eta^2)(1-\eta^2b^2)=\lambda^2, 1+\eta^2\rho^2-2\eta^2=-\tilde{\rho}+\lambda(1+\tilde{\rho}).\nonumber
\end{IEEEeqnarray}
Therefore, by multiplying numerator and denominator by $(1+\eta^2\rho^2-2\eta^2)^2$ in \eqref{eq:gaussian_developper_carre_rho_2} we obtain
\begin{IEEEeqnarray}{rCl}
\mathbb{E}[ \mathbb{E}[X|Z,V]^2] 
&=&
\dfrac{1}{\lambda^4}\Big[(1-\lambda)\tilde{\rho}^2\lambda^2+(\lambda-\tilde{\rho})\big[-\tilde{\rho}+\lambda(1+\tilde{\rho})\big]\lambda^2+2(1-\lambda)(\lambda-\tilde{\rho})\lambda\tilde{\rho}\lambda\Big] \nonumber \\*
&=& \dfrac{1}{\lambda^2}\Big[(1-\lambda)\tilde{\rho}^2+\big[\tilde{\rho}^2-\lambda\tilde{\rho}(2+\tilde{\rho})+\lambda^2(1+\tilde{\rho})\big]+2\tilde{\rho}\big[-\tilde{\rho}+(1+\tilde{\rho})\lambda-\lambda^2\big]\Big]\nonumber\\*
&=& \dfrac{1}{\lambda^2}\Big[1\cdot(\tilde{\rho}^2 + \tilde{\rho}^2 - 2\tilde{\rho}^2)+\lambda \cdot \big[-\tilde{\rho}^2-\tilde{\rho}(2+\tilde{\rho})+2\tilde{\rho}(1+\tilde{\rho})\big]
+\lambda^2\cdot\big[1+\tilde{\rho}-2\tilde{\rho}\big]\Big] \;
\nonumber\\*
&=& 1-\tilde{\rho} = \rho^2.\label{eq:end_of_long_computation_jointly_Gaussian}
\end{IEEEeqnarray}
This concludes the proof of \eqref{eq:rho_carre} and Claim \ref{claim:choice_of_b}. 
\vspace{10pt}
\subsubsection{Proof of Claim \ref{claim:Y_MMSE_is_a_good_choice}}
\label{app:subsubsec_proof_claim:Y_MMSE_is_a_good_choice}
\hfill\\
\indent We compute $I(X;V|Z)$ and $\mathbb{E}[d(X,Y)].$ By translation invariance of differential entropy we have \begin{equation*}
    h(X|Z,V) = h\big(X-\mathbb{E}[X|Z,V]\big|Z,V\big).
\end{equation*}Since uncorrelated Gaussian vectors are independent, we obtain $h(X|Z,V) = h(X-\mathbb{E}[X|Z,V]).$ Since $\mathbb{E}[X^2]=1$ we obtain \begin{equation*}
    \text{Var}\big(X-\mathbb{E}[X|Z,V]\big) = 1 - \mathbb{E}[ \mathbb{E}[X|Z,V]^2].
\end{equation*}The mutual information computation is concluded by using \eqref{eq:proved_Gaussian_Var_X_knowing_Z} and \eqref{eq:end_of_long_computation_jointly_Gaussian}. Also,
\begin{IEEEeqnarray}{rCl}
    \mathbb{E}[d(X,Y)]
    =
    \mathbb{E}[(X-\rho^{-1}\mathbb{E}[X|Z,V])^2] 
    =
    1 - 2\rho^{-1}\mathbb{E}\big[ \mathbb{E}[X|Z,V]^2 \big] + \rho^{-2}\mathbb{E}\big[ \mathbb{E}[X|Z,V]^2 \big]
    &=& 1-2\rho+1\nonumber\\*
    &=& \Delta.\label{eq:distortion_is_Delta}
\end{IEEEeqnarray}
Finally, we show that $I(Y;V) < \infty$ when $\eta \neq 0.$ By construction, $(Y,V)$ is a Gaussian vector. Since both $Y$ and $V$ have non-zero variance (equal to $1$), we only need to prove that the covariance matrix of $(Y,V)$ is non-singular, i.e. that $\mathbb{E}[YV]<1.$ By the Cauchy-Schwarz inequality, this is true if $Y$ and $V$ are not linearly dependent. From \eqref{eq:gaussian_proof_conditional_mean}, $Y$ is a linear combination of $Z$ and $V$ where the coefficient of $Z$ is non-zero since $|\eta|,b \in (0,1).$ 
We conclude by noticing that, by construction, $Z$ and $V$ are not linearly dependent.

\subsection{The uniform integrability constraint is non-degenerate}\label{app:subsec_example_not_satisfying_uniform_integrability}

We provide a non-degenerate example of a distribution $p_X$ and a distortion measure $d$ such that $(d,p_X)$ is not uniformly integrable.
Consider $\mathcal{X}=\mathbb{R} \times \{0,1\}$ --- which is Polish ---,
and let $p_X$ be the product of the Cauchy distribution with zero mean and scale $1$ and the Bernoulli distribution with parameter $1/2.$ Define the distortion measure
\begin{IEEEeqnarray}{c}
d: (x,y) \mapsto |x_1 - y_1| \text{ if } x_2 = y_2 = 0, \text{ and } d(x,y) = 1 \text{ otherwise.}
\end{IEEEeqnarray}
First, $(d,p_X)$ is not uniformly integrable.
Indeed, consider $Y$ distributed according to $p_X$ and independent from $X.$
Fix $\tau>0,$ and let $\xi \in\{0,1\}$ be independent from $(X,Y),$ having a Bernoulli distribution of parameter $\tau.$ Then, since $p_{X_1}$ is symmetric,
\begin{IEEEeqnarray}{c}
\mathbb{E}[d(X,Y)
\xi
]
\geq
\mathbb{E}[\xi]
\mathbb{E}[|\scalebox{0.98}{$X_1-Y_1$}|
\mathbf{1}_{X_2=Y_2=0}
]
\geq \tfrac{\tau}{4}
\mathbb{E}[|\scalebox{0.98}{$X_1-Y_1$}|
]
\geq \tfrac{\tau}{4}
\mathbb{E}[|\scalebox{0.98}{$X_1-Y_1$}|
\mathbf{1}_{X_1 Y_1 \leq 0}
]
\geq \tfrac{\tau}{8}
\mathbb{E}[|X_1|
] = \infty \nonumber.
\end{IEEEeqnarray}
This being true for any $\tau>0,$ then $(d,p_X)$ is not uniformly integrable.

We conclude by showing that this example is not degenerate, in that there exist
jointly distributed $X, Y$ such that $p_Y \equiv p_X,$ and $E[d(X,Y)] < \infty,$ with arbitrarily low $I(X;Y).$ Consider, the following coupling: fix a number $M \geq 1.$ If $|X_1| > M,$
then choose $Y_2 = 1 - X_2$ (so the bit is flipped), and $Y_1$ is drawn
from the distribution $p_{X_1}$ conditioned on $\{|X_1| > M\}.$ Otherwise, draw $Y_1$ from the
distribution $p_{X_1}$ conditioned on $\{|X_1| \leq M\},$ and draw $Y_2$ uniformly.
Then, $E[d(X,Y)] \leq 2M < \infty.$
We move to upper bounding $I(X;Y).$
Since we have the Markov chain $X_2-X_1-\mathbf{1}_{\{|X_1|>M\}}-Y_1,$ and $X_2$ is independent of $X_1,$ and $Y_2$ is independent of $(X_1,X_2,Y_1)$ knowing $\{|X_1| \leq M\},$
then
$$I(X;Y) = I(X_1;Y_1) + I(X_2;Y_1|X_1) + I(X_1,X_2;Y_2|Y_1) \leq H(\mathbf{1}_{\{|X_1|>M\}}) + 0 +Pr(|Y_1|>M)H(Y_2),$$
where the last term comes from writing conditional mutual information in integral form. This is valid when the latter is finite, by \cite[(7.32)]{2011GrayGeneralAlphabetsRelativeEntropy}. Moreover, the integrand is exactly the mutual infomration with respect to the conditional distribution, because $Y_2$ is finite, as detailed below:
\begin{IEEEeqnarray}{rCl}
I(X_1,X_2;Y_2|Y_1)
&=&
\int \ln\Big(\dfrac{dP_{Y_1,X_1,X_2,Y_2}}{d(P_{Y_1}\cdot P_{X_1,X_2|Y_1} \cdot P_{Y_2|Y_1}}(y_1,x_1,x_2,y_2)\Big)dP_{Y_1,X_1,X_2,Y_2}(y_1,x_1,x_2,y_2)\label{eq:using_integral_form}\IEEEeqnarraynumspace
\\
&=&
\int \ln\Big(\dfrac{dP_{X_1,X_2,Y_2|Y_1{=}y_1}}{d(P_{X_1,X_2|Y_1{=}y_1} \cdot P_{Y_2|Y_1{=}y_1})}(x_1,x_2,y_2)\Big)dP_{Y_1,X_1,X_2,Y_2}(y_1,x_1,x_2,y_2)\label{eq:derivative_of_conditional_distributions_is_density_of_joint}\IEEEeqnarraynumspace
\\
&=&
\int I_{P_{X_1,X_2,Y_2|Y_1{=}y_1}}(X_1,X_2;Y_2) dP_{Y_1}(y_1),\label{eq:re_using_integral_form}\IEEEeqnarraynumspace
\end{IEEEeqnarray}
where \eqref{eq:using_integral_form} follows from \cite[(7.32)]{2011GrayGeneralAlphabetsRelativeEntropy} because $I(X_1,X_2;Y_2|Y_1) < \infty;$ \eqref{eq:derivative_of_conditional_distributions_is_density_of_joint} holds because we know that the two Radon-Nykodim derivatives are equal $P_{Y_1}$-almost everywhere whenever the second exists, and because the latter is an information density, which exists when the corresponding mutual information is finite, by Proposition \ref{prop:information_density_exists};
and \eqref{eq:re_using_integral_form} follows again from \cite[(7.32)]{2011GrayGeneralAlphabetsRelativeEntropy}.
Hence,
$I(X;Y)$ can be made arbitrarily small by taking
$M$ large.

\subsection{Markov chains in converse proofs}\label{app:subsec:Markov_chains_precise_reference_for_converse}
The following results will be used during mutual information computations.
\begin{proposition}\label{prop:Markov_property_X_Z}
Let $A,B,C,D$ be subsets of $[n]$ such that $A \subseteq B,$ $C \subseteq [n] \backslash A$ and $D \subseteq [n] \backslash B.$ If $P_{X^n,Z^n,M,J,Y^n}$ is the distribution induced by a D-code, then
$$Z_A \independent (J,M,X_C,Z_D) | X_B.$$
\end{proposition}
In particular, one can choose $A = B = \{t\},$ $D = [n] \setminus \{t\}$ and $C=\emptyset.$
\begin{IEEEproof}
Since $P$ is induced by a D-code we have $J \independent (X^n, Z^n)$ and $Z^n \independent M | J, X^n.$ These imply inequalities between Shannon quantities with respect to $P$ as follows:
\begin{IEEEeqnarray}{rCl}
    I(Z_A; M, J, X_C, Z_D | X_B)
    &=& I(Z_A;Z_D) + I(Z_A; J| X_B, Z_D) + I(Z_A; M, X_C | J, X_B, Z_D) \nonumber \\*
    &\leq& 0 + I(Z_A, X_B, Z_D; J) + I(Z_A; M, X_{[n]\backslash A} | J, X_B, Z_D) \nonumber \\*
    &\leq& 0 + I(Z_A; X_{[n]\backslash A} | J, X_B, Z_D) + I(Z_A; M | J, X_B, X_{[n]\backslash A}, Z_D) \nonumber \\*
    &\leq& I(Z_A, Z_{D\cap A}, X_{B\cap A}; X_{[n]\backslash A}, X_{B \cap ([n] \backslash A)}, Z_{D \cap ([n] \backslash A)} | J) \nonumber \\*
    &&+ I(Z_A, Z_D; M | J, X^n) \nonumber \\*
    &=& 0. \nonumber
\end{IEEEeqnarray}
\end{IEEEproof}
Further similar conditional independence relations can be derived using the following lemma:

\begin{lemma}\label{lemma:basic_weak_union}
    For random variables $K,L,W$ we have $$W \independent (K,L) \text{ if and only if } W \independent K | L \text{ and } W \independent L. $$
\end{lemma}

Using standard techniques with a uniform index, Proposition \ref{prop:Markov_property_X_Z} implies the following:
\begin{corollary}\label{lemma:Markov_chain_X_T_Z_T}
Let If $P_{X^n,Z^n,M,J,Y^n}$ is the distribution induced by a D-code and $T$ be uniformly distributed on $[n]$ and independent of $(X^n,Z^n,M,J,Y^n).$ Let $(B(t))_{1\leq t \leq n},(C(t))_{1\leq t \leq n}$ be families of subsets of $[n]$ such that for all $t \in [n]$ we have $t \notin B(t)$ and $t \notin C(t).$ Then,
$$Z_T \independent (M, J, T, X_{B(T)}, Z_{C(T)}) | X_T.$$
\end{corollary}
In particular, one can choose $B(t) = \emptyset,$ and $C(t) = [n] \setminus \{t\}.$

As a direct implication of Markov chain $X^n-(M,J,Z^n)-Y^n$ we obtain:
\begin{corollary}\label{lemma:Markov_chain_Y_T_X_T}
Let $P_{X^n,Z^n,M,J,Y^n}$ be the distribution induced by a D-code or an E-D-code and $T$ be uniformly distributed on $[n]$ and independent of $(X^n,Z^n,M,J,Y^n).$ For any $t \in [n]$ let $V_t=(M,J,t,Z_{[n]\setminus\{t\}}).$ Then,
$$Y_T \independent X_T | V_T, Z_T.$$
\end{corollary}

\ifCLASSOPTIONcaptionsoff
  \newpage
\fi

\bibliographystyle{IEEEtran}

\bibliography{biblio}

\end{document}